\documentclass[a4paper,10pt,oneside,DIV15]{scrbook}
\usepackage[utf8]{inputenc}

\usepackage{float}
\usepackage{booktabs}
\usepackage{microtype}
\usepackage{commath}
\usepackage{graphicx}
\usepackage{listings}
\usepackage{longtable}
\usepackage{array}
\usepackage[T1]{fontenc}
\usepackage{lmodern}

\usepackage{algorithm}
\usepackage{algpseudocode}
\usepackage{dirtree}

\usepackage{textcomp}
\usepackage{hyperref}
\usepackage[euler]{textgreek}

\newcommand{\mb}[1]{\mathbf{#1}}
\newcommand{\atantw}[1]{\mathrm{atan2}\left(#1 \right)}
\newcommand{\asin}[1]{\mathrm{asin}\left(#1 \right)}

\newcommand{\apjs}{ApJS}

\newcommand{\aap}{A{\&}A}

\RequirePackage[strict]{changepage}
\RequirePackage{color}
\RequirePackage{framed}

\definecolor{formalshade}{rgb}{0.96,0.97,0.96}
\definecolor{darkblue}{rgb}{0.0, 0.3, 0.0}

\newenvironment{finequote}{%
\vspace*{2ex}
\begin{minipage}{\textwidth}
  \MakeFramed{\advance\hsize-\width\FrameRestore}%
  \noindent\hspace{-4.55pt}
  \begin{adjustwidth}{1.5ex}{1.5ex}%
  \vspace{0.5ex}\vspace{0.5ex}%
}
{%
  \vspace{1ex}\end{adjustwidth}\endMakeFramed%
  \end{minipage}
  \vspace*{1.5ex}
}

\definecolor{LightGray}		{gray}{0.9}
\definecolor{Gray}		{gray}{0.5}
\definecolor{DarkGray}     	{gray}{0.2}
\definecolor{Black}     	{gray}{0.0}
\definecolor{listinggray} 	{gray}{0.96}
\definecolor{DarkGreen}     	{rgb}{0.0,0.6,0.0}
\definecolor{DarkRed}     	{rgb}{0.6,0.0,0.0}
\definecolor{DarkBlue}     	{rgb}{0.0,0.0,0.6}
\definecolor{DarkCyan}     	{rgb}{0.0,0.5,0.5}
\definecolor{DarkDarkGreen}	{rgb}{0.0,0.4,0.0}

\lstdefinestyle{CA}{
        language=C,
        basicstyle=\ttfamily\footnotesize,
        keywordstyle=\ttfamily\bfseries\color{DarkBlue},
        stringstyle=\ttfamily\color{DarkRed},
        commentstyle=\ttfamily\color{DarkGreen},
        identifierstyle=\ttfamily\color{Black},
        backgroundcolor=\color{listinggray},
        emph={uint64_t, SGAIndividual, nonlinearModel, FitResult, dataIndex, Observation},
        emphstyle={\ttfamily\bfseries\color{DarkBlue}},
}

\lstdefinestyle{PY}{
        language=Python,
	emph={[1]1, 2, 3, 4, 5, 6, 7, 8, 9, 0,},
	emphstyle={[1]\color{DarkCyan}},
        basicstyle=\ttfamily\footnotesize,
        keywordstyle=\ttfamily\bfseries\color{DarkBlue},
        stringstyle=\ttfamily\color{DarkRed},
        commentstyle=\ttfamily\color{DarkGreen},
        identifierstyle=\ttfamily\color{Black},
        backgroundcolor=\color{listinggray},
	showtabs=true,
	tab={\tiny\color{Gray} >>}
}

\author{Robin Geyer}
\title{Investigation of Algorithms for Highly Nonlinear Model Fitting on Big Datasets}
\subtitle{~\\\textbf{Diplomarbeit}\\zur Erlangung des akademischen Grades Diplom-Informatiker}

\date{March 11th, 2014}

\begin{document}

\emergencystretch
\hfuzz
\vfuzz=3pt
\hbadness=10000
\vbadness=\maxdimen

\maketitle

\noindent\textbf{About this document}\\ This is the arXiv formated version of the Diploma Thesis ``Investigation of Algorithms for Highly Nonlinear Model Fitting on Big Datasets'' handed in by Robin Geyer (born October 17th, 1985 in Zschopau, Germany) at Technische Universität Dresden, Faculty of Computer Science, Center for Information Services and High Performance Computing on March 11th, 2014. Responsible Professor was Prof. Dr. Wolfgang E. Nagel, Tutors: Apl. Prof. Dr. habil. Sergei A. Klioner and Dipl. Inf. Thomas Wiliam.\\
For questions or other inquiries please contact Robin Geyer via E-Mail: \href{mailto:robin@robingeyer.de}{robin@robingeyer.de}.
\\[3ex]

\newpage

\noindent\textbf{Abstract}\\
This thesis investigates algorithms regarding their applicability for highly nonlinear model fitting on big datasets. Various mathematical methods are presented with which a model fit using the least squares criterion is possible. Special requirements regarding the processing of large data sets as a basis for such a model fit are discussed.\\

The specific example of the search for gravitational wave signals in simulated data of the ESA satellite mission Gaia is used to demonstrate how a model fit is possible, even with complex models and large amount of data. For this purpose, a highly parallel prototype of a future search software is implemented. The resulting prototype uses a hybrid algorithm which utilizes a linear search, an evolutionary algorithm and a classical iterative Gauss-Newton fit. The performance and behavior of its components are investigated in detail.\\

With the help of software presented in this work it has been possible for the first time to detect gravitational wave signals in simulated astrometric data, and to determine their parameters. Furthermore, it can be concluded from the runtime behavior of the software that such a search is also possible in real data of the Gaia mission.\\[5ex]

\noindent\textbf{Kurzfassung}\\
Diese Diplomarbeit untersucht Algorithmen auf ihre Eignung, eine stark nichtlineare Modellanpassung an große Datenmengen vorzunehmen. Es werden verschiedene mathematische Methoden vorgestellt, mit deren Hilfe eine Modellanpassung auf Grundlage der Summe der kleinsten Fehlerquadrate durchgeführt werden kann. Spezielle Anforderungen bezüglich der Verarbeitung großer Datenmengen als Grundlage für eine derartige Modellanpassung werden diskutiert.\\

Am konkreten Beispiel der Suche nach Gravitationswellensignalen in simulierten Daten der ESA-Satellitenmission Gaia wird gezeigt, wie es möglich ist, dass eine Modellanpassung auch bei komplexen Modellen und großen Datenmengen durchführbar ist. Für diesen Zweck ist ein hochparalleler Prototyp einer Suchsoftware implementiert worden. Der Prototyp nutzt einen hybriden Algorithmus aus linearer Suche, evolutionärem Algorithmus und klassischem iterativen Gauß-Newton Verfahren. Die Performance und das Verhalten der einzelnen Komponenten wurden untersucht.\\

Mit Hilfe der in dieser Arbeit vorgestellten Software ist es (zum ersten Mal) gelungen, Signale von Gravitationswellen in simulierten astrometrischen Daten zu detektieren, sowie deren Parameter zu bestimmen. Des Weiteren kann aus dem Laufzeitverhalten der Software geschlussfolgert werden, dass eine solche Suche auch in realen Daten der Gaia Mission möglich ist.

\newpage

\section*{List of Symbols}
\addcontentsline{toc}{chapter}{Symbols and Code Conventions}

The following symbols are used in the work.\\

\begin{center}
\begin{tabular}{lll}
\toprule
\textbf{Symbol}								&		& \textbf{Description}\\
\midrule
$\mb{A}$, ..., $\mb{Z}$				& ...	& Matrix\\
$\mb{I}$							& ...	& Identity matrix\\
$\mb{J}$							& ...	& Jacobian matrix\\
$\mb{H}$							& ...	& Hessian matrix\\
$\mb{E}$							& ...	& Unity matrix\\
$\mb{a}$, ..., $\mb{z}$				& ...	& Vectors\\
$\mb{e}$                                        & ...   & Unit vector\\
$a_i$								& ...	& Element $i$ from vector $\mb{a}$\\
$a_{ij}$							& ...	& Element $(i,j)$ from matrix $\mb{A}$\\
$\frac{\partial F}{\partial v}$		& ...	& Partial derivative of function $F$ with respect to variable $v$\\
$\|\mb{a}\|$						& ...	& Euclidean norm ($\sqrt{a_1^2 + ... + a_n^2}$) of vector $\mb{a}$\\
$\|\mb{a}\|_\infty$					& ...	& Maximum norm ($\max_{i=1, \dotsc, n} |a_i|$)\\
$\sim$								& ...	& Equality of distributions, or a random variable is from a distribution\\
$\approx$                                              & ... & Approximation, rounded values\\
\bottomrule
\end{tabular}
\end{center}
\section*{Pseudo-Code Conventions}

The following conventions for pseudo-code are used in this document.
\begin{itemize}
 \item Assignment is declared via the $\longleftarrow$ or $:=$ operator. The assignee (variable name) is on the left side of the arrow, the value to be assigned on the right side.
 \item Sets and arrays are assumed to be addressable. A set $A \longleftarrow \{11,22,33,44\}$ can be explicitly addressed by $A\left[i \right]$, where $i$ is a integer index. $\left[ \cdot\right]$ appended to the variable name is the index operator. In the example above $A\left[ 2\right]$ would yield $22$. The index starts with 1.
 \item It is assumed that there exists a function $\mathrm{Length}()$, which returns the number of elements in a set or array.
 \item Children of structures or classes can be accessed via the ``$.$'' operator. Let us assume that $B\longleftarrow(Params, Value)$ then $B.Params\left[i \right]$ would give the $i^{th}$ element of the $Params$ vector of $B$
\end{itemize}

\chapter{Introduction}
Modern scientific experiments produce enormous amounts of data. In most cases that data is highly complex in terms of structure and processing demands. But also the models and effects under investigation are highly complex. This situation leads to the problem that model fitting with such data and with such models is computationally very expensive.\\

One example of such an experiment is the ESA (European Space Agency) cornerstone satellite mission Gaia. It delivers high-precision astrometric, photometric and spectroscopic data of around one billion celestial objects. The raw data linked down to the Earth is around one petabyte in size and will be handled by selected processing sites \cite{gaia-overview}. Although the main goal of the mission is to generate a 6D map of our galaxy, the residues from the Gaia global solution can be used to investigate different effects. One example of such an investigation is the search for (extremely small) variations of star positions caused by gravitational waves (GW) passing through the location of the satellite. As in many other physical experiments, the input data in this case is noise-dominated and the process of model fitting is substantially nonlinear. The model of the gravitational wave is quite complex in terms of computational demands and the input data will be presumably over 50\,TB.\\

Hence, the goal of this thesis is to give an overview of algorithms and methods to perform a nonlinear model fit on such large and complex datasets, and to investigate possibilities to parallelize such a fitting algorithm. A prototype which implements selected methods is used to evaluate the practical feasibility of the GW search in simulated Gaia data. Furthermore, estimations of the feasibility and efficiency of such a search using real Gaia data will be given.

\section{Model Fitting}
The goal of model fitting, parameter estimation or parameter fitting is to find a set of parameters for a given model which make the model to resemble observational data as good as possible. Certain constraints to the parameters may apply.\\

Usually overdetermined systems of equations are used for parameter estimation, this means that one has much more data points than free parameters in the model. This work only covers parameter estimation in such overdetermined systems.\\

The relevance of parameter estimation for science cannot be overstated. Often experiments are conducted to prove theoretical models or to refine them. These models incorporate free parameters and some parameter estimation is needed to determine these parameters. This process also yield important statistical information with which the correctness and consistency of the model as well as the data can be analyzed.\\

A simple example for what parameter estimation can be used is the determination of the half-life of a radioactive isotope. Assume that we only know that such isotopes decay in some way, so that the activity gets less over time. A natural way to find out how they decay would be to measure the activity over time. Let us assume we have enough measurements in the right time span, now one can begin to use methods for parameter estimation to select our model and finding the right set of parameters. A first model might probably be a linear decay model. Somehow the method would find a set of parameters. But, the fit statistics would, without any doubt, tell us that the model is wrong. The same would happen if we apply simple polynomial models. If one would use an exponential decay model, the statistics would tell us that the model is (probably) correct. Additionally it yields parameters for that model (the half-life in this case) as well as their errors. In this way model fitting helps us first to rule out wrong models and second to find out the correct set of parameters, in this case the correct half-life.

\subsection*{Least Squares Fitting}
Least square fitting is probably the most widely used method for parameter estimation. In general, least square fitting is a method which tries to minimize the residues between a model function and observations. One residue $r_i$ is defined as:

\begin{equation}
 \label{eq:least-squares-r}
r_i = \dfrac{O_i - C\left( x_i;\mb{a} \right)}{\sigma_i}~~~~\text{.}
\end{equation}

Here $C$ is the model function with the parameter vector $\mb{a}$, and $O_i$ are the observed values for a specific $x_i$. The standard deviation, $\sigma_i$, is used to weight the data points. It is usually required that the errors of $O_i$ are normally distributed. The sum of the squared residues over $N$ data points is then defined as:

\begin{equation}
 \label{eq:least-squares-S}
S = \sum_i^N r_i^2~~~~\text{.}
\end{equation}

The methods of least squares fitting try to minimize $S$, which is the case when the gradient of the squared residues becomes zero:

\begin{equation}
 \label{eq:least-squares-partial}
\dfrac{\partial S}{\partial a_j} = 2\sum_i^N r_i \left( \dfrac{\partial r_i}{\partial a_j} \right) = 0 ~~~~ (j=0,..,M)~~~~\text{.}
\end{equation}

All the gradient-based methods presented in chapter \ref{chap:nonlinear-methods} try to minimize $S$ by using this property. 

\section{The ESA Gaia Mission}
\begin{figure}[H]
 \centering
 \includegraphics[keepaspectratio, width=0.6\textwidth]{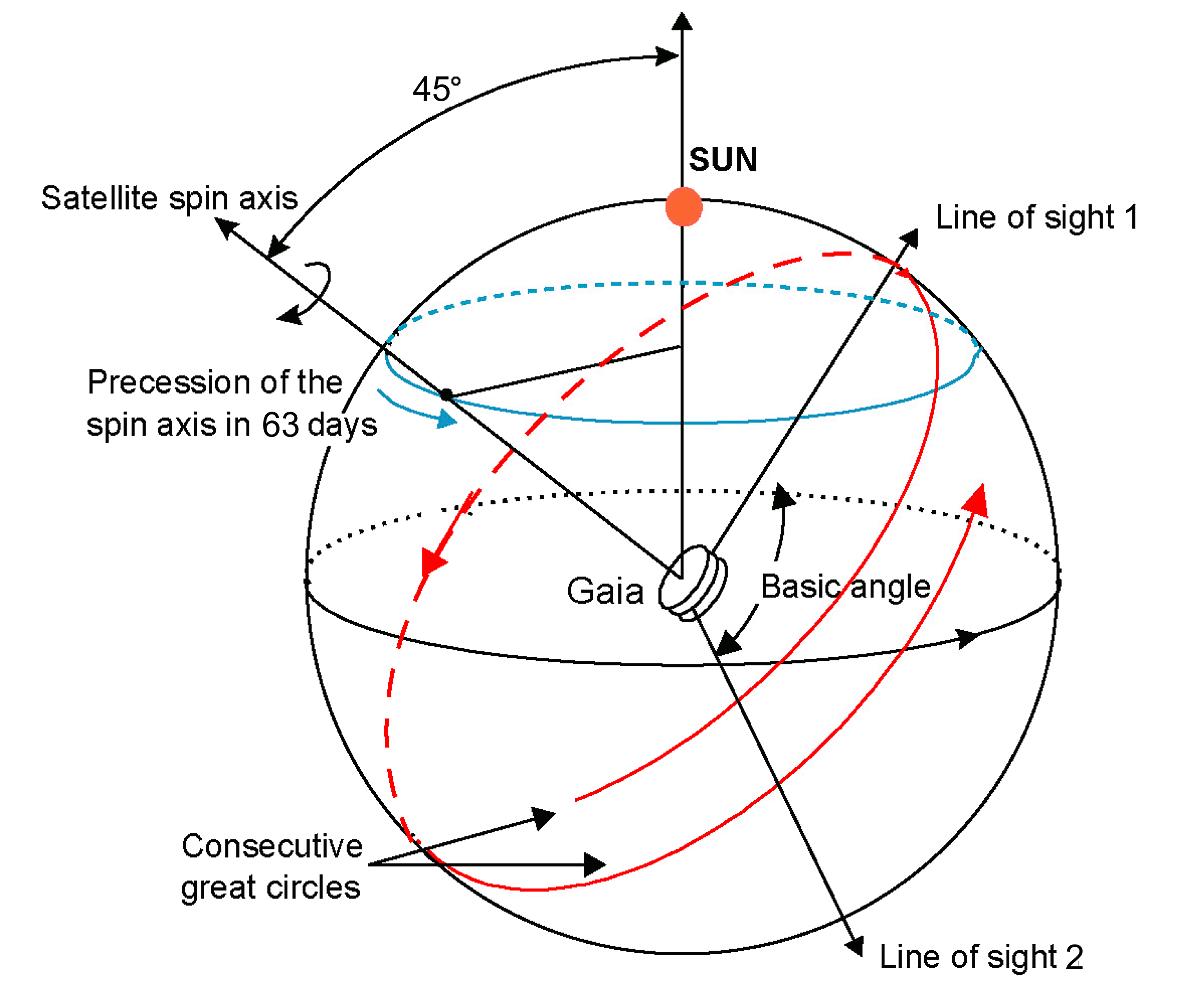}
 \caption[Basic orientation and scanning scheme for Gaia]{Basic orientation and scanning scheme for Gaia. The telescopes look in the direction of the ``line of sight'' and scan the sky along great circles. The precession of the spin axis around the position of the sun ensure that the whole sky is observed multiple times during the mission. (From: Jos de Bruijne, Gaia - Taking the Galactic Census; Scanning Law; white paper 2008)\label{fig:gaia-scann-law}}
\end{figure}

Gaia is mainly an astrometry mission which measures positions, distances and velocities of around one billion stars in our galaxy. The satellite consists of two telescopes which look in different directions with an angle of $106.6^\circ$ between them. This basic angle must be extremely stable to obtain the highest possible accuracy for measurements of absolute parallaxes \cite{suwGaiaParallax, gaia-overview}. Figure \ref{fig:gaia-scann-law} illustrates the angle between the two telescopes and the movement and orientation of Gaia. The satellite is placed on a Lissajous orbit around $L_2$. The spin axis of the satellite is slowly precessing in a $45^\circ$ angle around the Sun, the satellite rotates around this axis once every 6 hours. In this way the telescopes can scan the whole sky multiple times over the mission duration.

The result will be a star catalog with approximately one billion stars with completeness up to 20\,mag. The median parallax errors for stellar objects are planned to be 5-14\,µas for bright stars and 100-220\,µas for 20\,mag stars \cite{gaia-overview}.\\

In addition to the astrometric instruments, the satellite carries a spectrometer and photometric instruments \cite{esaGaiaPressSheet}. The mission is not limited to objects in the Milky Way, it will also measure certain parameters of extra-galactic objects like other galaxies and quasars. The mission goals are also not limited to simply cataloging celestial objects. Additional goals comprise, to name just a few: research of extra-solar planetary systems, star formation and asteroid detection. The payload of the satellite is highly complex and a description would certainly go far beyond the scope of this work. A picture of the payload structure as well as the assembled satellite can be found in Appendix \ref{sec:apx:gaiaSatPictures}. The reader can find additional information about the engineering of Gaia on the official ESA web page (\url{http://sci.esa.int/jump.cfm?oid=40129}).


\section{Detection of Gravitational Waves with Gaia}
\subsection*{Gravitational Waves}
``Gravitational waves are distortions of the space geometry that propagate through space with the speed of light'' \cite{esonline-gw}. They are predicted by the theory of general relativity and other gravitation theories.\\

It is presumed that the observation of GWs would allow fundamental insights in physics, astronomy and cosmology. Hence, substantial effort has been made to build detectors for GWs. So far, only earth based detectors like LIGO, VIRGO and GEO600 have been build. A space-based mission (LISA) is planed but is not started yet.

All of these detectors are solely dedicated to GW search and detection, and are sensitive for GWs with relatively high frequencies (earth based $>100$\,Hz \cite{geo600-freq, ligo-freq}, LISA $\approx 10^{-2}$\,Hz \cite{lisa-freq}).\\

If Gaia is suitable to detect GWs, it would be sensitive to much lower frequencies, probably waves with periods of years or days ($\approx 10^{-8}$\,Hz - $10^{-7}$\,Hz). It is important to stress that, in addition, Gaia would deliver the results from the main mission objectives too.\\

\subsection*{Detection with Gaia}
\begin{figure}[H]
 \centering
 \includegraphics[keepaspectratio, width=0.9\textwidth]{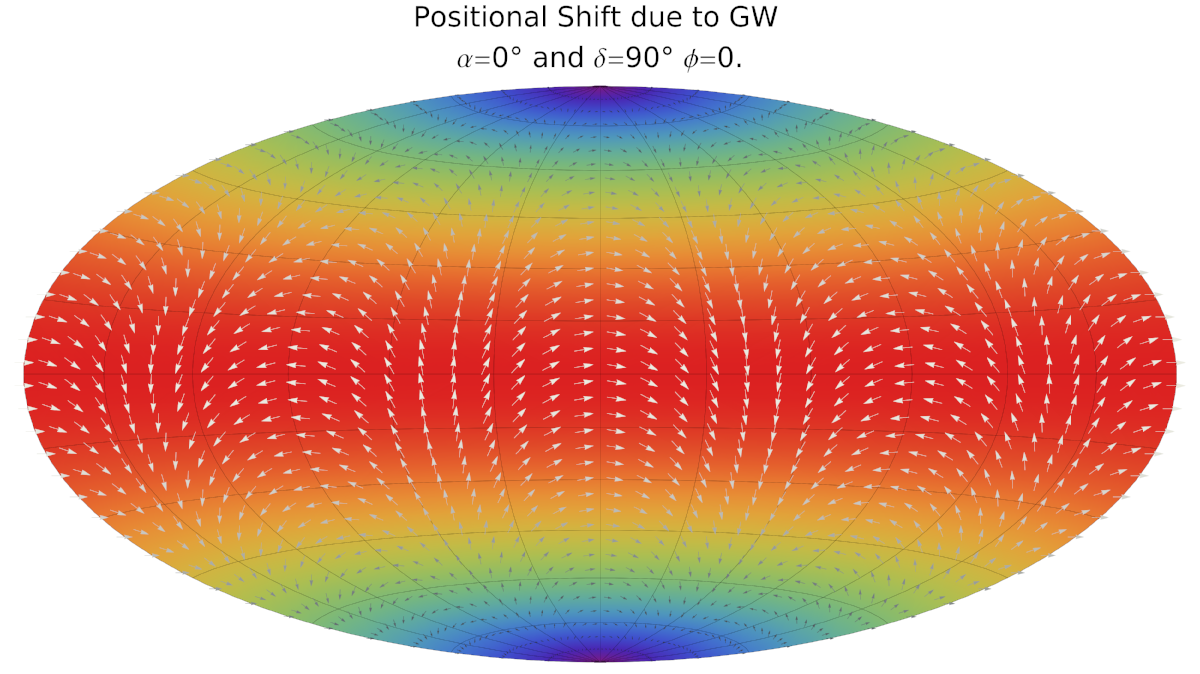}
 \caption[Variations in apparent star positions due to a GW]{All sky projection with the effects of an GW with the direction $\alpha=0^\circ$ and $\delta=90^\circ$ at $t=0$. The observer looks towards $\alpha=0^\circ$ and $\delta=0^\circ$. The black grid is the celestial coordinate grid, the rainbow colored background encodes the amount of change (larger change = red; fewer change = blue) and the vectors gives the direction as well as the amount.\label{fig:gw_change_star_pos}}
\end{figure}

Since a GW, which propagating through an observer influences the space-time at the point of observation, it also changes the apparent (incoming) direction of any photon which can be detected by the observer. If the observer measures positions of stars, their apparent positions will be slightly different than they would be without the presence of a GW. Figure \ref{fig:gw_change_star_pos} shows the theoretical variations in star positions due to a GW.

As already mentioned above, the main scientific goal of Gaia is to build a star catalog with the highest possible accuracy. In order to achieve this, complex processing of the raw data is necessary, from individual CCD\footnote{charge-coupled device, a semiconductor sensor used for capturing images} measurements to the astrometric catalog. The software and algorithm which is used to conduct this processing (AGIS) is not designed to detect and correct effects caused by a GW. Hence, traces of such a signal should persist in the residuals of the data processing. If such signal can be detected in the residuals, on the other hand, the information about such a signal could be used in AGIS to improve the catalog. A GW signal is expected to be very small and moreover highly nonlinear, in particular because of the unknown frequency.\\

The idea behind the study conducted in this work is to use these residues as input data and interpret them as observations of variations of star positions. A simple plane gravitational wave (PGW) model with different parameters is fitted to these observations, and statistically analyzed for goodness of fit. The parameters---and even the existence---of a possible GW in the data is absolutely unknown and cannot be constrained, except for certain instrument limitations. Hence, the parameter space has to be covered as wide as possible and care has to be taken to organize the search as efficient as possible.\\

Since Gaia is not yet in operational mode, the data used for this work has been simulated. The data has been produced using the AGISLab software, which has been designed to model the measurement and data processing of Gaia as realistically as possible.

\chapter{Methods for Nonlinear Parameter Estimation and Optimization}
\label{chap:nonlinear-methods}
This Chapter will give an overview of nonlinear parameter estimation and optimization algorithms, methods and strategies. A classification of the methods is given and basic working principles are shown.

\section{Why Parameter Estimation is an Optimization Problem}
\label{sec:why-lsq-optimization-prob}
Parameter estimation needs observational data to which the parameters of the model are fitted. Since measured data contains noise, the model will never exactly match the data. For this reason a measure of quality has to be found which gives a quantitative statement of the distance between model and observations. It is then assumed that, if this measure of quality has reached a (global) minimum, the parameters in the model are optimal. In other words, this quality function is a function depending on the model parameters and the observational data.

By variating the model parameters the quality function is optimized, hence parameter estimation is a optimization problem.

\bigskip

The quality function used in this work is the \textit{sum of squared residues}. This is also the most known and most used quality function for this purpose, the name Least-Squares-Fitting originates from it. Let us assume $N$ data points $(x_i,y_i)$ are given, so that an observation $O$ at a point $x_i$ (often, $x_i$ is for instance a time) yields the result $y_i$ so that $O(x_i) = y_i$. In addition a model $C(x;\mb{a}) = y$ is given, which has a parameter vector $\mb{a}$, then the sum of squared residues is:

\begin{equation}
\label{eq:sum-of-squared-residues-definition}
 S(\mb{a}) = \sum_{i=1}^{N}\left( \dfrac{O(x_i) - C(x_i;\mb{a})}{\sigma_i} \right)^2~~\text{.}
\end{equation}

The value $\sigma_i$ which can be ascribed to each data point is the corresponding error of the measurement. In general, the distribution of the errors are assumed to be Gaussian. In the following Sections, this ``error handling'' is omitted to prevent confusion. In principle all of the methods presented can be used with the measurement errors taken into account, some of them may require certain modifications.\\

This quality function is often also called fitness function, cost function, or objective function \cite{Madsen04immmethods}.

\section{Classification}
On a very high level, we can distinguish two different classes of minimizer methods, which can be used for the least square parameter estimation: global minimizers and local minimizers.\\

A \textbf{global minimizer} is defined by (\cite{Madsen04immmethods}):

\begin{equation}
 \label{eq:global-minimizer-definition}
\text{Find:}~~\mb{a^+} = \mathrm{argmin}_{\mb{a}}\left\lbrace S(\mb{a})\right\rbrace~\text{.}
\end{equation}\\

While a \textbf{local minimizer} is defined by (\cite{Madsen04immmethods}):

\begin{equation}
 \label{eq:local-minimizer-definition}
\text{Find}~~\mb{a^*}~~\text{so that:}~~S(\mb{a^*}) \leq S(\mb{a})~~\text{for all}~\mb{a}~\text{such that,}~~\|\mb{a}-\mb{a^*}\| < \delta~\text{ for some}~~\delta>0~~\text{.}
\end{equation}

Local minimizers always need a specific starting point from which they begin to search the region for a local minimum. The starting points of global optimizers are often chosen randomly or according to a method specific scheme.

\subsection{Gradient and Non-gradient Methods}
Another classification criterion, which correlates strongly with the local and global criterion, is the distinction between gradient and non-gradient methods. The main differences between the two is, as the name suggests, that gradient-based methods require knowledge about the gradient of the function under optimization, in our case the model function. In almost all gradient-based methods, this function is required to have continuous first derivatives. In some gradient-based methods also the second, or possibly even higher derivatives or partial derivatives are required \cite{sivaoptlecture}.

All gradient-based methods are designed to find a local optimum (or a locally optimal set of parameters if applied to a parameter estimation problem). On the other hand, many non-gradient methods exist which often attempt to find a global optimum. However, methods from both classes can diverge when applied to problems not suitable for them or a wrong configuration\footnote{Many of the methods require internal parameters, such as damping factors or parameters which control step-sizes. If these ``configuration parameters'' are chosen wrong, the method might not, or only insufficiently work.} has been chosen.

Some non-gradient methods require some sort of starting point and step size control, but they are usually quite insensitive to the selection of both. Gradient methods, on the other hand, always require a starting point and their result is quite sensitive to the selection of it, at least when many local optima exist.

Non-gradient methods usually require many more ``iterations'' to find a minimum of the quality function \cite{sivaoptlecture}. The set of ``points'' to probe is usually much higher compared to gradient methods. This usually makes non-gradient methods computationally more expensive, especially in cases where the starting point is near the desired minimum \cite{branham1990scientific}.

Gradient and non-gradient methods can be often combined. In the cases where the search space cannot be constrained and a suitable set of starting points cannot be given, a non-gradient method can be used to search for candidates of a global minimum. After a non-gradient method has found candidates for starting points, as second step, a gradient method can be used to find a exact fit for the candidates. This is called a hybrid approach.\\

However, according to the ``no free lunch theorem'' \cite{noFreeLunch} all approaches are equally bad from a statistical point of view if the set of all possible optimization problems are considered. It is also easy to see, that for a specific problem a specific optimization method might be found which outperforms any other strategy.

\section{Local Algorithms}
Many local algorithms exist and this Section can only give an overview of some of them. All of the methods have in common, that they are iterative methods. The content presented in this section is based on the publications \cite{Madsen04immmethods, NRC-1992} and \cite{branham1990scientific}, with some unifications in the notation.

\subsection{Gauss-Newton Method}
\label{sec:gauss-newton}
Detailed descriptions of the Gauss-Newton method (GN method) can be found in most book dealing with nonlinear parameter estimation, e.g. \cite{Madsen04immmethods, NRC-1992, branham1990scientific}. The Gauss-Newton method works by linearizing the quality function $\mb{S}(\mb{a})$ in the neighborhood of $\mb{a}$. The linearization is achieved by a Taylor expansion of function \ref{eq:sum-of-squared-residues-definition}. One can write $S$ in vector form in which $\mb{s}(\mb{a}) = \left( O(x_i) - C(x_i;\mb{a}) \right)$ and $\mb{x} = (x_i;\mb{a})$. Then $S$ can be expressed as the vector product of $\mb{s}(\mb{a})$.

\begin{align}
S &= \sum_{i=1}^{N}\left( O(x_i) - C(x_i;\mb{a}) \right)^2 \nonumber \\
  &= \mb{s}(\mb{a})^\top \mb{s}(\mb{a})
\end{align}

A Taylor expansion of $\mb{s}(\mb{a})$ can be written as:

\begin{equation}
\label{eq:taylor-of-sx}
 \mb{s}(\mb{a}+\mb{h}_{step}) = \mb{s}(\mb{a}) + \mb{J}(\mb{a})\mb{h}_{step} + O(\|\mb{h}_{step}\|^2)~\text{.}
\end{equation}

where $\mb{J}(\mb{a})$ is the Jacobian, a matrix containing the first partial derivatives of the function components (\ref{eq:jacobian}) and $\mb{h}_{step}$ is a step for which the linearization is in the error bounds of $O(\|\mb{h}_{step}\|^2)$. In the following a compact notation is used which implicitly assumes dependence of $\mb{J}, \mb{H}$ and $\mb{s}$ on $\mb{x}$ and $\mb{a}$.

\begin{equation}
\label{eq:jacobian}
\mb{J}(\mb{x})= \mb{J} =
\begin{pmatrix}
 \dfrac{\partial s_1}{\partial a_1}	& \dfrac{\partial s_1}{\partial a_2} & \cdots	& \dfrac{\partial s_1}{\partial a_M} \\
\dfrac{\partial s_2}{\partial a_1}	& \dfrac{\partial s_2}{\partial a_2} & \cdots	& \dfrac{\partial s_2}{\partial a_M} \\
\vdots								& \vdots							& \ddots	& \vdots							\\
\dfrac{\partial s_N}{\partial a_1}	& \dfrac{\partial s_N}{\partial a_2} & \cdots	& \dfrac{\partial s_N}{\partial a_M} \\
\end{pmatrix} 
\end{equation}

By solving

\begin{equation}
 \label{eq:gauss-newton-normal-eq}
(\mb{J}^\top \mb{J})\mb{h}_{step} = \mb{J}^\top\mb{s}
\end{equation}

for $\mb{h}_{step}$, one get the a refinement for the parameter vector $\mb{a}$ by $\mb{a}^{(k+1)} := \mb{a}^{(k)} + \mb{h}_{step}$. Iterations are done until $\|\mb{h}_{step}\|^2 < \varepsilon$, or another termination criterion is met, which one can chose freely. It is important to note, that the Jacobian $\mb{J}$ must be computed using $\mb{x}^{(k+1)}$ before the next iteration step.

\subsection{Levenberg-Marquardt Method}
\label{sec:levenberg-marquardt}
The Levenberg-Marquardt (LM) method is a ``damped Gauss-Newton method'' \cite{Madsen04immmethods}. The equation to solve for one LM step is:

\begin{equation}
 \label{eq:levmarq-normal-eq}
 (\mb{J}^\top \mb{J} + \mu \mb{I})\mb{h}_{step} = \mb{J}^\top\mb{s}~~~~\text{.}
\end{equation}

Where $\mb{I}$ is the identity matrix, $\mb{g} = \mb{J}^\top\mb{s}$, and $\mu \geq 0$. The parameter $\mu$ is called the ``damping parameter'' and has several effects (following list cited from \cite{Madsen04immmethods}, page 24 \& 25):

\begin{itemize}
 \item For all $\mu > 0$ $\mb{h}_{step}$ is a descent direction.
 \item For large values of $\mu$ one gets $$\mb{h}_{step} \simeq -\frac{1}{\mu}\mb{J}^\top \mb{s} = -\frac{1}{\mu}\mb{S}'(\mb{a})$$ i.e. a short step in the steepest descent direction. This is good if the current iterate is far from the solution.
 \item If $\mu$ is very small, then \eqref{eq:levmarq-normal-eq} is nearly equal to \eqref{eq:gauss-newton-normal-eq} which is a good step in the final stages of the iteration, when $\mb{a}$ is close to the local optimum. If $S(\mb{a^*}) = 0$ (or very small), then one can get (almost) quadratic final convergence.
\end{itemize}
 
The initial value of $\mu$ called $\mu_0$ can be chosen depending on the size of elements in $\mb{A}_0 =\mb{J}(\mb{a}_0)^\top \mb{J}(\mb{a}_0)$. One way is to set
\begin{equation}
 \label{eq:mu0-lm}
\mu_0 = \tau \cdot \max_i\left\lbrace a_{ii}^{(0)} \right\rbrace~\text{\cite{Madsen04immmethods},}
\end{equation}

and to set $\tau$ to a small value like $10^{-6}$ if $\mb{a}_0$ is assumed to be near $\mb{a^*}$. In other cases $\tau$ can also be set to values up to 1 \cite{Madsen04immmethods}.

During the iteration steps $\mu$ can be variated under the control of the gain ratio $\varrho$

\begin{equation}
\label{eq:lm-gain-ratio}
 \varrho = \dfrac{S(\mb{a}) - S(\mb{a}+\mb{h}_{step})}{L(\mb{0}) - L(\mb{h}_{step})}
\end{equation}

here $L(\mb{0}) - L(\mb{h}_{step})$ is the gain from the linear model $L$ in the linearized region of $\mb{a}$ \cite{Madsen04immmethods}.

\begin{align}
 \label{eq:lm-gain-linear}
L(\mb{0}) - L(\mb{h}_{step}) 
							 &= \frac{1}{2} {\mb{h}_{step}}^\top \left( \mu {\mb{h}_{step}} - \mb{J}^\top \mb{s} \right) 
\end{align}

Different methods exist to adjust $\mu$ between the iteration steps. One simple method is given in the following pseudo code:

\begin{algorithm}
\begin{algorithmic}
\If {$\varrho < 0.25$}
    \State $\mu := \mu \cdot 2$
\ElsIf{$\varrho > 0.75$}
	\State $\mu := \mu / 3$
\EndIf
\end{algorithmic}
\caption{Adjustment of the parameter $\mu$ in the LM method, from \cite{Madsen04immmethods}.\label{algo:lm-mu-control}}
\end{algorithm}

%

Iteration is done until one of three stopping criteria are met \cite{Madsen04immmethods}:
\begin{itemize}
 \item Gradient is near zero: $\| \mb{g} \|_\infty \leq \varepsilon_1$, where $\varepsilon_1$ is a small number chosen by the user.
 \item The change in $\mb{a}$ is small: $\| \mb{a}^{(k+1)} - \mb{a}^{(k)}\| \leq \varepsilon_2 (\|\mb{a}^{(k)}\| + \varepsilon_2)$ 
 \item A maximum number of iterations is reached: $k \geq k_{max}$
\end{itemize}

\subsection{Broyden's Update Rule and Secant LM Method}
\label{sec:broyden}
In \cite{Madsen04immmethods} a secant version of the Levenberg-Marquardt method is given, which does not require the Jacobian $\mb{J}$. According to \cite{Madsen04immmethods}, in practice the Jacobian $\mb{J}$ cannot always be computed. But one can also be interpret it in a different way. What if, it is to expensive to compute $\mb{J}$ in practical terms of computing power?\\


One way to evade this problem is to use a matrix $\mb{B}$ instead of the Jacobian, which is obtained by numerical differentiation by the finite difference method. The update of this matrix in each step is then done using the ``Broyden's Rank One Update'' rule. In cases where the partial derivatives are known, but are to complex to compute them \textit{every} step, it is also possible to compute them once at the beginning and update the ``Jacobian'' using Broyden's rule \cite{Gavin-LM}.\\

This rule reads as follows:

\begin{equation}
 \label{eq:broyden-rank1-update}
\mb{B}^{(k+1)} = \mb{B}^{(k)} + \frac{\left( \mb{s}(\mb{a}^{(k)} + \mb{h}) - \mb{s}(\mb{a^{(k)}}) - \mb{B}^{(k)}\mb{h}\right) \cdot \mb{h}^\top}{\mb{h}^\top \mb{h}}~\text{\cite{Gavin-LM}.}
\end{equation}

Gavin~\cite{Gavin-LM} makes the advantages of this method clear:

\begin{finequote}
For problems with many parameters, a finite differences Jacobian is computationally expensive.
Convergence can be achieved with fewer function evaluations if the Jacobian is
re-computed using finite differences only occasionally. The rank-1 Jacobian update equation
requires no additional function evaluations.\\[3ex]
--- {\footnotesize From: Henri P. Gavin; The Levenberg-Marquardt method for nonlinear least squares curve-fitting problems}
\end{finequote}

The computational expenses can be lowered further, as \cite{Madsen04immmethods} show in their secant version of the LM method. In this variant of LM the Broyden matrix $\mb{B}$ gets recalculated not entirely every step. This is done coordinate wise and cyclic, by defining an angle between $\mb{h}$ and the unit vector of the coordinate (when thought of the parameters as multi-dimensional coordinates) and checking the parameters in a round-robin fashion. If this angle is too large and the corresponding parameter is marked for update, the corresponding column in $\mb{B}$ will get updated.

\subsection{Powell's Dog Leg Method}
Powell's method is a trusted region method, which is based on the fact that a Gauss-Newton step is not necessarily in the same direction as the steepest descend \cite{dennis1996numerical}. As shown in section \ref{sec:gauss-newton} one can calculate a refinement $\mb{h}_{step}$ by solving equation \eqref{eq:gauss-newton-normal-eq}. Let us call the $\mb{h}_{step}$ from the Gauss-Newton step $\mb{h}_{GN}$ here. Furthermore we can calculate a $\mb{h}_{SD}$ which gives us the direction of the steepest descent.

\begin{equation}
 \label{eq:steepest-descend}
\mb{h}_{SD} = -\mb{J}^\top\mb{s}
\end{equation}

The length $\alpha$ of the step in the steepest descend direction can be calculated by \cite{Madsen04immmethods}:

\begin{equation}
 \label{eq:length-steepest-descend}
\alpha = - \dfrac{\mb{h}_{SD}^\top \mb{J}^\top \mb{s}}{\|\mb{J}\mb{h}_{SD}\|^2}~~\text{.}
\end{equation}

In this way one can chose between two directions in each iteration, $\mb{h}_{GN}$ and $\alpha\mb{h}_{SD}$. Powell suggests (in \cite{Powell01011964}) that the Dog Leg step $\mb{h}_{DLstep}$ is

\begin{equation}
\label{eq:dogleg-dlstep}
 \mb{h}_{DLstep} = \alpha\mb{h}_{SD} + \beta(\mb{h}_{GN} - \alpha\mb{h}_{SD})
\end{equation}

By solving \eqref{eq:dogleg-dlstep} for $\beta$, under the constraint that $\|\mb{h}_{DLstep}\| = \Delta$, one can calculate $\mb{h}_{DLstep}$. Here, $\Delta$ is the radius of the so called ``trust region'' from which one can assume that the linearized model resembles the underlying function to a certain accuracy. 

The complete algorithm can be found in \cite{Madsen04immmethods}. Although the Jacobian is used in \cite{Madsen04immmethods}, it has to be stressed, that this is not always necessary. It is common to use a approximate matrix $\mb{B}$ as in Section \ref{sec:broyden}. This matrix can be updated via the Broyden rank one update rule (see Eq. \eqref{eq:broyden-rank1-update}).

\subsection{Quasi-Newton and Other Local Methods}
Many more methods for local, gradient-based optimization and parameter estimation exist. Some of them shall be discussed in a much briefer way below.\\

Before additional methods are discussed, one important method, the ``Newton method'' has to be introduced. For this method in addition to the first partial derivatives of the parameters, also the second is used. These second derivative are stored in a matrix $\mb{H}$, the Hessian matrix. Consider again the Taylor expansion of the function $s(\mb{a})$ as in Eq. \eqref{eq:taylor-of-sx} ($\mb{h}_{step}$ is abbreviated as $\mb{h}$ here). This expansion can be developed further to the second degree \cite{grossmann1997numerik}:

\begin{equation}
 \label{eq:taylor-second-deg-sx}
 \mb{s}(\mb{a}+\mb{h}) = \mb{s} + \mb{J}\mb{h} + \frac{1}{2}\mb{H}\mb{h}^2 + O(\|\mb{h}\|^3)~~\text{.}
\end{equation}

This method is important due to its fast local convergence speed \cite{branham1990scientific}, but it has the disadvantage of needing information about the Hessian matrix. Either $\mb{H}$ is approximated or has to be computed from scratch in every step. One can circumvent this problem by using the fact that $\mb{H}$ can be constructed approximately in step $k$ by \cite{grossmann1997numerik, branham1990scientific}:

\begin{equation}
 \label{eq:quasi-newton}
\mb{H}^{(k+1)}\mb{h}^{(k)} = \mb{J}(\mb{a}^{(k+1)}) - \mb{J}(\mb{a}^{(k)})~~\text{with}~~\mb{H}~\text{symmetric \cite{grossmann1997numerik}.}
\end{equation}

This approximated Hessian matrix can be used in the Newtonian step ($\mb{h}^{(k)} = - \mb{H}^{-1}_{(k)}\mb{J}(\mb{a}^{(k)})$ and $\mb{a}^{(k+1)} = \mb{a}^{(k+1)} + \alpha \mb{h}^{(k)}$). The parameter $\alpha$ is the step size, which has to be controlled during the iterations by suitable methods. In Section \ref{sec:broyden} an update scheme for the Jacobian was discussed, such update schemes also exist for the Hessian matrix. It reads \cite{grossmann1997numerik}:

\begin{equation}
\mb{H}^{(k+1)} = \mb{H}^{(k)} + \frac{(\mb{J}(\mb{a}^{(k+1)}) - \mb{J}(\mb{a}^{(k)}) - \mb{H}^{(k)}\mb{h}^{(k)})\mb{h}^{{(k)}^\top}}{\mb{h}^{{(k)}^\top} \mb{h}^{(k)}}~.
\end{equation}

The main distinctions between different quasi-Newton methods is how the update of the matrices are done.

\subsubsection{Broyden-Fletcher-Goldfarb-Shanno Algorithm}
Often abbreviated BFGS, is a Quasi-Newton method with the update rule \cite{grossmann1997numerik}:

\begin{align}
\label{eq:bfgs}
 \mb{q}^{(k)} 	&:= \mb{J}(\mb{a}^{(k+1)}) - \mb{J}(\mb{a}^{(k)}) \nonumber \\
 \mb{H}_{BFGS} 	&= \mb{H} + \left( 1 + \dfrac{\mb{h}^\top \mb{H} \mb{h}}{\mb{q}^\top \mb{h}} \right) \dfrac{\mb{q}\mb{q}^\top}{\mb{h}^\top \mb{q}} - \dfrac{1}{\mb{h}^\top \mb{q}} \left( \mb{q}\mb{h}^\top \mb{H} + \mb{H}\mb{h}\mb{q}^\top\right)~.
\end{align}

By interchanging the vectors $\mb{q}$ and $\mb{h}$, and using $\mb{H}^{-1}$, one can get the inverse Hessian matrix directly from equation \eqref{eq:bfgs}.

\subsubsection{DFP}
The Davidon-Fletcher-Powell method is very similar to the BFGS method. Its update formula reads \cite{grossmann1997numerik}:

\begin{equation}
\label{eq:dfp-update}
\mb{H}_{DFP} 	= \mb{H} + \dfrac{\mb{q}\mb{q}^\top}{\mb{q}^\top\mb{h}} - \dfrac{\mb{H}\mb{h}\mb{h}^\top \mb{H}}{\mb{h}^\top \mb{H}\mb{h}}~.
\end{equation}

\newpage

\subsubsection{Other Methods}
Numberous other methods exist, some commonly used are:
\begin{itemize}
 \item Conjugate Gradient Methods (see \cite{grossmann1997numerik, shewchuk1994introduction})
 \item Hybrid GN-BFGS Methods (see \cite{fletcher1987hybrid})
 \item Nelder-Mead method, also called downhill simplex (see \cite{nelder1965simplex, branham1990scientific})
 \item Limited-memory BFGS (L-BFGS) (see \cite{lbfgs})
\end{itemize}

\section{Global Algorithms}
As the name suggests, gradient methods need information about the gradient of the objective function. If such information cannot be obtained (derivatives not known) because the objective function is a ``black-box'', one can use a non-gradient method. In the history of these methods, it was soon discovered that many of the non-gradient methods could be tuned to find the global optimum \cite{branham1990scientific, weise2009global}. Hence, many non-gradient methods are global methods and the names are used somewhat synonymously.

\subsection*{Non-gradient Methods}
With complex models and big datasets computing the Jacobian or the Hessian can be very time consuming. Additionally, one has to keep in mind, that all the methods discussed above are iterative, so the Jacobian or Hessian have to be calculated more than one time for a fit. According to Branham \cite{branham1990scientific}, another problem is noise in the data. It is known that differentiation is amplifying noise, and in some cases, the noisy data will not resemble the gradients given by the Jacobian sufficiently.\\

Another point of criticism of gradient based methods is, that they are not converging in all cases \cite{branham1990scientific}. Non-gradient methods exist which converge always and can be tuned so that they converge to a global minimum (especially genetic algorithms and other heuristic methods fall in this category) \cite{branham1990scientific}. So why using gradient methods at all?\\

The downside of non-gradient methods is that they do not use information which the gradient contains. If the quality function under optimization is at least locally smooth, a step in the direction given by the gradient is not completely wrong. Without this information many intermediate results are generated, which are not usable, and have to be ``thrown away'' since they do not contribute to the optimal solution. This often leads to much slower runtime, especially in the case where a local minimum has to be found.\\

Genetic algorithms and evolution strategies are heuristic and generic optimization algorithms. Generic means, that most of them are designed as black box tools, which can be used for a large field of optimization problems. However, in most cases some search parameters have to be set up according to the problem at hand.

\subsection{Evolutionary and Genetic Methods}
In the literature genetic algorithms, evolutionary algorithms and evolution strategies are often mixed. However, in this work two classes of methods are distinguished. Both, genetic algorithms (GA) and evolution strategies (ES), are manifestations of one parent class of methods: the evolutionary algorithms (EA). Although specific implementations sometimes use hybrid approaches, the differences between genetic- and evolutionary strategies are subtle but important.\\

Both methods have in common that they mimic biological evolution. A population of individuals is improved by a selection process and different kinds of mutation (randomization). In practice appropriate data structures (such as classes) are used to represent these individuals. The differences are, that GA using a coded form (often as binary string, analogous to the DNA) of the position of an individual in the search space, while ES store this information explicitly. Hence, the mutation in GA rely on operations on this coded ``genome'', like interchanging small sections of it between individuals. ES, on the other hand, using randomized functions to directly change the parameters of the individuals.\\

Another difference is, that in most GA no offspring is deleted, it only has a smaller probability to reproduce. In most ES, only a small part of the population is kept and act as the new parents for the next generation, in this way weaker individuals can not reproduce at all. Additionally GA usually operating on a larger population sizes that ES.

\subsection{(\textmu,\textlambda) Evolution Strategies}
Evolution strategies where pioneered in the 1960s by Rechenberg and Schwefel. One particularly intuitive and simple strategy is the $(\mu\overset{+}{,}\lambda)$-ES \cite{rechenberg1581evolutionsstrategie}. In this method two steps are repeated until a result is found. In the first step, $\mu$ parents generate $\lambda$ mutated offspring. At the start, the parameters of the individuals are distributed uniform and random over the search space. In the second step a selection, based on a fitness function, is done. The fittest members are the new parents, and the cycle starts again. The ``,'' or ``+'' indicates whether the parents are included in the selection process (+), or the parent are deselected in any case in the selection process (``,'' only).

\subsubsection{Graphical and Formal Description of ES}
Rechenberg introduced a graphical description of evolutionary strategies, which consist of so called game symbols. This graphical description can also be formalized. They can be found in \cite{rechenberg1581evolutionsstrategie}, since this book is only available in German, the graphical description will be repeated here.\\

A formal description is also available and can be found in \cite{rechenberg1581evolutionsstrategie} and \cite{beyerEScompIntro}, where the notation is described.

\subsubsection*{Game Symbols}
\begin{figure}[H]
\centering
\includegraphics[keepaspectratio,width=0.7\textwidth]{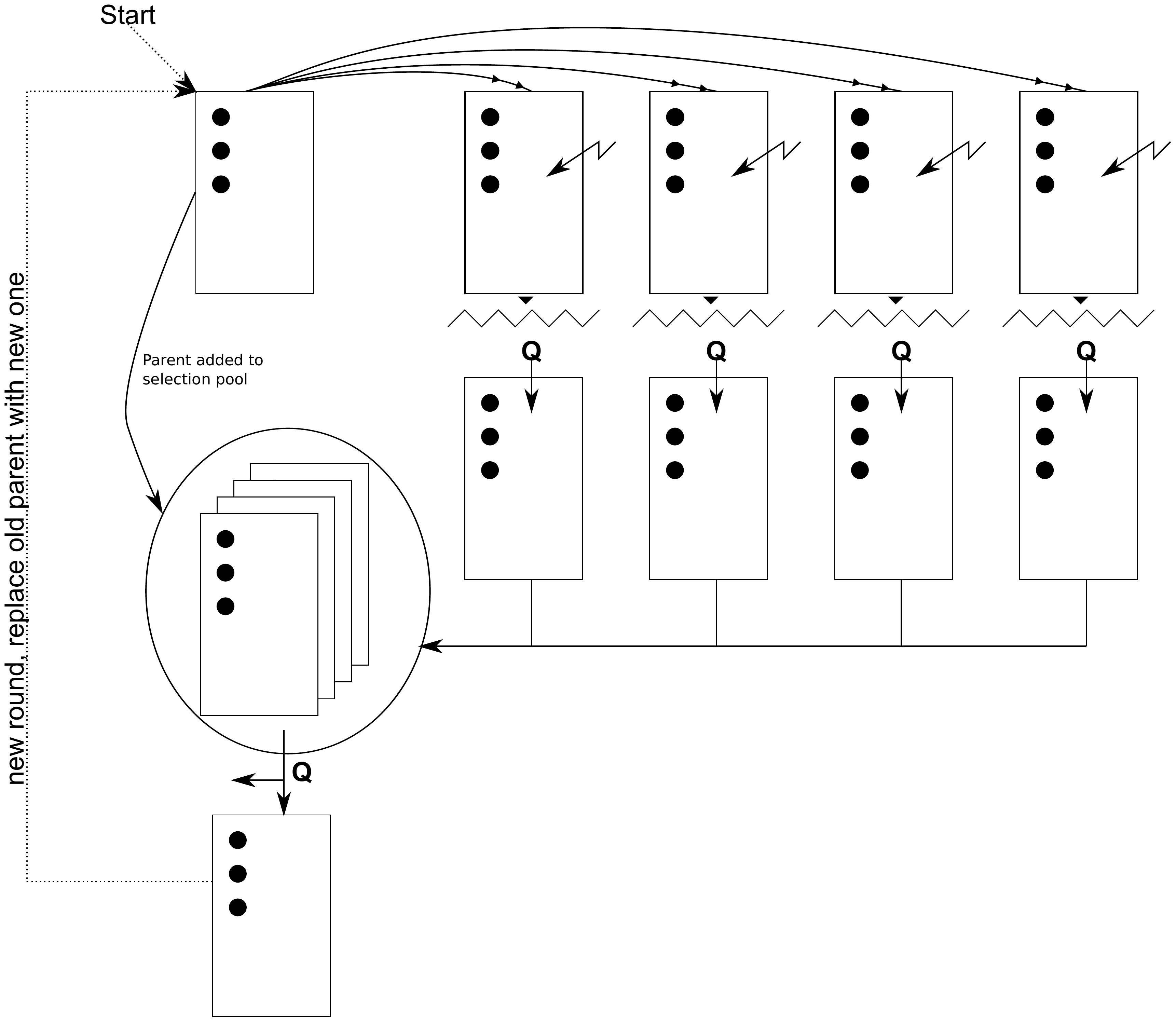}
 \caption[A example of a $(1+4)$-ES strategy in graphical description]{A example of a $(1+4)$-ES strategy in graphical description. One parent (Start) is duplicated four times, this offspring is mutated and then rated with the quality function $Q$. The parent is added to the selection pool in each round. After selection one single individual remains, which acts as the new parent. The new parent can also be the old parent, in case it is fitter than its offspring. This scheme is repeated until a certain quality is reached or another stopping criterion is met (e.g. maximum number of generations).\label{fig:14es}}
\end{figure}

Symbols can be used to represent a graphical description of an ES. Each of the symbols for individuals can be considered as a card. On this card the genetic parameters (the DNA) among other things (for instance the quality) can be noted. Imagine playing an iterative game with this cards. In this game you can copy, modify, rate and throw away these cards. Goal of the game is, to find the best parameters possible to create a ``super-card'' with the best set of parameters noted on it, rated by the quality function. This exactly matches the global optimum criterion from Eq. \eqref{eq:global-minimizer-definition}. The symbols can be found in appendix \ref{sec:apx-es-game-cards} a simple $(1+4)$-ES is shown in Figure \ref{fig:14es}.

%

\subsubsection{From Description to Algorithm}
The formal or graphical description can be transformed into an algorithm quite easy. It is advisable to modularize, at least, the quality function as well as the selection process. The pseudo-code in algorithm listing \ref{algo:modules-for-ES} shows functions which are necessary for a simple evolution strategy. This Listing also shows an implementation of a $(1+\lambda)$-ES in pseudo-code. In both listings $Parent(s)$ as well as $Children$ are ``Individuals'' (Game Cards) which consist of a addressable set of parameters ($Params$) and a set of step sizes ($StepSize$). In the simple scheme below, only one step size for all parameters exist. In a more advanced scheme one would certainly have a independent step size for each parameter. The function \textsc{Quality} in listing \ref{algo:modules-for-ES} can be any function of fitness. 

The variable $A$ in algorithm \ref{algo:modules-for-ES} on line \ref{line:A-step-size-param} is a step size factor which is used (line \ref{algo:makestep}) here like proposed in \cite{rechenberg1581evolutionsstrategie}. The step size control can also be linked against the quality gain, but should be kept independent in such simple schemes.

\begin{algorithm}
\begin{algorithmic}[1]
\State $\lambda \gets$ Number of Children per Round
\State $M \gets$ Number of Parameters in Model
\State $Parent \gets (Params=StartPoint; StepSize=StartStepSize)$
\State $Children \gets \{\}$
\State $A \gets 1.3$\label{line:A-step-size-param}
\State $z \gets \mathcal{N}(0,1) / M$

\vspace{1.2ex}

\For{$g \leq MaxGenerations$}
  \State \Call{Reproduction}{\,}
  \State $Best \gets $\Call{Selection}{$Children \cup Parent$}\label{line:unity-sets}
  \If{\Call{Quality}{$Best$} > $\varepsilon$}
        \State \textbf{Break}
  \EndIf
  \State $Parent \gets Best$
  \State $Children \gets \{\}$
\EndFor
%

\vspace{1.2ex}

\Function{Selection}{Set of Individuals $S$}
  \State $Best \gets S\left[ 1\right] $
  \For{$s$ in $S$}
	\If{\Call{Quality}{$s$} > \Call{Quality}{$Best$}}\label{line:selection-if}
	  \State $Best \gets s$
	\EndIf
  \EndFor
  \State \Return $Best$
\EndFunction

\vspace{1.2ex}

\Function{Reproduction}{\,}
  \For{$l \leq \lambda$}
	\State $Children \gets Children \cup \{Parent\}$
	\State $z \gets \mathcal{N}(0,1) / M$
	\State $Children\left[l\right].StepSize \gets Parent.StepSize \cdot A^{2\lfloor z + 0.5\rfloor -1}$
	 \For{$p \leq M$}
	  \State $z \gets \mathcal{N}(0,1) / M$
	  \State $Children\left[ l\right].Params\left[ p\right]  \gets Parent.Params\left[p\right] + Children\left[ l\right].StepSize\cdot z$\label{algo:makestep}
    \EndFor
  \EndFor
  \State \Return $z$
\EndFunction
\end{algorithmic}
\caption{$(1+\lambda)$-ES in pseudo-code, from \cite{rechenberg1581evolutionsstrategie}. Note that the difference between a $(1+\lambda)$ and a $(1,\lambda)$-ES just depends on line \ref{line:unity-sets}. The code shown here can be transformed to a $(1,\lambda)$-ES by passing only $Children$ to the selection function.\label{algo:modules-for-ES}}
\end{algorithm}

\subsubsection{ES Scheme with Recombination}
A particularly interesting ES scheme is the $(\mu/\rho_{\{I,D\}}\overset{+}{,}\lambda)$-ES. In this scheme $\rho$ parents recombine to a single new individual. This recombination works on the genotype, the parameters noted on the Rechenberg cards. The most ``natural'' scheme, hence, is the $(2/2_{\{I,D\}}\overset{+}{,}\lambda)$-ES where two parents recombine their genes during reproduction, and the two parents produce one child in one duplication/recombination step. The $\{I,D\}$ index denotes the type of recombination. $I$ marks intermediate, $D$ dominant recombination. In intermediate recombination the model parameters as well as the step sizes are generated from the geometric center of these parameters (see Eq. \eqref{eq:intermed-recomb} from \cite{beyer2001theory}). The dominant recombination randomly selects parts from the model parameter vectors or the step size vectors of the parents, and mixes them to the corresponding new vectors (see Eq. \eqref{eq:dominant-recomb} from \cite{beyer2001theory}). In both methods the $\rho$ parents of one child get randomly selected among the whole set of $\mu$ parents, $\mb{c}$ denotes the parameter or stepsize vector of the cild.
Hybrid variants are possible, for instance dominant recombination for the step sizes and intermediate recombination for the model parameter vectors.

\begin{equation}
 \label{eq:intermed-recomb}
\mb{c} := \left\langle \mb{x} \right\rangle_\rho := \frac{1}{\rho}\sum_{i=1}^\rho \mb{x}_i~~\text{with}~\mb{x}_i~\text{is parameter vector of the parent $i$}
\end{equation}

\begin{equation}
 \label{eq:dominant-recomb}
\mb{c} := \sum_{i=1}^{N} \left( \mb{e}_i^\top \mb{x}_{m_i} \right) \mb{e}_i~~\text{with}~~m := \mathrm{RandomSelect}\left( \{1,...,\rho\}\right)~\text{and}~N=\dim{\mb{x}}
\end{equation}

It is easy to see that recombination blurs the lines between genetic algorithms and evolution strategies. Recombination works on the genotype, as the genetic operators in the GA do, and with parent selection before recombination (if employed), like in GAs, only the strongest parents reproduce.

\subsection{The CMA Evolution Strategy}
A more complex ES scheme is the one proposed by Hansen and Ostermeier \cite{HansenOstermeierCMA, hansen1997convergence, hansen2005cma}. It employs a covariance matrix adaption (CMA) to detect (what could be called) the gradient on in which the evolution progresses.\\

The ES-CMA has been implemented by a number of groups in different software packages \cite{MuellerpCMALib, shark08, Keijzer2001, Gagne2006BEAGLE}. ES-CMA can be considered as the most commonly used ES based black-box optimizers used today.\\

The main loop of the algorithm consists of four steps \cite{hansen2005cma}. First, a new population with $\lambda$ individuals is sampled from a multi-variant normal distribution under the use of a covariance matrix. In the second step, selection and recombination is done. A weight is ascribed to each individual, and the individuals are sorted according to this weight. The $\mu$ fittest individuals are kept, the rest is deleted. In this way the mean of the search distribution is moved with respect to weight coefficients of each individual.\\

The third step is step-size control. Under the usage of an evolution path and a ``backward time horizon'', the step size for the next step is computed. In the last step, the covariance matrix is updated. This step mainly uses two sources of information, the evolution path and the mean of the search distribution. This two sources are combined and form the new covariance matrix.\\

These steps get repeated until the stopping criterion is met. The full algorithm can be looked up in \cite{hansen2005cma} and on the webpage of Hansen \cite{cmaeswebpage}. Further improvements are conceivable, and are done e.g. by Jastrebski \cite{JastrebskiActiveCMA1688662} using an ``Active CMA'', where also information from unsuccessful offspring is used.

\subsection{Genetic Algorithms}
While in most ES, the population size is quite small (typically $\lambda$ is set to 10 - 20) GAs profit from large population sizes. It is not uncommon to use (starting) population sizes of 1000 in GAs \cite{melanie1999introduction, coley1999introduction}.

Simple schemes, proposed in \cite{gen-algo-web, RajpaulGA, CharbonneauGA, goldberg1989genetic}, work as follows:
\begin{enumerate}
 \item A large starting population is generated at random points of the search space. The location information is stored in a coded form (often as a binary string) in the genotype. This string is often called chromosome.
 \item \label{GAenumeratestart} The genotype is decoded and a function of fitness measures the quality of the individuals.
 \item New offspring is generated by randomly selecting and recombining pairs of parents. Better parents are favored for reproduction. In this step the genetic information of the parents is recombined randomized by genetic operators (like recombination in ES). The fundamental operator is ``crossover''.
 \item Mutations of the genetic information of the new individuals are applied. A random point in the genotype string is flipped with a very small probability.
 \item The individuals with the lowest fitness are replaced by the new offspring.
\begin{itemize}
 \item Note that in this point the GA differs substantially from the ES, in particular the $(\mu,\lambda)$-ES. ``Old'' individuals still get included in the new round. Only the weakest ones get replaced. 
\end{itemize}
 \item Go to \ref{GAenumeratestart} or terminate if the quality threshold, or other termination criterion, is reached.
\end{enumerate}

In the algorithmic sketch above is a ``genetic operator'' is used to recombine the parents genetic information to form offspring. A genetic operator defines how the genetic information is modified during mutation and recombination and can be understood as a function which operates on one or more genome. Different operators exist, and the different application and combination is one of the main distinctions between different GAs. The following list contains some examples for genetic operators. It is assumed that the genotype is stored in binary format and that the operator uses the chromosome of two individuals to combine them:

\begin{itemize}
 \item Crossover: This is the most important operator. Randomly selects a fixed cutting position in the binary string. Cuts at this position and exchange the tails of the string. \cite{coley1999introduction, gen-algo-web}
 \item Inversion: The binary string get inverted. This is a mutation operator.
 \item Gene duplication: A (small) part of the binary string is inserted again after itself. This is also a mutation operator and it leads to a longer genotype string.
 \item Sexual differentiation: The sex of an individual is coded in the genetic information. Only individuals from different genders can reproduce. \cite{goldberg1989genetic}
 \item Deletion: Parts of the binary sequence are deleted.
\end{itemize}

Further operators are: dominance, dipolity, translocation, genetic memory, crowding \cite{kenneth1975jong}, mating restrictions (like local mating, incest prevention) \cite{Smith92populationdiversity, eshelman1991preventing} and many more.

\subsection{Simulated Annealing}
Simulated annealing (SA) is a heuristic optimization method, which mimics the physical process of slowly cooling a molten bath or red-hot metal. During the cooling phase, the atoms or molecules have enough time to arrange in stable crystals or formations. When a material is very hot, the atoms in it can move around randomly in a large area. As the material cools down, the freedom of this movement is reduced, the atoms are locked more or less in place. The idea of SA is to mimic this process. SA probes a relatively large number of parameter points (molecules or atoms) with high probability for large step sizes in the beginning, but ``cools down'' this probability over the time. In this way the parameter sets with the best quality getting more and more locked to their place, hopefully to a global optimum.\\

Thomas Weise gives a convenient description of the physical processes mimicked by SA \cite{weise2009global}:

\begin{finequote}
In physics, each set of positions of all atoms of a system $(pos)$ is weighted by its
$E(pos)$ Boltzmann probability factor $e^{-\frac{E(pos)}{k_{B^\top}}}$ where $E(pos)$ is the energy of the configuration $pos$, $T$ is the temperature measured in Kelvin, and $k_B$ is the Boltzmann's constant $k_B~=~1.380650524 \cdot 10^{-23}\,J/K$.

The Metropolis procedure was an exact copy of this physical process which could be used
to simulate a collection of atoms in thermodynamic equilibrium at a given temperature. A
new nearby geometry $pos_{i+1}$ was generated as a random displacement from the current
geometry $pos_i$ of an atom in each iteration. The energy of the resulting new geometry is
computed and $\Delta E$, the energetic difference between the current and the new geometry,
was determined. The probability that this new geometry is accepted [is] $P(\Delta E)$.\\

[..]\\

Thus, if the new nearby geometry has a lower energy level, the transition is accepted.
Otherwise, a uniformly distributed random number $r = \mathrm{random}_u() \in [0, 1)$ is drawn and the
step will only be accepted in the simulation if it is less or equal the Boltzmann probability
factor, i. e., $r \leq P (\Delta E)$. At high temperatures $T$, this factor is very close to 1, leading to the
acceptance of many uphill steps. As the temperature falls, the proportion of steps accepted which would increase the energy level decreases. Now the system will not escape local regions anymore and (hopefully) comes to a rest in the global minimum at temperature $T = 0\,K$.\\

[..]\\

It has been shown that Simulated Annealing algorithms with appropriate cooling strategies will asymptotically converge to the global optimum.\\[3ex]
--- {\footnotesize From: Thomas Weise; Global optimization algorithms--theory and application; Self-Published}
\end{finequote}

A simple algorithm is given in algorithm listing \ref{algo:SA-simple}. The algorithm consists of a main loop (line \ref{SA:while}) which checks for termination criteria. Inside each loop iteration a neighbor $\mb{x}_{tmp}$ of the parameter set $\mb{x}$ is randomly selected from the parameter space. This random selection (line \ref{SA:GetNeighbor}) has to satisfy the constraint that the probability $p(\mb{\Delta})$ for the distance $\mb{\Delta}$ from $\mb{x}$ to $\mb{x}_{tmp}$ is appropriate for the current temperature $t$. After a neighbor is selected, three things can happen: The quality of the new neighbor $\mb{x}_{tmp}$ is better than the old parameter set (line \ref{SA:if1}), than this new version is used. Second, with a certain probability (line \ref{SA:E}), depending on the temperature, take the neighbor even if the quality is not as good as the previous parameter set. Third, nothing happens and the old parameter set is kept for the next round. The last important step (line \ref{SA:cooling-schedule}) is the cooling schedule. The function \textsc{Temperature}$(t)$ regulates how the temperature cools down, this is called \textit{cooling plan} and represents a crucial parameter for the success of any SA application. The cooling plan presented here is just an example and might be far from optimal. Finding a optimal cooling plan is a nontrivial task and involves a certain amount of experimentation.

\begin{algorithm}[htb]
\begin{algorithmic}[1]
\State $t \gets$ Initial Temperature
\State $\mb{x} \gets$ Initial Parameter
\While{$i \leq i_{max} \wedge \Call{Quality}{\mb{x}} > \epsilon$}\label{SA:while}
  \State $\mb{x}_{tmp} \gets$ \Call{GetNeighbor}{$\mb{x},t$}
  \If{\Call{Quality}{$\mb{x}$} $<$ \Call{Quality}{$\mb{x}_{tmp}$}}\label{SA:if1}
	\State $\mb{x} \gets \mb{x}_{tmp}$
  \ElsIf{$e^{-(\Call{Quality}{\mb{x}_{tmp}} - \Call{Quality}{\mb{x}})/t} > \Call{Random}{0.0,1.0}$}\label{SA:E}
	\State $\mb{x} \gets \mb{x}_{tmp}$
  \EndIf
  \State $i \gets i+1$
  \State $t \gets$ \Call{Temperature}{$t$}\label{SA:cooling-schedule}
\EndWhile

\vspace{1ex}

\Function{Temperature}{$t$}
  \State \Return $t / (1+a\cdot t)$ with $a > 0$ and sufficiently small
\EndFunction

\vspace{1ex}

\Function{GetNeighbor}{$\mb{x},t$}
  \State \Return $\mb{x}+\mb{\Delta}$ with $\mb{\Delta}_i$ chosen so that $p(\mb{\Delta}_i)$ is appropriate to $t$\label{SA:GetNeighbor}
\EndFunction
\end{algorithmic}
\caption{A simple simulated annealing algorithm. \cite{weise2009global, sa_pres}\label{algo:SA-simple}}
\end{algorithm}

\subsection{Other Global and Non-Gradient Algorithms}
As with local methods, also in case of global and non-gradient methods, more publications and algorithms exist than can be covered here. In this Section it is attempted to sum up two additional methods and give a very brief description of them.\\

\textbf{Neural networks (NN)} are software implementations of networks of artificial neurons. The neurons are connected knots in a graph-like structure. The connection topology is set up according to the task which the NN is supposed to solve. In a learning phase the NN is trained. During this phase neurons can get added or removed, threshold parameters are tuned and the weight of the connections is modified. The learning process itself is often a high dimensional nonlinear optimization problem in itself. Although neuronal networks are often used in classification, signal processing and detection systems, they can also be used as optimizers and parameter estimators \cite{cochocki1993neural}.\\

\textbf{Quantum annealing search} is a meta-heuristic optimization strategy, which is similar to SA but the effects which are mimicked come from quantum mechanics.\\

Other methods exists, like: random search \cite{zabinsky2009random}, random optimization, parallel tempering \cite{earl2005paralleltempering}, swarm algorithms \cite{Pontani2010particleswarm} and many other.

\chapter{Application to Big Datasets}
This Chapter focuses on general challenges which arise when large amounts of data have to be fitted. Although the term ``Big Data'' often refers to solving graph-related problems on public and business data, model fitting on large scientific datasets poses similar challenges. Hence, model fitting on large amounts of scientific data can be considered as a sub-field of ``Big Data''. Both fields have to cope with two main challenges, data handling in the sense of storage and IO, and to provide massively parallel and scalable processing to compute results.

\section{IO and Input Data Format}
\label{sec:io_big_data_input}
An input data format for a nonlinear fitter, like the one presented in this work, should satisfy the following general requirements:

\begin{enumerate}
 \item The data should be stored as compact as possible\label{bigdat_compact}
 \item The data should be stored as structured as feasible\label{bigdat_structured}
 \item The data should be stored in a way that it can be partitioned easily\label{bigdat_part}
 \item The data format should fit in the environment of the user and be portable\label{bigdat_env}
\end{enumerate}

The points above might seem obvious, but are important to mention and worth discussing. The points \ref{bigdat_compact} and \ref{bigdat_structured} are sometimes conflicting. Consider for instance a flat binary format\footnote{Consider dumping a table or matrix as binary data with \texttt{fwrite()} in a specific way, possibly compressed.} and XML\footnote{Extensible Markup Language} for storing tabular, numeric data. One can easily see that the binary format will be possibly the most compact way to store this kind of data but is not very structured. Whereas the XML format offers a great amount of structure but introduces also a lot of redundancy with its tags and fields. Compromises have to be found to leverage advantages and disadvantages between structure and size.\\

Point \ref{bigdat_part} refers both to the problem of parallel file reading as well as distribution in the processing computer\footnote{computer can be anything parallel from a loose cloud to a massively parallel multiprocessor system}. Especially in loosely coupled computers, the data have to be already distributed among different processing nodes since no single storage medium exists to hold all of the data, or is not connected fast enough to read the data more than once. But also for tightly coupled computers with fast IO, the data has to be stored in a way that minimized (quasi) random access and can benefit as much as possible from burst access.\\

Often point \ref{bigdat_env} dictates the format, since the data is collected by instruments which output only one special format. Historical reasons also often play a role, especially when data is reprocessed. In such cases an investigation of benefits from conversion and unification can be useful.\\

Let us consider the case where the data format can be chosen more or less freely, be it because of a successful conversion or design decision from the start. In the following, two methods are presented for this case. The first one is ASCII output combined with the transparent writing and reading routines of \texttt{libbzip2} \cite{libbzip2}. This is a very pragmatic approach but has proven to be very easy to use and shows a reasonable performance. The second method is to use HDF \cite{hdf5web}, the Hierarchical Data Format, and the corresponding routines.

\subsection{Transparently Compressed ASCII}
It is often claimed that ASCII files are not usable for exchanging and storing large amounts of data. Arguments are, that they are too big and parsing them is too slow and computational expensive. However, in practice this is often not true as the following quote from the renown Usenet group \texttt{comp.lang.c} emphasizes it:

\begin{finequote}
\textbf{Q:} How can I write data files which can be read on other machines with different word size, byte order, or floating point formats?\\

\textbf{A:} The most portable solution is to use text files (usually ASCII), written with \texttt{fprintf} and read with \texttt{fscanf} or the like. [...]. Be skeptical of arguments which imply that text files are too big, or that reading and writing them is too slow. Not only is their efficiency frequently acceptable in practice, but the advantages of being able to interchange them easily between machines, and manipulate them with standard tools, can be overwhelming.\\[3ex]
--- {\footnotesize From: \texttt{comp.lang.c} FAQ list, The C FAQ, \url{http://c-faq.com/misc/binaryfiles.html}}
\end{finequote}

When dealing with multiple terabytes of data, the argument regarding the size might become valid. Fast storage which can hold many terabytes is expensive and therefore a more space-efficient storage format might be desirable.

One solution for this is to use compression and IO functions of the \texttt{libbzip2}. This library offers---among others--functions which can be used together with the standard C library functions to write bzip2 compressed, portable ASCII files. Other compression libraries have similar APIs, and can also be used. These functions are: \texttt{BZ2\_bzReadOpen}, \texttt{BZ2\_bzRead}, \texttt{BZ2\_bzReadClose}, \texttt{BZ2\_bzWriteOpen}, \texttt{BZ2\_bzWrite}, \texttt{BZ2\_bzWriteClose} \cite{libbzip2}. All these function require typeless pointers to an N-byte large piece of memory as input or output parameters, for the writing and the reading functions respectively. One can easily fill this memory with ASCII encoded characters by using standard C library functions like {sprintf}, or converting it to other data types using functions like \texttt{strtok, strtol, atoi} and others. It is also possible to write wrappers for functions like \texttt{fprintf} to make outputting of compressed data completely transparent to the user.\\

\subsubsection*{Performance}
For files over 2\,GB in size, Bzip2 is approximately three to four times faster for decompressing than for compression, as own experiments showed. While reading (e.g. during parsing) compressed ASCII data, approximately a factor of 3.6 more time is necessary compared to reading the same data as uncompressed plain text\footnote{all measurements tested with bzip2 1.0.6, highest possible compression ratio, build with icc, one thread used for compression/decompression, on server with Intel Xeon CPU E5-2690 @ 2.90GHz, RAID5 SATA2 7200rpm HDD. The HPC installation used for this test showed equivalent results.}.
An experiment with large, compressed ASCII files (the input data of the prototype) has shown, that in this way a hardware IO bandwidth of 7.5\,MB/s can be achieved. This corresponds to 34\,MB/s of uncompressed data, since the effective size after decompression is higher. This is a rather low IO bandwidth, considering that a modern SATA2 hard drive should deliver 120\,MB/s. For ASCII files, like the ones used in this work, bzip2 shows compression ratios around 1:3.5. If the bandwidth of the IO fabric is saturated during input file reading (see also Section \ref{sec:proto-IO-perf}), this disadvantage can turn to an advantage, since much less data has to be transfered.\\

To illustrate this, take the example of the HPC system used in this work (compare to Section \ref{sec:taurus}). Under ideal conditions the IO bandwidth of the fast file system is 20\,GB/s. When 4000 Cores read data, 5\,MB/s remain for each core. Since the data used in this thesis is compressed, at a ratio of 1:3, the effective bandwidth would be approximately 15\,MB/s.

\subsection{Hierarchical Data Format (HDF)}
\label{sec:hdf5}
The HDF format is a storage as well as a meta data format which stores data in a structured binary format. A corresponding software library allows access to this data model from a variety of language like C/C++, Fortran, Java and some others. It provides features like data compression, parallel IO, MPI-IO bindings and on the fly partitioning and sub-setting \cite{hdf5web}. The most recent version is HDF5, which differs substantially from its predecessor HDF4. In the following, only HDF5 is discussed and referred to as HDF.\\

As the name suggest, HDF is hierarchical organized. It basically consist of two objects which are arranged in a tree-like structure. The HDF documentation explain this as follows:

\begin{finequote}
HDF5 files are organized in a hierarchical structure, with two primary structures: groups and datasets.
\begin{itemize}
 \item HDF5 group: a grouping structure containing instances of zero or more groups or datasets, together with supporting metadata.
 \item HDF5 dataset: a multidimensional array of data elements, together with supporting metadata.
\end{itemize}

--- {\footnotesize From: The HDF5 Documentation; \url{http://www.hdfgroup.org/HDF5/doc/H5.intro.html\#Intro-FileOrg}}
\end{finequote}

The data in a dataset can be either atomic (integer, double, ...) or a user-defined compound type. The corresponding library functions help to declare and use user-defined data types. The internal control structures in the files are organized as tables and B-trees and the files can be stored either in chunks or contiguous on the storage medium. The meta data structure has equal design freedom as the data itself, although HDF assumes that the meta data can be stored in small key-value-like structures. Nevertheless it is possible to generate very complex meta data and store them in a HDF dataset itself and point to this dataset in the key-value attribute descriptor. \cite{hdf5intro}

\subsection*{Performance}
Performance gains compared to other format and methods are achieved by the performance-optimized library infrastructure and bindings to low level hardware drivers. Furthermore, optimal partitioning and compression is possible. The performance of HDF has not been investigated for this work.

\section{Parallelization and Memory Organization}
Since most of the nonlinear fitting methods are based on basic linear algebra, it should usually not be a problem to parallelize them. In all local gradient methods investigated in this work, the matrices and vectors can be blocked and partial results can be exchanged in collective operations among the nodes. However, simpler methods like the Gauss-Newton or Levenberg-Marquardt require much less global collective data exchange of intermediate results, than more complex methods like BFGS. A general quantitative statement about trade-off of different methods, regarding parallelization, is difficult to make. Global data exchange or reduction almost always contains waiting time for some processors, which directly decrease scalability, hence one would usually tent to use a method which requires a minimal amount of these operations.\\

Parallel file IO is necessary, especially for reading observational data. In most cases this data will be stored on large NAS storage devices, which are connected via an IO fabric to the computation units. As already elaborated in the previous Section, using such global file systems can be done efficiently using special high-performance data formats like HDF. But also pragmatic approaches like plain text files or compressed ASCII can saturate the IO bandwidth of HPC installations, as can be seen in section \ref{sec:proto-IO-perf}.\\

Another strategy would be to store the data on the local disks of the compute nodes, if available. Local storage is mainly available in general purpose clusters. Machines like the Cray XK, XE and XC, IBM BlueGene series and SGI UV systems regularly do not have local disks. Moreover, the local storage in almost all HPC cluster installations is much smaller than the global file system.

\section{Selection of Suitable Algorithms}
The selection of a suitable mathematical method or algorithm must be made specifically for the problem at hand. in general it can be stated that with a nonlinear problem the input data has to be accessed multiple time, since gradient as well as non-gradient methods will require multiple ``iterations'' to converge. For each of these iterations the sum of least squares has to be computed, this requires accessing the data. In most cases, it will be therefore of advantage to store the observations in memory to avoid IO operations. In this way the data can be accessed fast. However, this in-memory certainly comes with the downside of a high memory usage.\\
Many gradient-based methods need the complete Jacobian for the fit to update it with approximations to avoid the expensive re-computation of all partial derivatives in each iteration. With many observations (or measurements) and complex models with multiple parameters this Jacobian will also be large. This conflicts with memory requirement of the input data, which is also stored in memory, in many cases it will not be possible to store the input data \textit{and} the full Jacobian in the memory. It is probably often the case that not even the Jacobian alone can be stored in memory.\\

A solution for this dilemma can be to use hybrid algorithms. The parameters of which the partial derivatives are easy to compute are fitted using a simple gradient-based method which does not require the full Jacobian but only the normal matrices. These normal matrices can be computed block-wise to safe memory, but must be computed in every iteration. The parameters of which the partial derivatives are computational expensive, on the other hand, can be fitted using a non-gradient method. In this way the computation of these partial derivatives can be avoided. But it comes probably with the cost that more ``non-gradient iterations'' are needed in which the model have to be computed. An optimal ration between gradient-based and non-gradient methods in such a hybrid algorithm is problem dependent and might require tests and experimentation.

\section{Relation to the Gaia-specific Task}
As mentioned in the introduction, the prototype implemented in this thesis is specifically designed for the detection of GWs in simulated Gaia data. In this case the, input data format is comma separated ASCII text. The usage of HDF was not possible since it would have required too many modifications in the AGISLab software (compare to Item \ref{bigdat_env} in the list in Section \ref{sec:io_big_data_input}). A conversion was not deemed necessary for the prototypical purposes.\\

Because of the reasons elaborated in the previous section (also compare to Section \ref{sec:in-memory}), the decision has been made to store the observational data in memory during computation. The amount of memory available in the HPC systems used for the experiments is not sufficient for storing the complete Jacobian. The gradient based fitting method have been chosen accordingly (see Section \ref{sec:Selection_of_Fitting_Algorithms}).
\label{chap:fitting-large-datasets}

\chapter{Description and Architecture of the Prototype}
Besides of the theoretical descriptions of different methods and algorithm, a prototype which implements selected methods is the main part of this work. The prototype has been implemented to show that it is computationally and practically feasible to conduct a highly nonlinear parameter estimation on large datasets. The focus for this practical part of the work is solely on the search for GW signals in simulated Gaia data.\\

This following Chapter describes the architecture and functionality of this prototype. In the first Section, selection criteria for the implemented fitting algorithms are elaborated. In the following, the overall architecture is presented, followed by a performance analysis.

\section{Selection Criteria of Algorithms for the Prototype}
\label{sec:Selection_of_Fitting_Algorithms}
Constraints regarding the algorithms which can be used arise from two facts. First, with many data points, limitations concerning the amount of intermediate results occur. Second, the model to fit poses constraints regarding the order of partial derivatives and computational expenses. In the Gaia specific task, the following constraints apply:

\begin{enumerate}
 \item Although second (partial) derivatives of the model could be provided, they would be computational complex and their contribution to the solution is small.
 \begin{itemize}
  \item Therefore, the algorithm must not need second partial derivatives (Hessian matrix).
 \end{itemize}
 \item The matrix containing the first partial derivatives (Jacobian matrix) can not be stored completely, it must be computed and added block-wise.
 \begin{itemize}
  \item \label{item:complete-jacobian}Hence, update schemes, which require the complete matrix, can not be used.
 \end{itemize}
 \item The behavior of an approximated Jacobian matrix is unknown for the Gaia-specific model.
 \begin{itemize}
  \item Therefore, algorithms which use an approximated Jacobian should not be used. This point coincides with point \ref{item:complete-jacobian}, since most of the update schemes which  approximate the Jacobian need access to the full Jacobian.
 \end{itemize}
 \item The highly nonlinear nature of the problem and the unknown limits to the parameter space requires an global optimization.
 \item The global search or optimization part should not require nonlinear parameter tuning itself. For some global optimizers, internal ``configuration'' parameters---like the input weights of neural networks---have to be tuned, which requires a nonlinear parameter estimation itself.
\end{enumerate}

The considerations listed above lead to the decision to implement a hybrid approach for the prototype. The parameters, for which a gradient-based approach is not feasible, are fitted using a global optimization algorithm. As a quality function, the statistics of the linear local fit of the other (linear) parameters is used. If a ``good'' set of parameters has been found, all parameters can be fitted using a local fitting algorithm.\\

Regarding the algorithm for the local method, the selection is straight-forward, since all the methods, which need the complete Jacobian stored for updates, are ruled out. Even for the small datasets considered here, this would not be feasible due to limitations in memory size. This leaves us the choice between the Levenberg-Marquardt (see Section \ref{sec:levenberg-marquardt}) and the Gauss-Newton method (see Section \ref{sec:gauss-newton}). Finally, the Gauss-Newton method was chosen because it can be used directly if only a linear fit is required and it does not use an additional parameter (damping parameter) which has to be controlled.\\

In a first attempt the Simple Genetic Algorithm (SGA) by Goldberg \cite{goldberg1989genetic} was selected as the global optimization algorithm. It is a straight-forward implementation of the biological principles of genetic evolution and has a minimal set of parameters, for which even a default exists. The second algorithm considered was simulated annealing (SA), the decision not to implement it was based on experiences with simple testing examples which showed promising results with evolutionary algorithms. Other algorithms like neural networks require a very complex nonlinear parameter tuning for themselves and are hence not suitable. Evolutionary strategies (ES) have not been chosen in the beginning, because they often have a small population size, and this was considered as a disadvantage in case of a noisy search landscape. A selection from many individuals was considered more advantageous, to prevent premature convergence to local optima.\\

However, numerical experiments showed that an evolutionary algorithm is not feasible to search the frequency space (see Section \ref{sec:freq-search}). The merit function for the frequency has an extremely narrow optima, and lot of local optima. The probability to place an individual close enough to the global optimum (right frequency) to ``get grip'' on the objective function in subsequent steps, is too low. In this way the global optimizer get stuck in a local optima almost certainly.\\

On the other hand, a global optimizer can be used to circumvent the computation of the highly complex derivatives for the direction of the wave. For the search of the GW directions, the SGA algorithm showed a very bad performance. A simple $(1,\lambda)$-ES \cite{rechenberg1581evolutionsstrategie} has been implemented instead. A general problem with classical GA is to find a sufficient chromosome encoding, which is working correctly with the crossover and selection operation. Goldberg suggested a BCD binary string coding for multiple parameters (page 82 in \cite{goldberg1989genetic}) when multiple numbers have to be encoded. However, in this case the SGA algorithm with binary coding lead to divergence and the effect of many ``cloned'' individuals. Cloned individuals have the same chromosome and lead to unnecessary reevaluation, which is computational expensive in this case. Since every individual has to get its fitness value, the objective function has to be computed for every individual. But this computation results in the exact same value for two cloned individuals. One could work around this issue with lookup tables or such, but this was deemed to just mitigates the effects of a deeper problem.\\

The $(1,\lambda)$-ES, on the other hand, does not need a chromosome, and the whole coding problem does not exist. At the start, the search space is uniformly covered by $\lambda$ random individuals. In every generation afterwards, the best individual is selected and the parameters are variated using a Gaussian distributed random number and a decreasing step size. Problems with local optima can occur if they are further away from each other than the distance which can be covered with a certain probability of the step size and the Gaussian random number. In the GW direction search this problem has not observed.

\section{Software Architecture of the Prototype}
The software architecture of the prototype resembles a toolbox which the user can use to build his own experiments. The local (gradient-based) as well as the global (evolutionary) algorithm can be used independently from each other and from the Gaia-specific task. The Gaia-specific routines are separated and some independent mathematical tools are provided. In the following the prototype will be called ``GaiaGW'' (for Gaia Gravitational  Waves) for simplicity.

\begin{figure}[H]
\centering
 \includegraphics[keepaspectratio, width=0.4\textwidth]{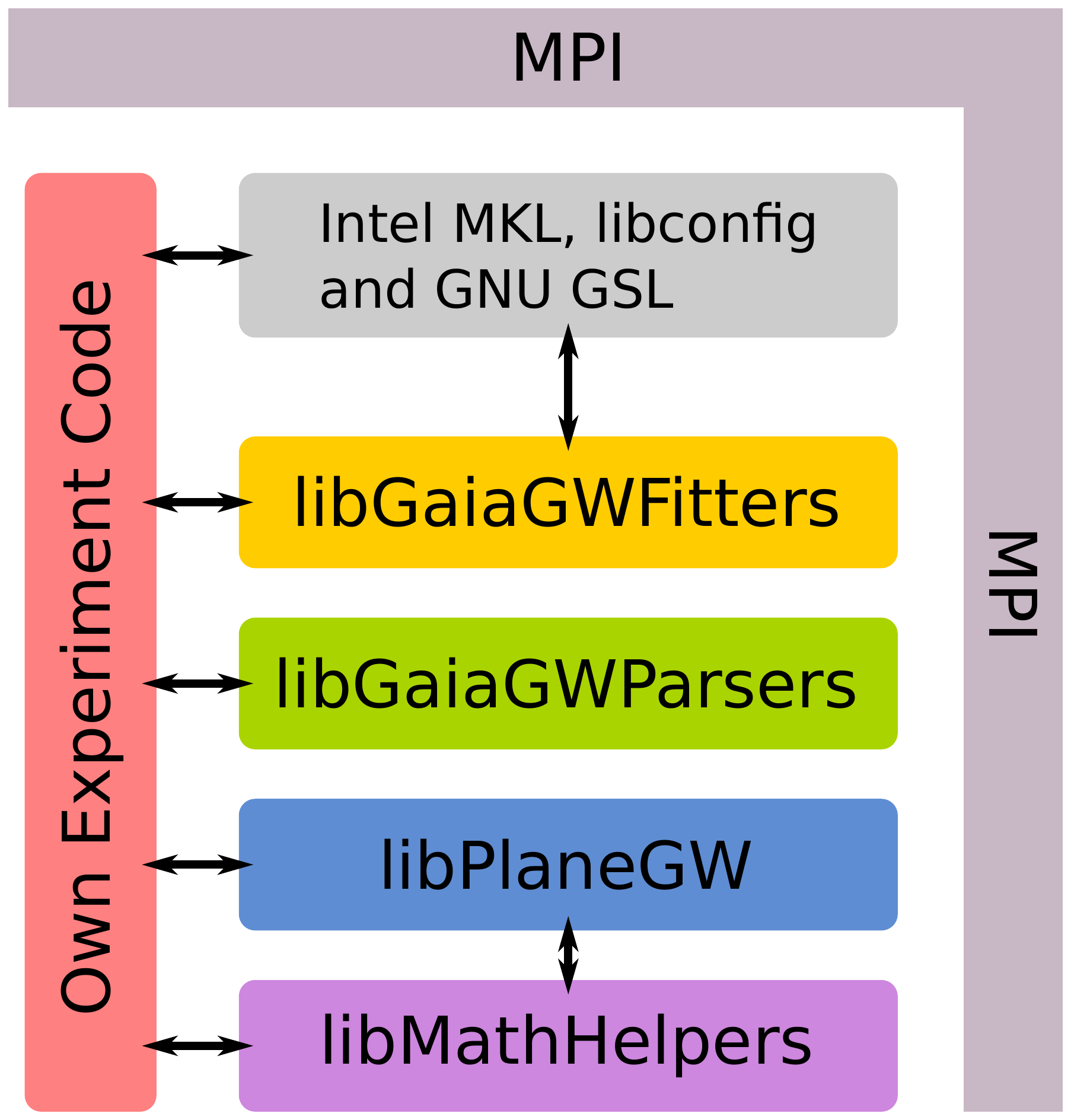}
 \caption[Architecture of the prototype]{Architecture of the prototype. The software is structured in independent libraries (colored boxes) and uses some external libraries (gray). MPI is used for communication and can be accessed throughout all parts of the software. The user has to provide glue code to conduct his experiments.\label{fig:proto-architecture}}
\end{figure}

As Figure~\ref{fig:proto-architecture} shows, the toolkit is organized in four libraries. These libraries make use of the following external libraries: libconfig\footnote{\url{http://www.hyperrealm.com/libconfig/}}, the Intel MKL\footnote{\url{http://software.intel.com/en-us/intel-mkl}}, the GNU GSL\footnote{\url{http://www.gnu.org/software/gsl/}} and a MPI\footnote{\url{http://www.mpi-forum.org/docs/docs.html}} library. The user has to provide glue-code to set up, control and conduct the experiment.

The four libraries in Figure~\ref{fig:proto-architecture} are documented using the doxygen\footnote{\url{http://www.stack.nl/~dimitri/doxygen/}} toolkit, this documentation can be found on the attached DVD. The following list provides a short overview of the libraries:

\begin{itemize}
 \item[] \texttt{ libGaiaGWFitters}
 \begin{itemize}
  \item Provides the generic API for the fitting routines, the Gauss-Newton least squares fitter and the SGA and ES global optimizers
  \item The Gauss-Newton method is parallelized using MPI, the SGA and ES optimizer require a pointer to an parallelized quality function
  \item Both methods can be compiled for GaiaGW-specific and standalone use
  \item Functions to check fit statistics are provided
 \end{itemize}
 \item[] \texttt{libGaiaGWParsers}
 \begin{itemize}
  \item GaiaGW-specific library
  \item Contains parsers for configuration, satellite attitude and data files, and state machines for the parser
 \end{itemize}
 \item[] \texttt{libPlaneGravitationalWave}
 \begin{itemize}
  \item GaiaGW specific library, could be used standalone to compute variations of star positions due to a plane gravitational wave (PGW)
  \item Implements the model of a PGW as well as the first partial derivatives
  \item Provides an interface suitable for the Gauss-Newton fitter
 \end{itemize}
 \item[] \texttt{libMathHelpers}
 \begin{itemize}
  \item Standalone library with mathematical functionality needed for GaiaGW
  \item Contains routines for unit conversions, trigonometry, linear algebra, number lists and vectorial spherical harmonics
 \end{itemize}
\end{itemize}

\subsection{In-Memory Computing}
\label{sec:in-memory}
In this prototypical implementation, all input data is read to and kept stored in the main memory distributed over the nodes. The decision for this in-memory storage has been made because of the IO to computation ratio of the problem. The software run for multiple hours (3-24\,h for the experiments presented in Section \ref{sec:gw_experiments}), the reading and parsing of the input data is done in seconds (5-120\,s for smaller and larger data sets respectively). Hence, the time spend for computation ($t_{\mathrm{Comp}}$) is high compared to the time necessary for IO ($t_{\mathrm{IO}}$). The speedup if the IO would be overlaid with the computation tends to 1 in this case.

\begin{equation}
 \label{eq:speedup-hiddenIO-fit-sumtime}
 S_{\mathrm{HiddenIO}} = \dfrac{t_{\mathrm{Comp}} + t_{\mathrm{IO}}}{t_{\mathrm{Comp}}} \approx 1~~\text{with}~~\dfrac{t_{\mathrm{IO}}}{t_{\mathrm{Comp}}} \rightarrow 0
\end{equation}

It is also worth noting that the average period in which the data of each observation is needed (to compute the model) is substantially less than the time needed to read the input data. The data is needed once every Gauss-Newton iteration, in the experiments from Section \ref{sec:gw_experiments} one iteration takes 0.05\,s for small and up to 3\,s for large datasets. Hence, streaming the data is also not an option.\\

This consideration led to the design decision to use pure in-memory computing. The only intermediate results written to disk are log files. Considering the final, real dataset size of about 50\,TB, this should not be a problem, since 50\,TB of main memory is available in many large HPC installations and will be easier to find in the future.

\section{Parallelization}
\label{sec:proto-parallelization}
To obtain optimal portability, the message passing interface (MPI) was selected for parallelization. This allows GaiaGW to run on large shared memory as well as distributed memory machines. The type of parallelization can be classified as coarse grained, since data exchange is only necessary once every iteration during the local gradient-based fit.\\

In the first program step, the data is read from the input files in parallel. Each MPI process read the part of the data assigned to it. In this way each MPI process also has a unique set of data points.\\

\begin{figure}[H]
 \centering
 \includegraphics[keepaspectratio,width=0.6\textwidth]{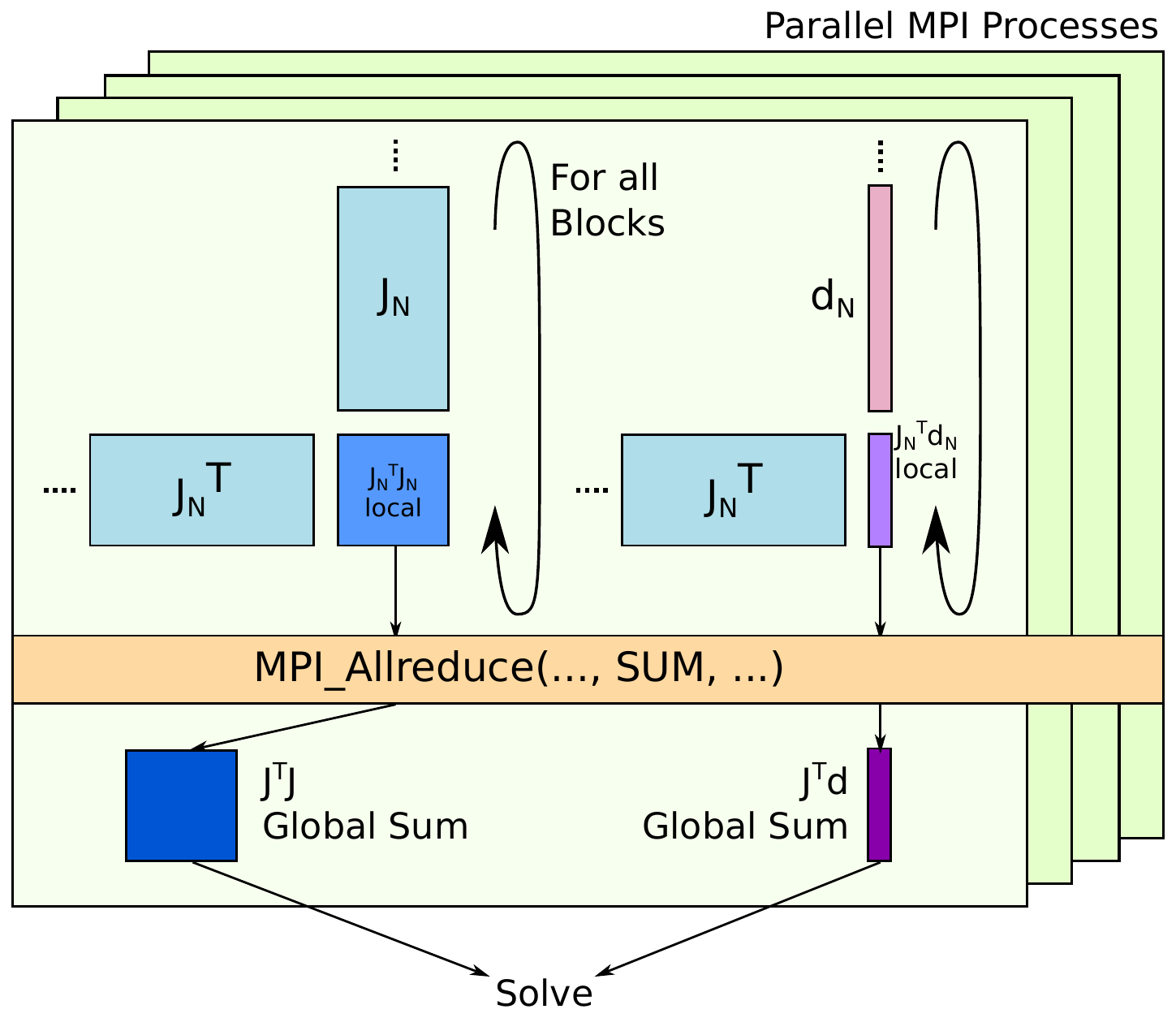}
 \caption[Parallelization scheme for the local method]{Parallelization scheme for the local gradient-based method. The local normal matrices are computed block-wise and summed up in a final reduction operation. These ``global'' normal equations are then solved and the result can be used by the processed which need them.\label{fig:gauss-newton-parallelization}}
\end{figure}

The local fitting routine can now work as shown on Figure~\ref{fig:gauss-newton-parallelization}. With its unique dataset, each process can compute the partial derivatives (i.e. Jacobian $\mb{J}$) and the model function (or $O - C = \mb{d}$) for these points. Each process can also independently compute a local version of $\mb{J}^\top \mb{J}$ and of $\mb{J}^\top \mb{d}$, which can be reduced over all processes (with the sum operator) to form the final set of normal equations. Since the full Jacobian cannot be stored in memory, the computation is divided into small blocks ($\mathrm{J_N}$ in the picture) and only the intermediate product is stored and updated ($\mathrm{J_N^TJ_N}$local), the same is done for the data vector $\mb{d}$. After all local blocks have been summed up to the local normal matrices, a global data exchange (MPI\_Allreduce) is necessary to sum up all the local results and form the global normal matrices. After the global exchange every process has the exact same set of normal matrices and can solve them independently or in parallel. For the Gaia-specific task, the system of normal equations is very small (7 parameters at most). Hence, the independent computation (each process does the same in this case, and get the same result) of the results induces less overhead than the usage of a parallel solver. For use-cases which yield larger normal matrices, the use of a parallel solver can be considered.

Since it is not possible to store the whole Jacobian for the data held by one process, the local normal matrices have to be computed block-wise. This can also be beneficial from the performance point of view as Section \ref{sec:blocking_performance} shows.

Both genetic algorithms, the ES as well as the SGA, are parallelized via their fitness functions. As can be seen in Section \ref{sec:profiling} the runtime spent in the ES algorithm itself\footnote{generating individuals, mutation and stepping etc.} is negligible compared to the time spend to compute the fitness of one individual. The fitness function has to be provided by the user, in this way the user also can control its parallelization.

\section{Generic Fitting Routines API}
The \texttt{libGaiaGWFitters} provides three fitting functions which have a generic interface (see code Listing \ref{lst:fittingAPIs}). When using the local routine, all the user has to do, is to provide one function for generating the Jacobian matrix and one function for computing the model values ($C$). For the global genetic routines a pointer to the user-written fitness function has to be provided.\\

All these routines need to have access to special data structures: \texttt{nonlinearModel}, \texttt{dataIndex} and \texttt{FitResult}. The \texttt{nonlinearModel} describes the properties of the model to fit (see Appendix \ref{sec:apx:model}). It contains the number of parameters to fit, a bit-mask for fixing parameters, parameter factors for the scaling of the design matrices and the starting and resulting values before and after the fit.\\

The \texttt{dataIndex} describes the data, it is a wrapper for two arrays with knowledge about its length (see Appendix \ref{sec:apx:index}). It contains a field size which counts the number of elements in the arrays, a pointer to an array of Gaia observations (which can be \texttt{NULL} for standalone purposes) and a pointer to an array of data values.\\

Results of the fit are stored in a data structure called \texttt{FitResult}, it contains some statistical parameters which are computed during the Gauss-Newton step. Appendix \ref{sec:apx:fitresult} shows all the fields in this structure.

\begin{lstlisting}[style=CA,caption={Interfaces for the genetic as well as the Gauss-Newton routine. These APIs are generic and not limited to the Gaia specific application.},label={lst:fittingAPIs}]
int [ES|SGA]Fit(
    double (*[ES|SGA]objfunc)([ES|SGA]Individual*, void*),
                               // function ptr to objective function
    void *infoPtr,             // generic ptr, handed through to
                               // objective function
    nonlinearModel *model,     // ptr to model struct
    FitResult *fres            // ptr to fit result, not all
                               // fields filled for ES and SGA
);
 
int nonlinearGaussNewtonFit(
    void (*ModelFct)
         (nonlinearModel*, dataIndex*, uint64_t, uint64_t, double*),
                              // function ptr to the function
                              // which gives C (from (O-C)^2)
    void (*JacFct)
         (nonlinearModel*, dataIndex*, uint64_t, uint64_t, double*),
                              // function ptr to the function
                              // which gives J from JtJx = Jtd
    nonlinearModel *model,    // ptr to model struct
    dataIndex *idx,           // ptr to data index
    double *data,             // ptr to O
    FitResult *fres           // output statistics
);
\end{lstlisting}
\vspace*{3ex}

The function which computes the partial derivatives and the model for the local method must provide a specific interface which is shown in Listing~\ref{lst:jacmodAPI}. The parameters are the following: \texttt{*model} is a pointer to the description of the model to be fitted, \texttt{*idx} is a pointer to the index which holds the data points, \texttt{start} and \texttt{blocksize} control the beginning and size of the current block to compute and \texttt{*J} or \texttt{*d} are the output matrix or vector respectively. The Jacobian matrix is stored in a linear vector in ``Row Major'' order.

\begin{lstlisting}[style=CA,caption={Interfaces for the local fitting routines Jacobian and model computing functions. Both functions must provide this interface.},label={lst:jacmodAPI}]
void ModelFct(
    nonlinearModel *model,  // ptr to model description
    dataIndex *idx,         // ptr to index of observed data
    uint64_t start,         // block size controll
    uint64_t blocksize,     
    double *d               // output plain vector containing C
);

void JacFct(
    nonlinearModel *model, // ptr to model description
    dataIndex *idx,        // ptr to index of observed data
    uint64_t start,        // block size controll
    uint64_t blocksize,    
    double *J              // Jacobian matrix (#Obs X #Params),
                           // organized "obs1-p1, obs1-p2, ...,
                           // obsN-p1, ..., obsN-pM
);
\end{lstlisting}
\vspace*{3ex}

The objective function for the global algorithm must provide an interface which is shown in listing~\ref{lst:sgafitnessAPI}. Since this function computes the fitness of one individual in the genetic algorithm a pointer to an individual (\texttt{*indv}) must be provided. Furthermore a typeless pointer to an arbitrary information structure (\texttt{*ptrToInfo}) is passed through by the \texttt{ESFit} function (\texttt{ESFit::*infoPtr == ESobjfunc::*ptrToInfo}), this pointer can also be \texttt{NULL} if no additional information is needed.

\begin{lstlisting}[style=CA,caption={Interfaces for the global methods objective function.},label={lst:sgafitnessAPI}]
double [ES|SGA]objfunc(
    [ES|SGA]Individual *indv,
    void *ptrToInfo
);
\end{lstlisting}
\vspace*{3ex}

The interface for the SGA routines is the same, the naming is \texttt{SGA\_\_\_} instead of \texttt{ES\_\_\_}, so for SGA the type of the individual is not \texttt{ESIndividual} but \texttt{SGAIndividual}.\\

The model and index structures are explained in the Appendices \ref{sec:apx:model} and \ref{sec:apx:index}. Further and detailed documentation can be found in the doxygen documentation of the source code.

\section{Input File Parser}
The input files for the Gaia-specific part are structured, comma separated values (CSV) text files. One requirement to the data format has been that only minimal modifications of AGISLab are necessary to write it. CSV can easily be exported by AGISLab and it is portable. Furthermore the parser needed in GaiaGW can be kept simple. Listing \ref{lst:inputfile} shows the first four lines from such an input file. Real data would be much more compressed and likely come from a database interface. 

\begin{lstlisting}[style=CA,breaklines=true,caption={First four lines from an input file for the Gaia specific part of this work. The first line is comment, the data is in lines 2-4.},label={lst:inputfile}]
ContainerIndex-SourceIndex-ObsIndex#obsTime;residual[AL];residual[AC];obsSigma[AL];obsSigma[AC];k;propSrcDir;theta;gaiaPosition
0-0-0#64573246697715972;-1.979716501843143E-9;1.2061773663196718E-9;9.90325283218615E-10;2.750229060875767E-9;(0.5397663091766982, -0.23673623577577715, 0.807841745732715);(-0.5398280045489584, 0.23668047208729226, -0.8078168602085664);-1.40217753247108;(-6.939805186073785E10, 1.2061023251521103E11, 5.232626010291127E10)
0-0-1#64573252535520600;1.3505108142908284E-9;2.972511824357593E-9;4.906033437162371E-10;2.5057071326127255E-9;(0.53976630917618, -0.23673623577635478, 0.8078417457328921);(-0.5398280045022432, 0.2366804719947158, -0.8078168602669077);-1.4021755274796537;(-6.939821059678459E10, 1.2061015717170638E11, 5.232622767370489E10)
0-0-2#64573257390517459;5.087992249741546E-10;2.523816225452613E-12;4.4088159650238184E-10;2.298455091427712E-9;(0.5397663091757489, -0.23673623577683509, 0.8078417457330391);(-0.5398280044633926, 0.23668047191772498, -0.8078168603154273);-1.402173860043525;(-6.939834260917903E10, 1.2061009451232286E11, 5.232620070397295E10)
\end{lstlisting}
\vspace*{3ex}

\begin{figure}[htb]
 \centering
 \includegraphics[keepaspectratio, width=0.9\textwidth]{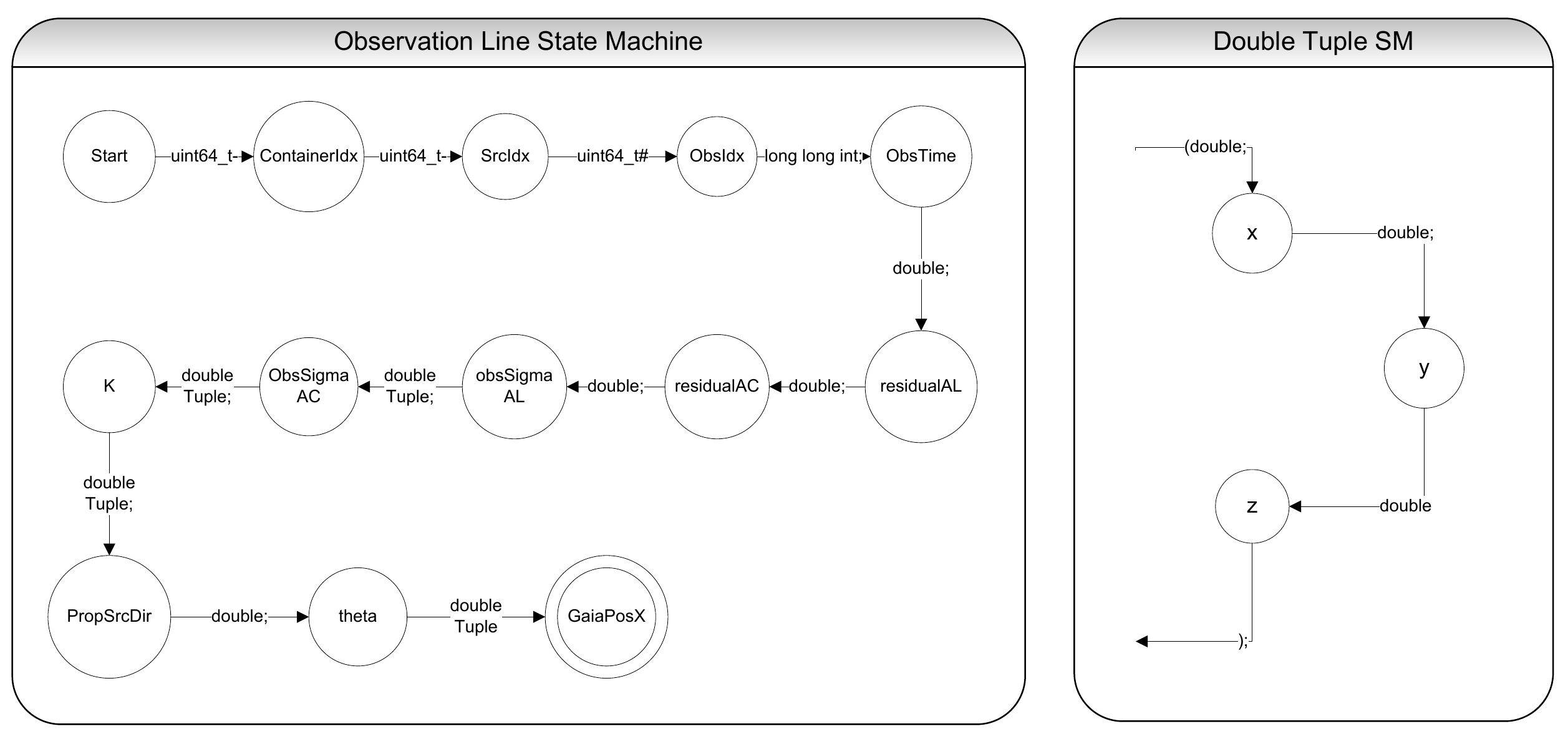}
 \caption[State Machine for Input Reader]{The state machine chart for the input reader of the observation files. The ``double tuple'' is parsed in a sub state machine on detection.\label{fig:sm-input-reader}}
\end{figure}

With this type of structured files it is advisable to implement the reading of the files as a state machine (SM). This has the advantage that data fields can be added or removed easily by inserting or deleting states in the SM. The SM in GaiaGW works on one line of the input data and transitions back to the start when a line was read and a new line is available. Figure \ref{fig:sm-input-reader} shows the state chart for the state machine used for the input files shown in Listing \ref{lst:inputfile}.

The ingestion of the character stream is done with the standard \texttt{strtok} and \texttt{strtod, strtol} functions.

\subsection*{Parallelization of File IO}
File reading is inherently parallel since the input is split into many\footnote{typically >10,000 files, significant more than CPU cores used for computation} files of equal size. Each MPI process performs the input reading and parsing on its set of files. Care is taken that each process reads approximately the same amount of data. In this way 128 MPI processes can saturate the IO subsystem of the Taurus cluster (machine descriptions see Section \ref{sec:taurus}), and achieve $\approx 20$\,GByte/sec reading bandwidth.

\section{Performance Analysis}
The performance analysis conducted of the prototype consists of three parts. In a first step profiling using VampirTrace has been used to determine hot-spots in the code. The second part focuses on the IO performance and is used to make sure that the prototype can saturate the IO bandwidth of the HPC installation used for the tests. Since the data format of future real mission data will be completely different, these measurements have only limited use. The third part covers VampirTrace/Vampir \cite{muller2007vampirtrace, vampir08} measurements of the Gauss-Newton as well as the ES routines and their communication behavior. They can give valuable insights regarding scalability weak-points and optimization potentials.

\subsection{The Test Machine and Environment}\label{sec:taurus}
All performance measurements have been conducted on the ZIH machine Taurus, unless explicitly stated otherwise. Taurus has 3 partitions (sometimes called islands), from which only the ``Sandy'' partition has been used for performance measurements. The machine can be classified as a typical general purpose, high performance Linux cluster. Table \ref{tab:taurus-sandy-architecture} gives a brief overview of this partition of the system:

\begin{table}[H]
\centering
 \begin{tabular}{ll}
  \toprule
  Component & Quantity \\
  \midrule
  Nodes                 & 270\\
  CPUs per Node         & $2~\times$ Intel Xeon E5-2690 (8 cores) 2.90GHz (Sandybridge) \\
  Hyperthreading        & Disabled \\
  Memory per Node       & 228 Nodes with 32\,GB \\
                        & 28 Nodes with 64\,GB \\
                        & 14 Nodes with 128\,GB \\
  Memory Type           & DDR3 1600\,MHz\\
  Interconnect          & ConnectX-3 Infiniband FDR (56\,Gb/s)\\
  IO Subsystem          & 20\,GByte/s Lustre\\
  Operating System      & bullx Linux Server release 6.3 (V1)\\
  MPI Implementation    & bullxmpi-1.2.4.3\\
  \bottomrule
 \end{tabular}
 \caption[ZIH Taurus HPC System Specifications, ``Sandy'' Partition]{Description of the ``Sandy'' partition of the Taurus HPC system. Taurus is a typical general purpose, distributed memory, computing cluster which is based on the x86\_64 architecture and Infiniband.\label{tab:taurus-sandy-architecture}}
\end{table}

\subsection{Profiling}
\label{sec:profiling}
The following profiles have been generated using 128 Cores and a dataset with 300,000 stars (79\,GB size). Table \ref{tab:vt_trace_300k} shows the profiles of three different runs, obtained with the profiling facilities of VampirTrace. They cover five independent fits each with a single iteration per fit. In the first case only the four amplitudes of the GW model have been fitted\footnote{The model is described in section \ref{sec:PGW-Model}, in this context it is only important to know that basically three scenarios are relevant.}. The second case includes the frequency in addition, and the third case is a fit over all seven parameters, including the directions. The times shown are in seconds and being the summed inclusive times. The run times from multiple calls are summed up, and top level functions include the times from low level functions too.\\

\begin{figure}[H]
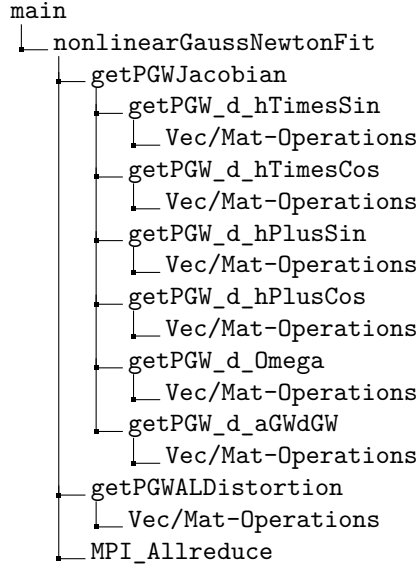

\centering
\begin{minipage}{0.4\textwidth}
 \dirtree{%
 .1 main.
 .2 nonlinearGaussNewtonFit.
 .3 getPGWJacobian.
 .4 getPGW\_d\_hTimesSin.
 .5 Vec/Mat-Operations.
 .4 getPGW\_d\_hTimesCos.
 .5 Vec/Mat-Operations.
 .4 getPGW\_d\_hPlusSin.
 .5 Vec/Mat-Operations.
 .4 getPGW\_d\_hPlusCos.
 .5 Vec/Mat-Operations.
 .4 getPGW\_d\_Omega.
 .5 Vec/Mat-Operations.
 .4 getPGW\_d\_aGWdGW.
 .5 Vec/Mat-Operations.
 .3 getPGWALDistortion.
 .4 Vec/Mat-Operations.
 .3 MPI\_Allreduce.
 }
\end{minipage}
 \caption[Simplified call tress of functions in profile.]{Simplified call tree of the functions in the profiling analysis.\label{fig:calltree}}
\end{figure}

To understand the call hierarchy better, Figure \ref{fig:calltree} shows a simplified call tree of the functions shown in the profile.

Table \ref{tab:vt_trace_300k} is extensively simplified for clarity, a more comprehensive profile can be found in Appendix \ref{apx:tab:vt_trace_300k}.

\begin{table}[H]
 \centering
 \begin{tabular}{lrrr}
 \toprule
                                & \textbf{Incl. Time}            & \textbf{Incl. Time}          & \textbf{Incl. Time}\\
  \textbf{Function}                      & \textbf{4 Parameters}              & \textbf{5 Parameters}            & \textbf{7 Parameters} \\
  \midrule
    vecVecDot3d                 &582.12                &682.19                &967.06\\
    vecScale3d                  &606.23                &718.68                &1046.72\\
    vecVecAdd3d                 &489.31                &587.89                &979.08\\
    matMatMul3d                 &626.96                &628.31                &2006.11\\
    matScale3d                  &328.50                &513.60                &883.27\\
    matVecMul3d                 &261.71                &316.46                &474.36\\
    matTranspose3d              &218.12                &216.58                &472.74\\
    localTriad3d                &-                     &-                     &136.16\\
    matMatAdd3d                 &56.47                 &111.33                &494.79\\
    \midrule
    $\sum$-Vec/Mat-Operations   &3169.42               &3775.04               &7460.29\\
    \midrule
    getPGW\_d\_hTimesSin        &935.53                &934.45                &932.74\\
    getPGW\_d\_hTimesCos        &936.50                &943.21                &938.63\\
    getPGW\_d\_hPlusSin         &961.40                &961.18                &955.10\\
    getPGW\_d\_hPlusCos         &989.17                &996.33                &1004.8\\
    getPGW\_d\_Omega            &-                     &1264.44               &1281.78\\
    getPGW\_d\_aGWdGW           &-                     &-                     &7932.27\\
    getPGWJacobian              &6326.12               &7787.81               &16182.40\\
    \midrule
    getPGWALDistortion          &2660.99               &2656.30               &2649.33\\
    \midrule
    nonlinearGaussNewtonFit     &9977.3 7              &11753.20              &20645.40\\
    MPI\_Allreduce              &856.07                &1166.40               &1651.42\\
  \bottomrule
 \end{tabular}
 \caption[Profiling results for three different work scenarios]{Shortened profiling results for three different work scenarios. The test setup was: 128 cores, 300k star dataset, one single fit with one iteration per scenario.\label{tab:vt_trace_300k}}
\end{table}

One can clearly see the following effects:
\begin{itemize}
 \item Most obvious to see is an increase of runtime when fitting additional parameters.
 \item The four derivatives of the four amplitudes (\texttt{getPGW\_d\_h\dots}) require the same computation time.
 \item The derivatives for the frequency are slightly more expensive with a factor of around 1.3 compared to the amplitudes (\texttt{getPGW\_d\_Omega}).
 \item To compute the derivatives of the two directions takes almost a factor of 9 more time than to compute them for one amplitude (\texttt{getPGW\_d\_aGWdGW}).
 \item Approximately half of the time for derivative computation (\texttt{getPGWJacobian}) is spend in routines which calculates generic 3D matrix/vector operations.
 \item To compute the derivatives takes 2.4, 3.0 and 6.2 times longer, for 4, 5 and 7 parameters respectively, than to compute the model.
 \item The matrix operations in the Gauss-Newton step alone (building $\mb{J}^\top \mb{J}$, $\mb{J}^\top \mb{d}$ etc.) are minor compared to the functions which compute the derivatives, the model and the MPI communication.
\end{itemize}

Those observations can be taken as a basis for some recommendations regarding future code. One major recommendation is to put effort in the efficient implementation of the 3D matrix and vector operations used to compute the derivatives. As the performance counter measurements in Section \ref{sec:vampirtrace} show, the current implementation only uses 50\% of the theoretical peak performance. A promising approach to increase the performance could be to use libraries which are specialized in small scale linear algebra operations. One of such libraries is the Intel Integrated Performance Primitives (IPP)\footnote{\url{http://software.intel.com/en-us/intel-ipp}} function collection.\\

The second recommendation is to use an MPI implementation together with an interconnect which can perform MPI reduction operations more efficiently, preferably in hardware. Since reductions are almost equal to sequential time, a further parallelization would not be successful. However, for larger datasets this effect is attenuated by weak scaling, which would correspond to Gustafson's law.\\

Another observation can be made when profiling the version which uses the $(1,\lambda)$-ES to find the directions of the wave too. In this case the flat profile looks the same for the Gauss-Newton part. The time spent in the ES itself for generating the individuals and selecting them is negligible. In all experiments the ratio of the time spend in the Gauss-Newton part compared to the time spend in the ES is approximately $1:10^{-7}$. These are, in absolute numbers, around 50-150\,µs per MPI rank per ES generation with 10 individuals.\\

\subsection{IO Performance}
\label{sec:proto-IO-perf}
The IO performance is of minor importance for the final purpose of the software implemented in this work. The input file format will be different for a final software. Furthermore will the time for reading the input data ($\approx 50$\,TB for a real dataset) be only a very small fraction of the overall runtime of the software. It is, however, necessary to implement the input and parsing of the data files as efficiently as possible. For evaluating this, two questions are important: first can the bandwidth of the IO fabric be saturated? Second, are there any delays in the IO of some processes?\\

To answer both of the questions a VampirTrace IO trace has been produced. Therefore a version of the code has been manually instrumented with user defined counters. The reason for that is that VampirTrace does not support IO events from the C function \texttt{getline}, which is used in GaiaGW.\\

\begin{figure}[H]
\centering
\includegraphics[keepaspectratio, width=0.95\textwidth]{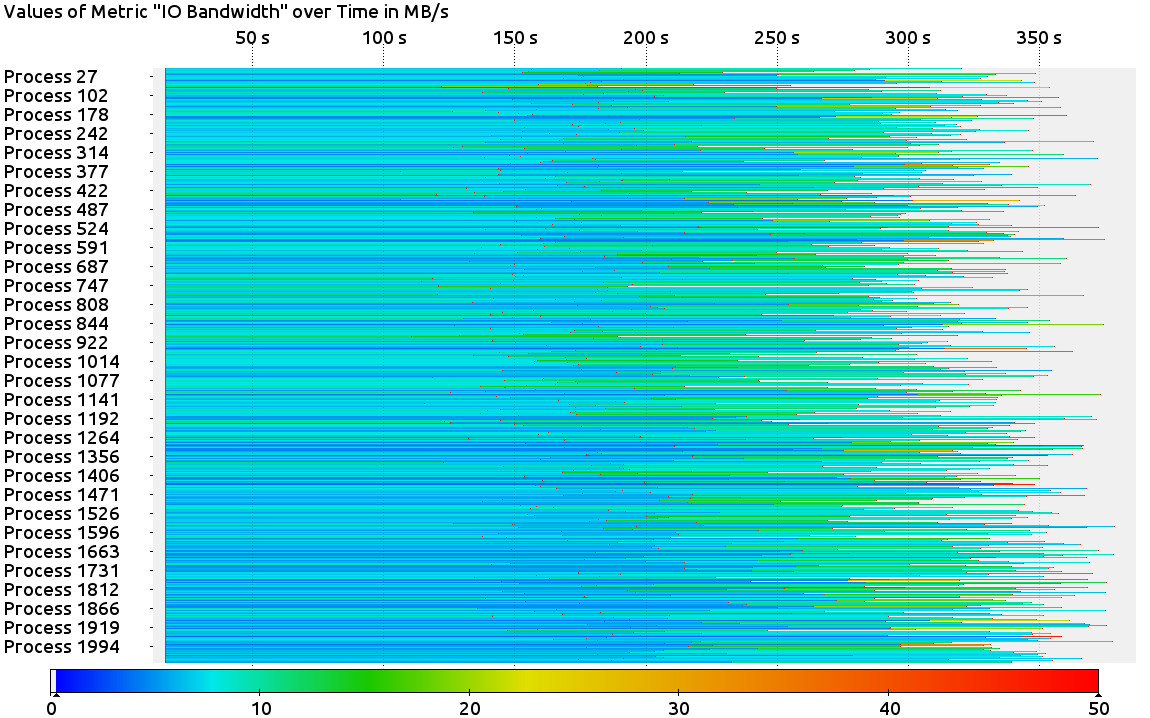}
\caption[]{Vampir IO trace\label{fig:vampirIO}}
\end{figure}

The result of the IO tracing test can be seen in Figure \ref{fig:vampirIO}. In this test, 2048 processes read 1.2\,GiB of data each, and simultaneously parsing it. The figure shows the program flow of the input parsing over time for each parallel process. The color coding is correlated with the IO bandwidth which each process uses. It is easy to see that the IO bandwidth can be saturated with the input parser. The bandwidth per process does not exceed 50\,MB/s at any time. But, one can also see that there are significant variations in bandwidth, and therefore, significant variations in reading time per process.

These variations of used (and hence available) bandwidth per process is in agreement with the behavior of user-space file system operations. In addition it is worth noting that in the current configuration of the Lustre installation of the Taurus system a first-come first-serve policy is active. A solution can be the use of direct-IO mechanisms, in this case no file handles have to be managed by the operation system, the Lustre controller and the software can directly access the object storage targets. This would also reduce the variance of reading times since effects like cacheing and sheduling in the operation system of the nodes is circumvented.\\

Another approach can be to use specialized file formats like HDF (see Section \ref{sec:hdf5}), which provides optimized reading APIs. If optimally configured, this also should lower the reading bandwidth variance, since these APIs can use direct IO in the background.

\subsection{Vampir Trace Analysis}
\label{sec:vampirtrace}

Trace analysis has been conducted for the Gauss-Newton method as well as the evolution strategy. The two main goals of this analysis are to find imbalances between computation and communication, and to find weak spots in the floating point performance.\\

\begin{figure}[H]
\centering
\includegraphics[keepaspectratio,width=\textwidth]{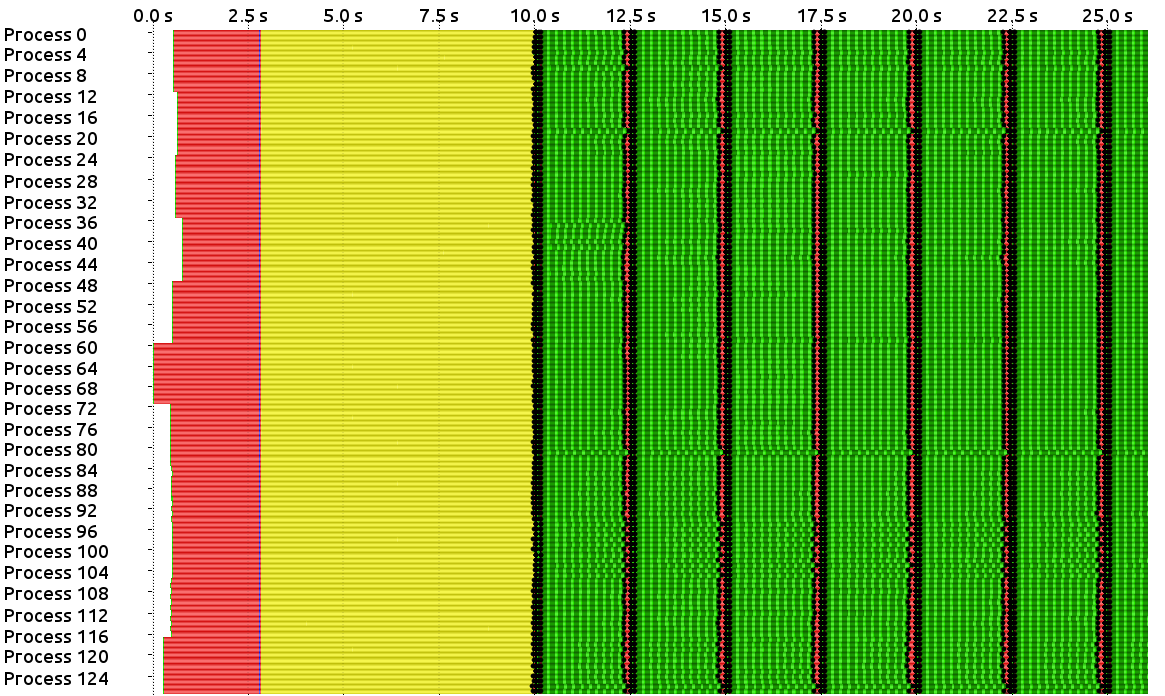}
\caption[General VampirTrace/Vampir overview of the prototype]{General VampirTrace/Vampir generated overview of the prototype. The color coding is: MPI is red, input parsing is yellow, green colors are the Gauss-Newton step. Wherein the darker green corresponding to the computation of the derivative (amplitudes only) and lighter green of the model.\label{fig:vamp_general}}
\end{figure}

The general behavior can be seen best in the example shown in Figure \ref{fig:vamp_general}. This Vampir screen-shot shows a run in which the Gauss-Newton method is used to compute fit statistics which are then fed into the ES as the quality function. The trace starts with the MPI initialization (red), followed by the parsing of the input data (yellow). After all data is read and some global parameters are exchanged, the ES starts. The Figure shows the evaluations of the fitness function (green colors) for the first 6, and parts of the 7th, individual. The fitness function used here is the post fit $\chi^2$ value of a fit of the amplitudes. Hence, the 6 green parts show one iteration of the Gauss-Newton fitter each. The time which is spent in the ES as such---for the generation of the individuals and selecting them---cannot be seen, since the duration is in the range of micro-seconds.\\

The black lines and dots marking the collective operations in the beginning and the end of each Gauss-Newton iteration. The Figure also shows some communication slack (also red between the black collective markers). This slack is caused (for instance) by processes 19, 82 and 126. They need longer time for computation of the model and the derivatives, since they have to compute them for a small excess of observations. For future implementations this can be mitigated by introducing an exchange step before computation which levels out input imbalances.\\

Larger Vampir traces, up to 4096 cores, have been produced and investigated. The communication slack is of equal amount and additional imbalances could not be found.

Hardware counter have been investigated using VampirTrace and Vampir with the corresponding PAPI \cite{mucci1999papi, browne2000portable} counters enabled during trace. Table \ref{tab:PAPI_TCM} shows representative results from one process for different counters, distinguishing between the functions which compute the model and the derivatives respectively. To investigate the floating point performance, it has been tried to use the PAPI \texttt{PAPI\_FP\_OPS} counter. The floating point performance reported by this counter was found to be unreasonably low (see last row of table \ref{tab:PAPI_TCM}). Independent tests showed that this counter reports wrong results on the Sandy-Bridge platform. After considering alternatives, it was found that the \texttt{PAPI\_TOT\_INS} counter gives a fairly representative measure of performance. This counter gives the successfully retired instructions per second, the theoretical maximum is 11.6\,G/s on this platform.

\begin{table}[H]
\centering
 \begin{tabular}{cll}
 \toprule
 \textbf{PAPI Counter}  & \textbf{Model}    & \textbf{Derivatives}\\
 \midrule
  \texttt{PAPI\_TOT\_INS}  & 5.5\,G/s      & 6.0\,G/s\\
  \texttt{PAPI\_L1\_TCM}   & 13.3\,M/s     & 1.9\,M/s \\
  \texttt{PAPI\_L2\_TCM}   & 926\,k/s      & 122\,k/s\\
  \texttt{PAPI\_L3\_TCM}   & 304\,k/s      & 88\,k/s\\
  \texttt{PAPI\_STL\_ICY}  & 727\,M/s      & 863\,M/s\\
  \texttt{PAPI\_BR\_INS}   & 345\,M/s      & 401\,M/s\\
  \texttt{PAPI\_BR\_MSP}   & 14.1\,k/s     & 2.4\,M/s\\
  \texttt{PAPI\_TLB\_DM}   & 281\,k/s      & 30.7\,k/s\\
  \texttt{PAPI\_TLB\_IM}   & 9.2\,k/s      & 1.7\,k/s\\
  (\texttt{PAPI\_FP\_OPS}   & 862\,M/s     & 823\,M/s)\\
 \end{tabular}
\caption[PAPI counter readings]{PAPI counter readings for the functions which compute the model and the derivatives.\label{tab:PAPI_TCM}}
\end{table}

The level 2 and level 3 cache misses are very low compared to the number of instructions per second. Assuming that every L3 cache miss leads to a complete stall, than it would take 150\,ns to fetch the data from memory in the worst case \cite{dram-access}. With 300\,k misses per second these are 45\,ms, in which the Sandy-Bridge core could theoretically compute 1.044\,GFLOP. One can hence state that only very minor performance problems can originate from memory accesses. Level 1 cache misses are higher by nature but still under 1\% of the instructions.\\

The ratio between the total number of branch instructions (\texttt{PAPI\_BR\_INS}) and the miss-predicted branches (\texttt{PAPI\_BR\_MSP}) is under 1\%, which is a very good value. Transaction look-aside buffer misses (data misses \texttt{PAPI\_TLB\_DM} and instruction misses \texttt{PAPI\_TLB\_IM}) are very low. The number of stall cycles (\texttt{PAPI\_STL\_ICY}) is still in a tolerable range, but can be optimized. It can explain a part of the gap to the maximal possible performance.\\

Since no integer counters are available on the systems, one can only speculate about how large the amount of floating point operations, compared to the total operations, is. The relatively high amount of branch instructions most probably originates from the calls of functions from the math library. The issue-to-retire latency of such calls and the integer operations associated with them can be a performance weak-spot. It should be considered to use in-lining and compiler features like IPO (interprocedural optimization) in the future code to mitigate such problems.

\subsection{Scalability}
\begin{figure}[htb]
\centering
\includegraphics[keepaspectratio, width=0.8\textwidth]{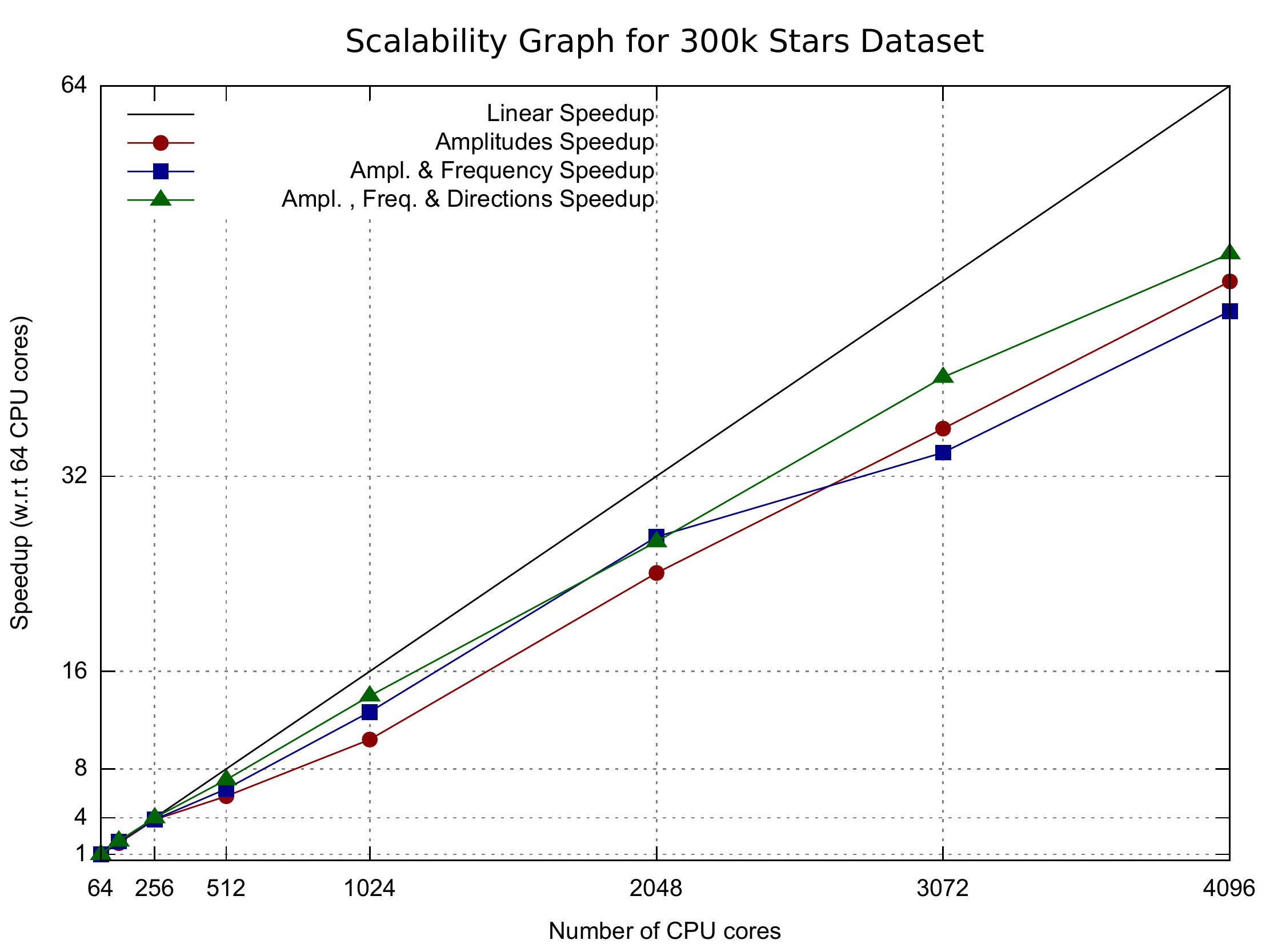}
\caption[Scalability test for a dataset with 300,000 stars]{Scalability test for the dataset with 300k stars. Three different scenarios are shown: only the four amplitudes are fitted, the amplitudes and the frequency is fitted and all seven parameters including the direction is fitted. The speedup is shown with respect to 64 CPU cores.\label{fig:300kspeedup}}
\end{figure}

\begin{figure}[htb]
\centering
\includegraphics[keepaspectratio, width=0.8\textwidth]{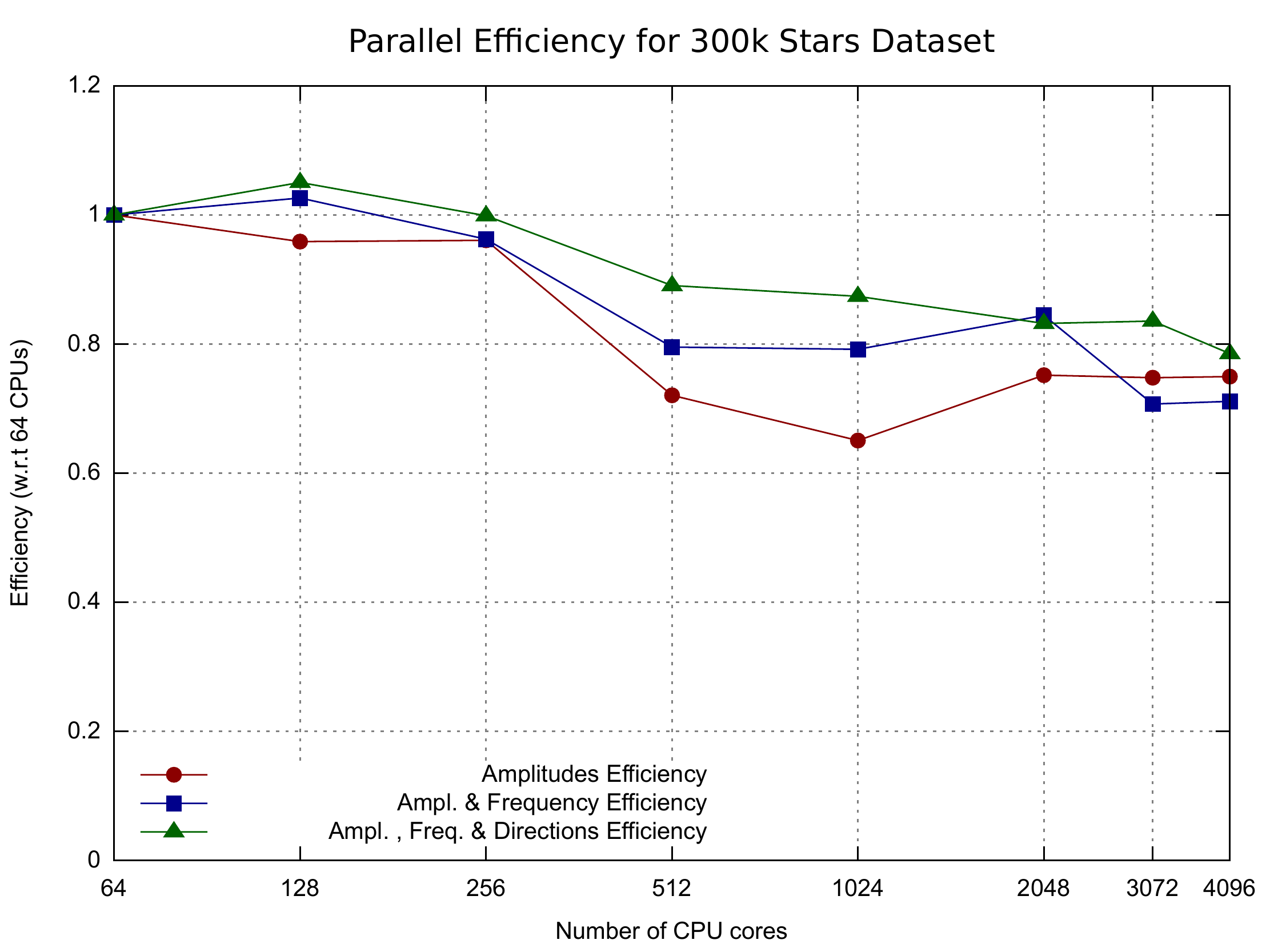}
\caption[Parallel efficiency plot]{Plot of the parallel efficiency for the dataset with 300k stars. Three different scenarios are shown: only the four amplitudes are fitted, the amplitudes and the frequency is fitted and all seven parameters including the direction is fitted. The efficiency is normalized to 64 CPU cores.\label{fig:300kefficiency}}
\end{figure}

The scalability has been tested with two datasets. One consist of 300,000 stars and has a size of 79\,GiB , the other consists of approximately 18,485,300 stars and is 4381\,GiB in size. In the following, the datasets are referred to as ``300k stars dataset'' and ``18M stars dataset'', respectively. The larger dataset could not be produced by the AGISLab software and has been manually created by copying the 300k dataset multiple times and concatenating these copies to one large dataset. It has to be stressed, that this dataset is not usable for Gaia-specific purposes and is only used to investigate the computer performance. Figures \ref{fig:300kspeedup} and \ref{fig:300kefficiency} show the speedup and the parallel efficiency of the prototype when applied to the 300k stars dataset. Each of the graphs show three different scenarios, in which: only the four amplitudes are fitted, the amplitudes and the frequency is fitted and all seven parameters including the direction is fitted. Each test consist of 30 fits with 5 iterations each, with 4096 CPU cores one iteration takes on average 0.123\,s, 0.144\,s and 0.222\,s respectively for the three scenarios. Due to memory limitations, 64 is the smallest number of CPU cores which can be used to process the 300k stars dataset. Hence, the speedup and efficiency must be given with respect to 64 CPU cores

One can clearly see, that for this relatively small dataset, no perfect linear speedup can be achieved in general. As the VampirTrace/Vampir analysis has already shown, this is mainly due to the ratio of time for computing versus time for collective operations and imbalances in the computational phase. However, even this small dataset still scales with over 70\% efficiency on 4096 CPU cores. 

According to Gustafson's law the speedup can be increased by weak-scaling. This method works very well in this case, as Table \ref{tab:bigdata_eff_speed} shows. The minimal amount of cores usable for the 18\,M stars dataset is 2048 cores, since only 2\,GB of memory are available. With this dataset the performance increase from 2048 to 4096 cores is almost linear. It should hence be possible to efficiently weak-scale to a final dataset with hundreds of millions of stars and hundred of thousands of cores.

\begin{table}[htb]
\centering
\begin{tabular}{clll}
\toprule
 \textbf{Number of CPU cores}         & \textbf{Runtime}               & \textbf{Runtime}               & \textbf{Runtime}\\
                        & \textbf{4 Parameters}          & \textbf{5 Parameters}          & \textbf{7 Parameters} \\
 \midrule
 2048                   &410\,s                 & 390\,s                & 734\,s \\
 4096                   &204\,s                 & 217\,s                & 366\,s \\
 \midrule
 Speedup with           &                       &                       & \\
 4096 cores              & 2.0                   & 1.8                   & 2.0\\
 \bottomrule
\end{tabular}
\caption{Speedup and parallel efficiency for the 18M stars dataset \label{tab:bigdata_eff_speed}}
\end{table}

\subsection{Influence of the Blocking on the Performance}
\label{sec:blocking_performance}
\begin{figure}[htb]
 \centering
 \includegraphics[keepaspectratio, width=0.8\textwidth]{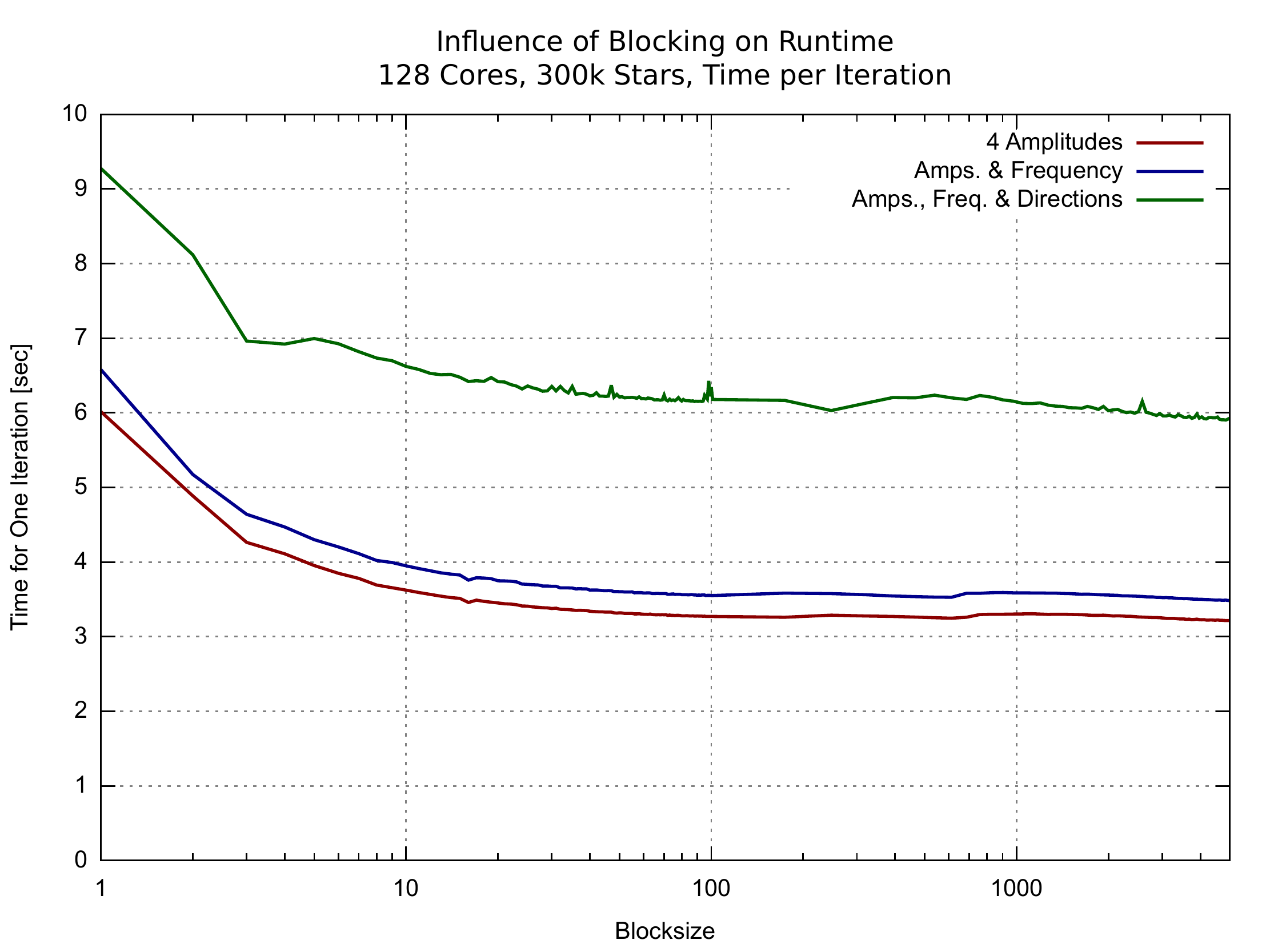}
 \caption[Block size influences performance]{The block size in which the design matrices are computed influences the runtime.\label{fig:blocking_performance}}
\end{figure}

As figure \ref{fig:gauss-newton-parallelization} shows, the computation of the normal matrix as well as the right-hand sides of the normal equations are organized block-wise (see Section \ref{sec:proto-parallelization}). This is done for performance reasons, Figure \ref{fig:blocking_performance} shows how the block size influences the performance. The runtime increases for block sizes smaller than 30, and stays more or less constant for block sizes larger than 100.\\

However, it is possible that the block size has an influence on the numerical stability of the Gauss-Newton method. Especially the consecutive summation of the local normal matrices can become a problem with a very high observation count due to rounding errors. In deep investigation of numerical stability has not been done during this work but will become necessary for future implementations.

\subsection{Speed Depending on the Number of Stars Processed}
\label{sec:stars_vs_speed}
\begin{figure}[htb]
 \centering
 \includegraphics[keepaspectratio, width=0.8\textwidth]{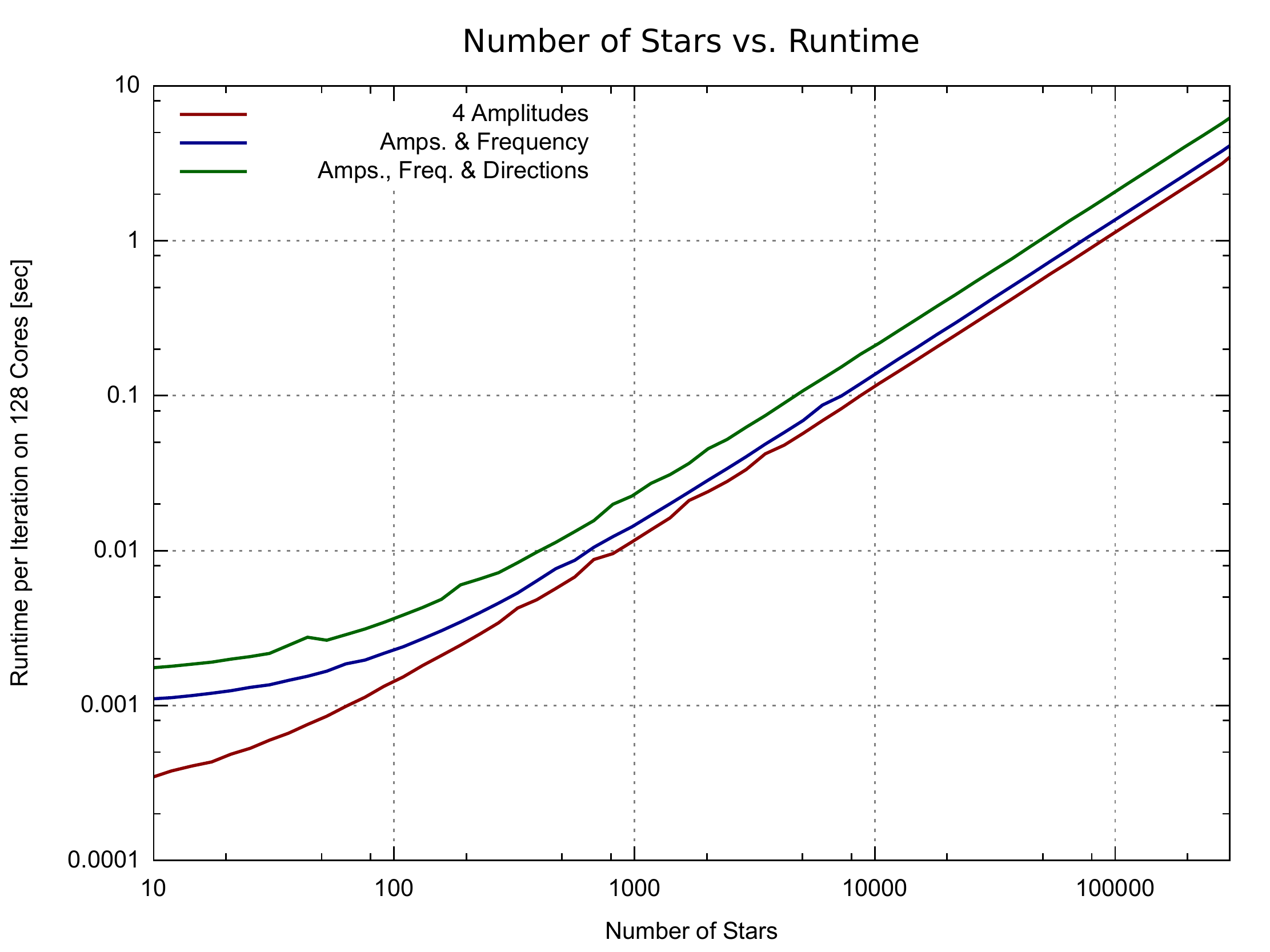}
 \caption[Number of Stars vs. Runtime]{Number of stars vs. runtime. The runtime increases linear with the number of stars. For very few stars, some overhead can be seen. A block size of 5000 has been used for the this analysis.\label{fig:stars_vs_runtime}}
\end{figure}

Figure \ref{fig:stars_vs_runtime} shows how runtime depends on the number of stars. The run times shown in this figure are for one iteration on 128 cores. Except some overhead for very few stars, the runtime increases linear with the number of stars.

\label{chap:prototype}

\chapter{Application to Simulated Gaia Data and Gravitational Waves}
\label{chap:Application_to_Gaia-Data_GW}
\section{Data Representation}
\label{sec:input_data}

\begin{figure}[H]
 \centering
 \includegraphics[keepaspectratio, width=0.5\textwidth]{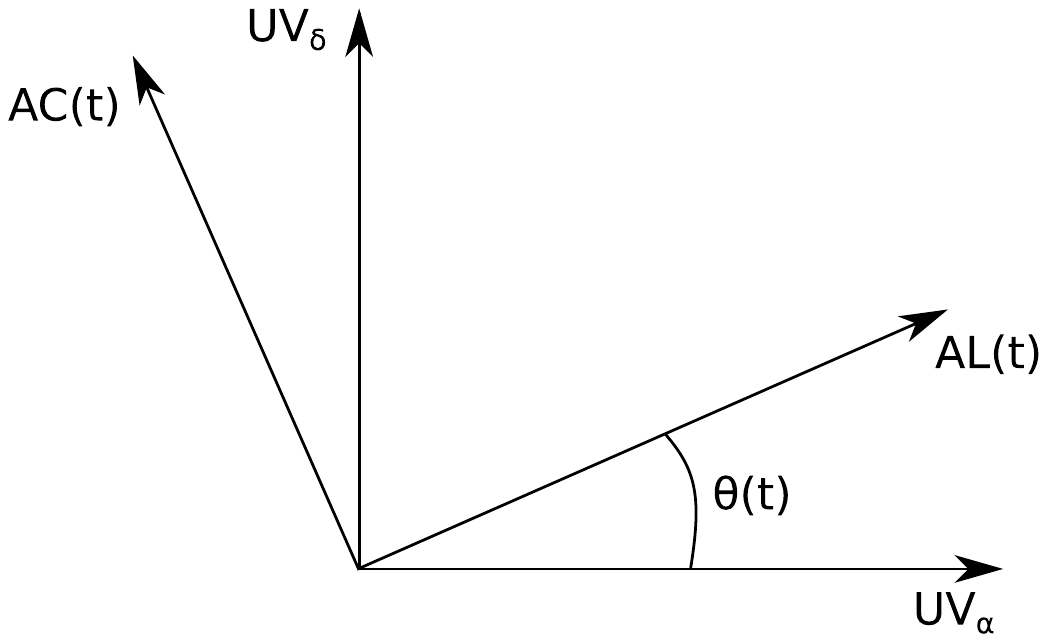}
 \caption[Illustration of the vectors of the scanning direction and the celestial coordinates.]{Illustration of the vectors of the scanning direction and the celestial coordinates. Note that $\mb{UV}_\alpha \cdot \mb{UV}_\delta = 0$ and $\mb{AC} \cdot \mb{AL} = 0$ at any time.\label{fig:UV_vs_ALAC}}
\end{figure}

Before the input data can be explained, a brief and greatly simplified description of the measurement principle is given. The satellite is scanning the sky constantly while spinning around its own axis and moving around the sun (see Figure \ref{fig:gaia-scann-law}). While doing so, the stars enter, sweep through and leave the field of view in a (almost) horizontal line. The light from both telescopes is projected on the astrometric instrument. This instrument can only measure along this direction of motion (along scan) and perpendicular to it (across scan). Due to technical reasons, the measurement in along scan direction is roughly three times more accurate than in across scan. But this also means that each of the measurements yields the positional change of the star in a different direction in celestial coordinates, since both scanning directions, with respect to the celestial directions, change with the motion of the satellite. Figure \ref{fig:UV_vs_ALAC} illustrates this, the direction of the celestial coordinates ($UV_\alpha$ and $UV_\delta$, for right ascension and declination, respectively) are fixed at any time, while the direction of the measurements changes over time.

The input data is organized in single observations, a sample of the input data can be found in Listing \ref{lst:inputfile} and is repeated here for clarity. Each observation consist of the following values:

\begin{lstlisting}[style=CA,breaklines=true,label={lst:inputfilesmall},basicstyle=\scriptsize\ttfamily]
ContainerIndex-SourceIndex-ObsIndex#obsTime;residual[AL];residual[AC];obsSigma[AL];obsSigma[AC];k;propSrcDir;theta;gaiaPosition
0-0-0#64573246697715972;-1.979716501843143E-9;1.2061773663196718E-9;9.90325283218615E-10;2.750229060875767E-9;(0.5397663091766982, -0.23673623577577715, 0.807841745732715);(-0.5398280045489584, 0.23668047208729226, -0.8078168602085664);-1.40217753247108;(-6.939805186073785E10, 1.2061023251521103E11, 5.232626010291127E10)
\end{lstlisting}

\begin{itemize}
 \item \texttt{ContainerIndex, SourceIndex, ObsIndex}: are of organizational purpose, they specify the source and provide a consecutive number for each observation per source.
 \item \texttt{obsTime}: The time in nanoseconds from J2010.0. It is represented by $t$ in the PGW model and the following mathematical descriptions.
 \item \texttt{residual[AL]} and \texttt{residual[AC]}: are the residuals from the Gaia solution in along (AL) and across scan (AC) direction in radian. Here they are interpreted as observations and are represented by $r_{AL}$ and $r_{AC}$ in the following formulas.
 \item \texttt{obsSigma[AL]} and \texttt{obsSigma[AC]}: are the errors of the observations in AL and AC direction in radian, represented by $\sigma_{AL}$ and $\sigma_{AC}$.
 \item \texttt{k}: is the GREM\footnote{Gaia Relativity Model} corrected direction from the star to Gaia, represented by $\mb{k}$
 \item \texttt{propSrcDir}: direction to the star, this is currently not used
 \item \texttt{theta}: is the angle (in radian) between the unit vector in $\alpha$ and the along scan (AL) direction, denoted as $\theta$
 \item \texttt{gaiaPosition}: is the BCRS\footnote{Barycentric Celestial Reference System} coordinates of the satellite at the time of observation, $\mb{x}_{obs}$ in the formulas
\end{itemize}

\section{PGW Model and Derivatives}
\label{sec:PGW-Model}
The model for the gravitational wave is described in detail in \cite{klionerPGW013}. The essential equations from this paper will be repeated here to give the reader an overview of the model.\\

The deflection of a light ray due to a PGW ($\delta\mb{u}$) is \cite{klionerPGW013}:
\begin{equation}
\label{eq:deltaUPGW}
 \delta u^i = \dfrac{u^i + p^i}{2(1 + \mb{u} \cdot \mb{p})} h_{jk} u^j u^k - \dfrac{1}{2} h_{ij}u^j~~~\text{.}
\end{equation}

Note that $\delta\mb{u}\cdot \mb{u} = 0$ so that $\delta\mb{u}$ gives a change of direction. Here $\mb{u} = -\mb{k}$ is the vector from the observer to the observed star at the time of observation, and:

\begin{align}
\label{eq:PGWhij}
 h_{ij} &= (\mb{P}\mb{e}^+\mb{P}^\top)_{ij} 
 \left( 
 h_c^+ \cos \left( \Omega (t-\frac{1}{c} \mb{p}\cdot \mb{x_{obs}}) \right) +
 h_s^+ \sin \left( \Omega (t-\frac{1}{c} \mb{p}\cdot \mb{x_{obs}}) \right)
 \right) \nonumber \\
 &+ (\mb{P}\mb{e}^\times\mb{P}^\top)_{ij} 
 \left( 
 h_c^\times \cos \left( \Omega (t-\frac{1}{c} \mb{p}\cdot \mb{x_{obs}}) \right) +
 h_s^\times \sin \left( \Omega (t-\frac{1}{c} \mb{p}\cdot \mb{x_{obs}}) \right)
 \right)
\end{align}

In Formula \eqref{eq:PGWhij}, $\mb{p}$ is the direction of propagation of the GW and $\Omega$ is its frequency. The amplitudes as well as phases are parametrized by the four independent parameters $h_c^+, h_s^+, h_c^\times$ and $h_s^\times$. The matrices $\mb{e}$ are defining the polarization of the GW and $\mb{P}$ is a rotational matrix which is defined as \cite{klionerPGW013}:

\begin{equation}
 \mb{P} = \mb{R}_z \left( \frac{\pi}{2} - \alpha_{GW} \right) \mb{R}_x \left( \frac{\pi}{2} - \delta_{GW} \right)
\end{equation}

For the 4 amplitudes $h_{s/c}^{+/\times}$ and $\Omega$, it is easy to see that the derivatives are $\frac{\partial h_{ji}}{\partial h_{s/c}^{+/\times}}$ and $\frac{\partial h_{ji}}{\partial \Omega}$ inserted in \eqref{eq:deltaUPGW}. The derivative with respect to $\alpha$ and $\delta$ are much more complex due to the occurrences of both parameters in \eqref{eq:deltaUPGW} and \eqref{eq:PGWhij}, these derivatives have been provided as an implementation in Java in the framework of AGISLab. For this work a reimplementation in C has been made, both versions are given as source code on the attached DVD.

\subsection{From Input File to the Fit}
\label{sec:input_to_fit}
This model gives us the deflection in celestial coordinates. As shown in section \ref{sec:input_data}, these results have to be transformed into along and across scan directions. To do so one has to first compute the celestial coordinates of the direction of the observed star:

\begin{align}
\label{eq:alphadelta_from_u}
 \alpha &= \atantw{u^2, u^1}\nonumber \\
 \delta &= \asin{u^3}~~\text{.}
\end{align}

Now the unit vectors of the celestial coordinates at the star position can be computed by:

\begin{equation}
\label{eq:EHV_in_alphadelta}
 \mb{e}_\alpha = \begin{pmatrix} - \sin \alpha\\ \cos \alpha\\ 0 \end{pmatrix}~~~~~\mb{e}_\delta = \begin{pmatrix} - \cos \alpha \sin \delta\\ -\sin \alpha \sin \delta \\ \cos \delta \end{pmatrix}
\end{equation}

The variation of the star position $\delta\mb{u}$ or its derivatives (both generalized as $\mb{x}$ here) can be projected to the along- and across scan directions by:

\begin{align}
\label{eq:deltau_to_ALAC}
 \mathrm{projAL}(\mb{x}) &= \mb{x} \left( \mb{e}_\alpha \sin \theta + \mb{e}_\delta \cos \theta \right)\nonumber \\
 \mathrm{projAC}(\mb{x}) &= \mb{x} \left( -\mb{e}_\alpha \cos \theta + \mb{e}_\delta \sin \theta \right)
 \end{align}

Equations \eqref{eq:alphadelta_from_u} and \eqref{eq:EHV_in_alphadelta} can be used for both the deflection as well as the derivatives. In this way the normal equations ($\mb{J}^\top \mb{J}\delta = \mb{J}^\top \mb{d}$) for the $M$ observations can be build using the Jacobian:

\begin{equation}
\label{eq:jacobian_PGW_ALAC}
\mb{J} =
\begin{pmatrix}
 \mathrm{projAL}\left(\dfrac{\partial \delta \mb{u}_0}{\partial h_s^+}\right)     & \mathrm{projAL}\left(\dfrac{\partial \delta \mb{u}_0}{\partial h_c^+}\right) & \cdots   & \mathrm{projAL}\left(\dfrac{\partial \delta \mb{u}_0}{\partial \delta_{GW}}\right) \\
 \mathrm{projAL}\left(\dfrac{\partial \delta \mb{u}_1}{\partial h_s^+}\right)     & \mathrm{projAL}\left(\dfrac{\partial \delta \mb{u}_1}{\partial h_c^+}\right) & \cdots   & \mathrm{projAL}\left(\dfrac{\partial \delta \mb{u}_1}{\partial \delta_{GW}}\right) \\
\vdots                                                          & \vdots                                                        & \ddots        & \vdots                                                        \\
\mathrm{projAL}\left(\dfrac{\partial \delta \mb{u}_M}{\partial h_s^+}\right)     & \mathrm{projAL}\left(\dfrac{\partial \delta \mb{u}_M}{\partial h_c^+}\right) & \cdots   & \mathrm{projAL}\left(\dfrac{\partial \delta \mb{u}_M}{\partial \delta_{GW}}\right) \\

 \mathrm{projAC}\left(\dfrac{\partial \delta \mb{u}_0}{\partial h_s^+}\right)     & \mathrm{projAC}\left(\dfrac{\partial \delta \mb{u}_0}{\partial h_c^+}\right) & \cdots   & \mathrm{projAC}\left(\dfrac{\partial \delta \mb{u}_0}{\partial \delta_{GW}}\right) \\
 \mathrm{projAC}\left(\dfrac{\partial \delta \mb{u}_1}{\partial h_s^+}\right)     & \mathrm{projAC}\left(\dfrac{\partial \delta \mb{u}_1}{\partial h_c^+}\right) & \cdots   & \mathrm{projAC}\left(\dfrac{\partial \delta \mb{u}_1}{\partial \delta_{GW}}\right) \\
\vdots                                                          & \vdots                                                        & \ddots        & \vdots                                                        \\
\mathrm{projAC}\left(\dfrac{\partial \delta \mb{u}_M}{\partial h_s^+}\right)     & \mathrm{projAC}\left(\dfrac{\partial \delta \mb{u}_M}{\partial h_c^+}\right) & \cdots   & \mathrm{projAC}\left(\dfrac{\partial \delta \mb{u}_M}{\partial \delta_{GW}}\right) \\
\end{pmatrix} 
\end{equation}

and the data vector $\mb{d}$

\begin{align}
\mb{d} &= ( r_{AL_0} - \delta u_{AL_0};  r_{AL_1} - \delta u_{AL_1};  \cdots;  r_{AL_M} - \delta u_{AL_M}; \nonumber\\
  & ~~~~~~~~~r_{AC_0} - \delta u_{AC_0};  r_{AC_1} - \delta u_{AC_1};  \cdots;  r_{AC_M} - \delta u_{AC_M} )^\top
\end{align}

with $\mathrm{projAL}(\delta \mb{u}) = \delta u_{AL}$ and $\mathrm{projAC}(\delta \mb{u}) = \delta u_{AC}$.\\

It is important to note, that, because of their higher accuracy, \textit{only along scan measurements have been used in all experiments}. A second reason for this decision is a simplification: it is easier to debug and explore certain effects if only one direction is used. The software is, however, equipped with the ability to fit AC data too, the routines for the PGW computation can output this data and the fitting routines are prepared to use them. The usage of the AC measurements in addition to the AL measurements might yield slightly better detection accuracy or sensitivity.

\newpage

\section{Experiments}
\label{sec:gw_experiments}
The following Sections present experiments which have been conducted with the 300k stars dataset. The order of these experiments also represent the approach with which the search strategy for a PGW with completely unknown parameters has been developed.

\subsection{Dependence of the Formal Fitting Errors on the Frequency}
The mission duration of Gaia is planned to be 5 years. This limits the detection to waves with periods shorter than around 10 years. The upper limit for the frequency (shortest period) cannot be given beforehand and is depending on the future performance of the instrument. For technical reasons one can speculate about shortest periods around the spin period of the satellite or even around 100\,sec.\\

\begin{figure}[H]
 \centering
 \includegraphics[keepaspectratio, width=0.9\textwidth]{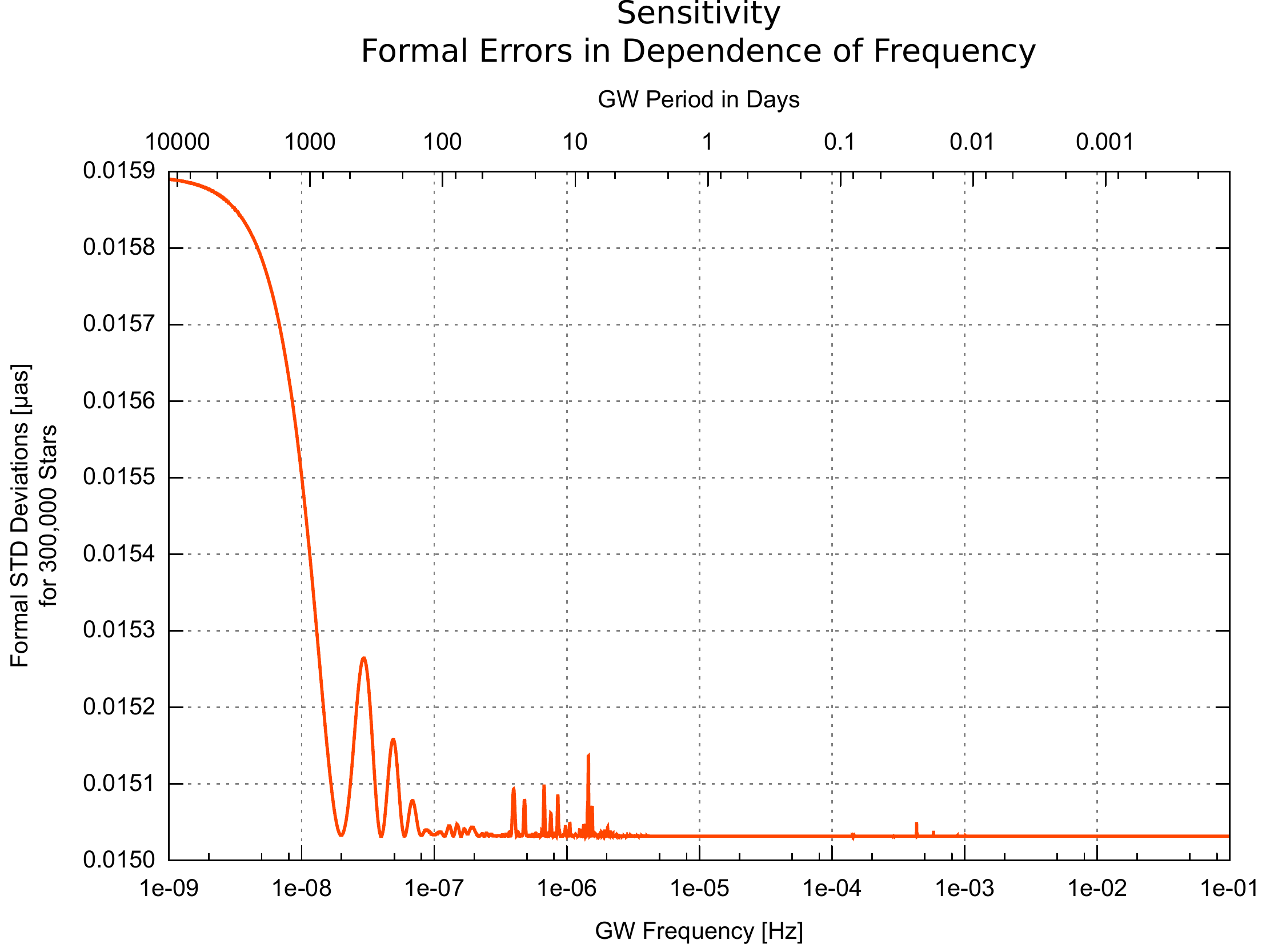}
 \caption[Sensitivity Curve for Simulated Data]{Sensitivity curve for the dataset with 300k stars. The errors increase for very low frequencies; due to the Gaussian nature of the noise in this simulated dataset no other features (limit for high frequencies) can be found.\label{fig:sensitivity_curve}}
\end{figure}

Limits in the detectable wave period should become visible when the amplitudes from the GW model are fitted to data which does not contain a wave signal while the frequency is fixed and the directions have no influence. The formal errors of the fit can be plotted with respect to the frequency. In such a plot one main feature should be visible: an increase of the standard errors for frequencies below the limit of 10\,yr. In this case the timespan of the data is not sufficient to support the model function accurately enough. Depending on the noise in the data, two other features might be visible. First, increasing errors above an instrument specific frequency, and second, a certain frequency at which the errors are lowest. Figure \ref{fig:sensitivity_curve}, shows the result of such a test.

One can easily see the increase of the errors for long GW periods (i.e. low frequencies). The other features are missing, no upper limit seems to exist, and the curve is flat without an optimal point. The simulated data uses a Gaussian noise model, which does not produce these features.

\subsection{Fitting the Amplitudes}
\label{sec:amplitudes_fitting}
\begin{figure}[H]
 \centering
 \includegraphics[keepaspectratio, width=0.9\textwidth]{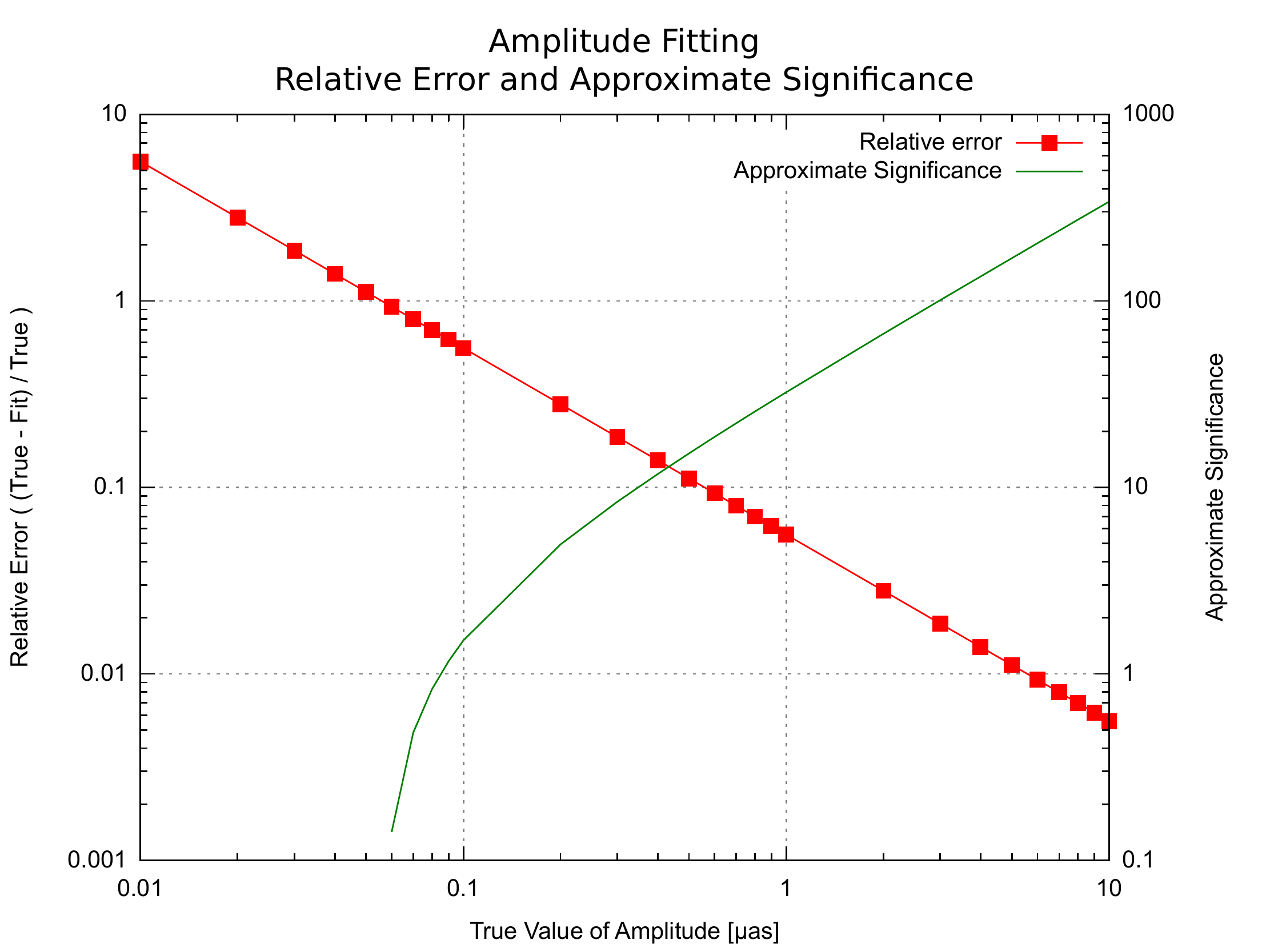}
 \caption[Result of an experiment, in which only the $h_s^+$ amplitude has been fitted]{Result of an experiment, in which only the $h_s^+$ amplitude has been fitted. All other parameters have been set to their true values. The error for the $h_s^+$ amplitude decreases exponentially with increasing amplitude (left y-axis), while the SNR increases (right y-axis).\label{fig:amplitudes_fitting}}
\end{figure}

The four amplitudes of the wave are linear parameters and can be fitted with one Gauss-Newton iteration (linear fit). If all other parameters (frequency and directions) are known and set to true values, such a fit can give a hint on how sensitive the fit is with respect to the amplitudes. Figure \ref{fig:amplitudes_fitting} shows the relative errors from such an experiment.

In this experiment a GW with a fixed $h_{s}^+$ amplitude is added to the data, the other three amplitudes are set to zero. All model parameters are set to true values of the GW, the amplitudes are set to zero and only the $h_{s}^+$ is fitted from the data. One can see that the relative error

\begin{equation}
\eta = \dfrac{h_{s_{true}}^+ - h_{s_{Fit}}^+}{h_{s_{true}}^+} 
\end{equation}

is decreasing exponentially with higher amplitude. The signal to noise ration (SNR) of the fitted amplitude with respect to the errors for this amplitude from the fit can be interpreted as an approximate statistical significance. It can be defined\footnote{Compare to Equation \eqref{eq:statsig}} as

\begin{equation}
 s = \dfrac{h_{s_{Fit}}^+}{\sigma_{h_{s_{Fit}}^+}}
\end{equation}

and it is over $3.0$ for amplitudes higher than 0.15\,\textmu as. This is a first indicator for the sensitivity, and has been used as a guiding for additional experiments.

\subsection{Feasibility of Frequency Determination}
\label{sec:freq-search}
\begin{figure}[H]
 \centering
 \includegraphics[keepaspectratio, width=0.8\textwidth]{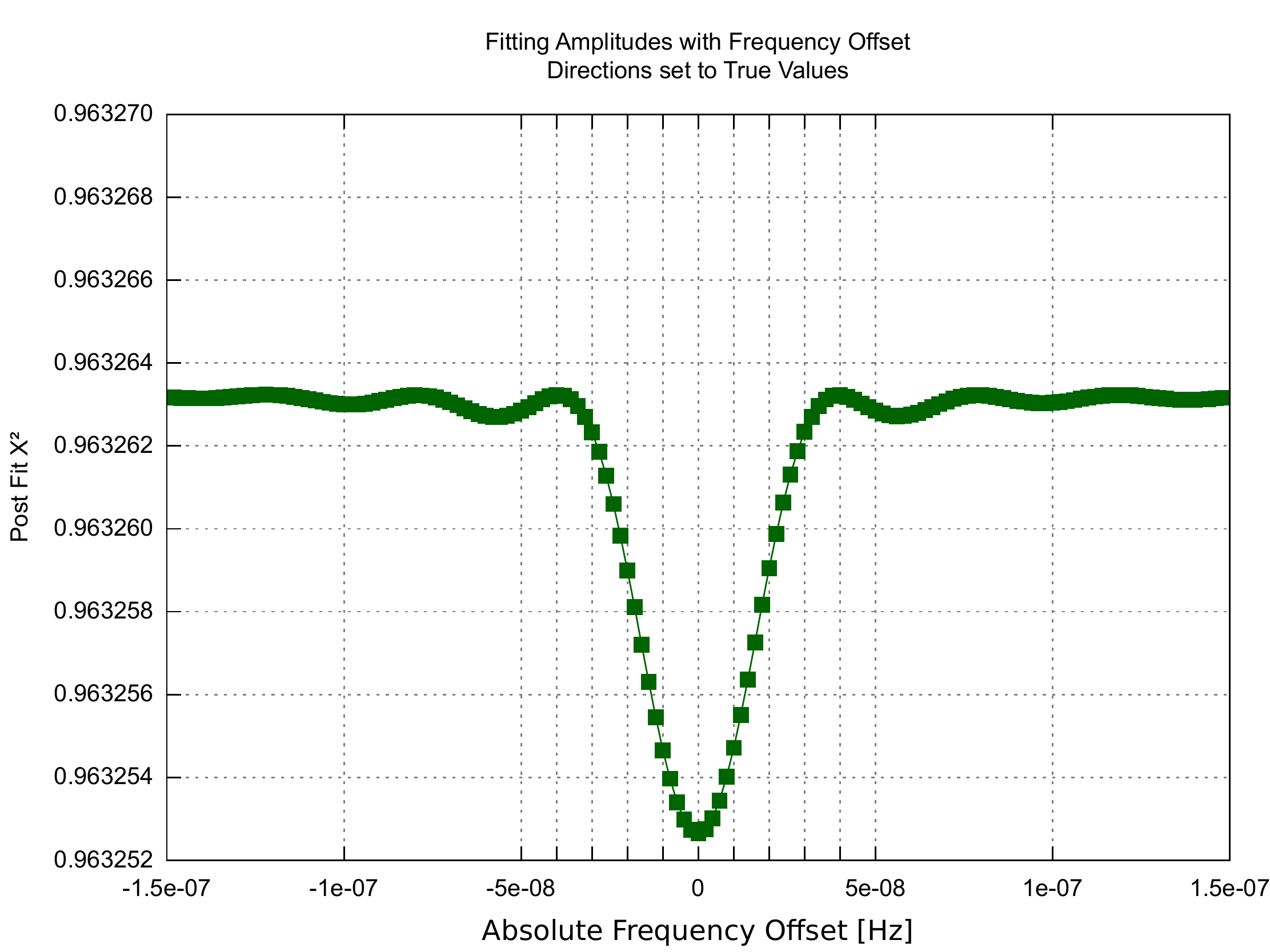}
 \caption[Post fit $\chi^2$ for an amplitude fitting as a function of the frequency offset]{Post fit $\chi^2$ for an amplitude fitting as a function of the frequency offset with respect to the true frequency of the wave. The width of the pit as well as the magnitude is independent of the frequency of the wave.\label{fig:freq_determination}}
\end{figure}

For the particular subject of this work, no FFT-like divide and conquer algorithm exists at the moment for the frequency determination. The frequency space can be probed with a a linear search at this time. Therefore it is necessary to determine the step size with which the frequency space is swept. Figure \ref{fig:freq_determination} shows how the $\chi^2$ goodness-of-fit measure behaves when the frequency is fixed to a slightly wrong value. An 1.5\,\textmu as GW with a fixed frequency is added to the data. The amplitudes are fitted for different model values of the frequency. Values for the directions are set to the true values. As one can see, the $\chi^2$ value has its minimum at the right frequency (0 offset), and is increasing with increasing offset from the true frequency. The full width at half maximum (FWHM) of the pit shown in Figure \ref{fig:freq_determination} is $\approx 3.6\cdot 10^{-8}$\,Hz one could argue that this is the maximum step size for the search in the frequency space. Since the magnitude of the peak will be smaller for weaker GWs, the step size should be also smaller than the FWHM. Otherwise it might be possible that the $\chi^2$ value of the evaluated point is in the range of the noise. For further experiments a step size of $1\cdot 10^{-8}$\,Hz has been chosen to make sure that no unwanted side-effects are occurring. In reality even smaller step sizes might be necessary since the phase shift by a frequency offset of $10^{-8}$\,Hz from the true frequency is in the worst case $10^{-8} s^{-1} \cdot 2.5\,yr = 45^\circ$, if the wave is ``centered'' in the middle of the observations.

\begin{figure}[H]
 \centering
 \includegraphics[keepaspectratio, width=0.8\textwidth]{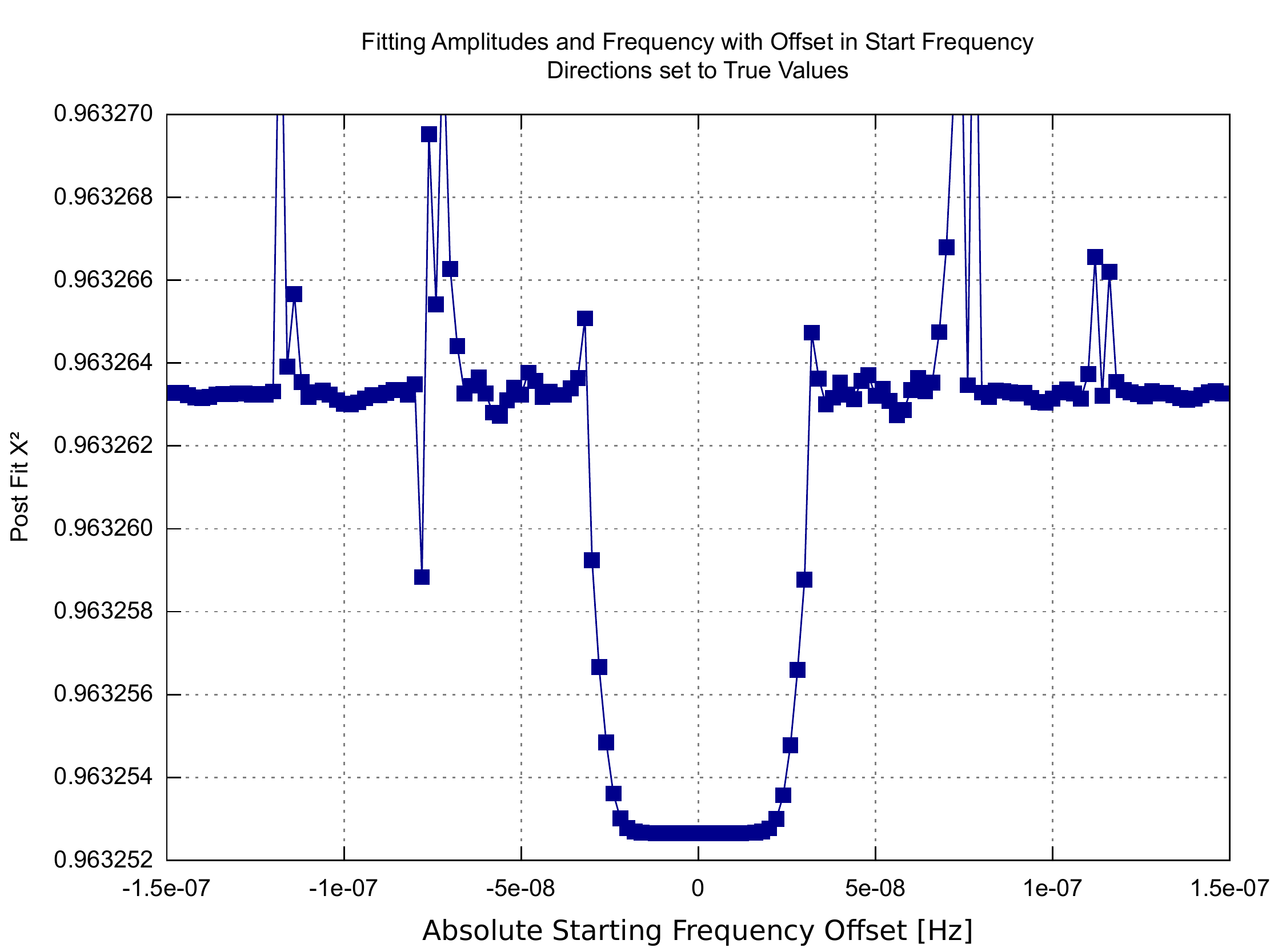}
 \caption[Post fit $\chi^2$ with the frequency is fitted in addition to the amplitudes]{Post fit $\chi^2$, with the same scenario as in figure \ref{fig:freq_determination}, but here the frequency is fitted in addition to the amplitudes (3 iterations). For offsets in the frequency around $\pm 2\cdot 10^{-8}$ from the true frequency, the fit converges to the true values.\label{fig:freq_determination_omfit}}
\end{figure}

The frequency can also be fitted in the range of the peak shown in Figure \ref{fig:freq_determination}. As shown in Figure \ref{fig:freq_determination_omfit}, the $\chi^2$ values improves in a U-shaped region. In this region the frequency has been fitted to the correct one, with a relative error of 0.0001. The offset in the starting frequency can be bridged in a range of $\approx 4\cdot 10^{-8}$\,Hz.

A use of the property shown in Figure \ref{fig:freq_determination_omfit} was considered but not omitted. The main reason is, that multiple iterations are necessary and therefore the performance gains are low. A second reason is, that it only works reliably if the directions of the wave are known approximately.

\subsection{Directions}
\label{sec:directions_fit}
\begin{figure}[H]
 \centering
 \includegraphics[keepaspectratio,width=0.9\textwidth]{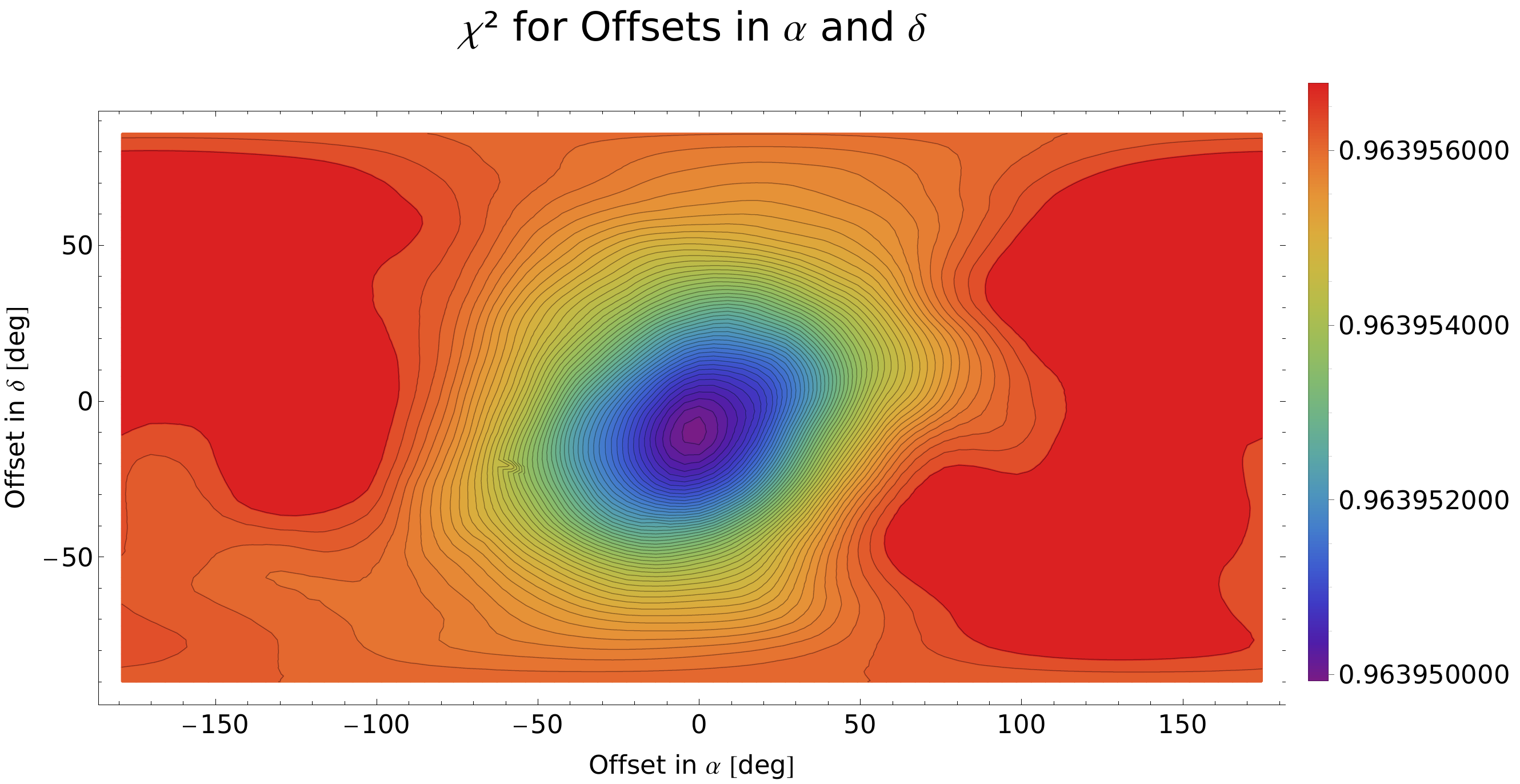}
 \caption[Dependence of $\chi^2$ to offsets in $\alpha$ and $\delta$]{When the frequency is known well enough, the post fit $\chi^2$ value for an amplitude fit is rather insensitive to some moderate offsets in $\alpha$ and $\delta$.\label{fig:ad_off}}
\end{figure}

The last two parameters of the wave which have to be determined represent its direction. With the given GW model, the $\chi^2$ objective function is rather insensitive to changes in directions. As Figure \ref{fig:ad_off} shows, an offset from the true direction of 10 or 20 degree has only a limited influence on the $\chi^2$ value, if the frequency is known.\\

For this figure a GW with $h_s^+ = 1.5$\,µas has been added to the data. The direction of the wave has been set to $\alpha = 180^\circ$ and $\delta = 0^\circ$. The frequency is set to the true frequency and only the amplitudes have been fitted for certain $(\alpha,\delta)$-point in the range of $-180^\circ \leq \alpha \leq 180^\circ$ and $-90^\circ \leq \delta \leq 90^\circ$ (80\texttimes 80 equally sampled points). The behavior shown in Figure \ref{fig:ad_off} is independent of the true frequency of the wave, the peak does not change size or magnitude.

\begin{figure}[H]
 \centering
 \includegraphics[keepaspectratio,width=0.9\textwidth]{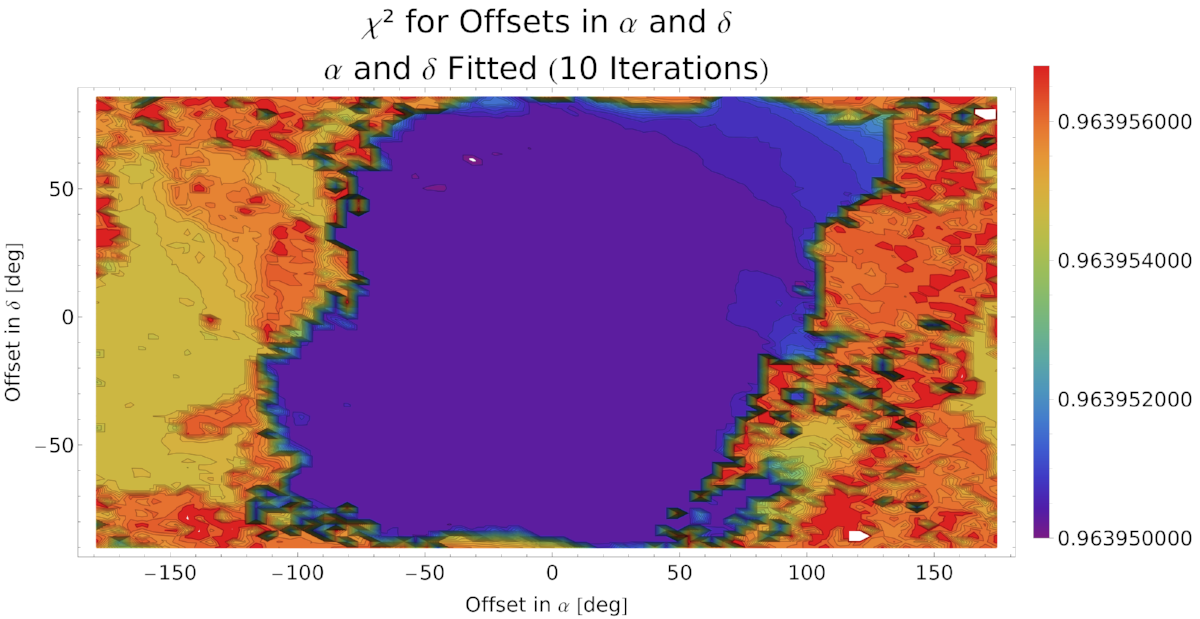}
 \caption[Dependence of $\chi^2$ to offsets in direction with $\alpha$ and $\delta$ fitted]{Post fit $\chi^2$ value as in figure \ref{fig:ad_off}. But in this case $\alpha$ and $\delta$ have been fitted with 10 iterations. This figure shows that the amplitudes as well as the directions can be fitted in a quite broad range.\label{fig:ad_off_ad_fitted}}
\end{figure}

The direction can also be fitted with the Gauss-Newton fitter together with all other parameters. Assuming that the frequency is known well enough, fitting the directions require at least 3 iterations (offset $10^\circ$) and up to 10 iteration (offset $50-60^\circ$), possibly more with higher frequency offsets. With offsets greater than $110^\circ$ (in particular in $\alpha$) the fit might not converge at all. This can be seen in Figure \ref{fig:ad_off_ad_fitted}, in this plot the directions have been fitted in 10 iterations (the frequency was also set to true values).

Also in this case (compare to Figure \ref{fig:freq_determination_omfit}) it is not feasible to use these features during the search phase (in the fitness function of the ES). However, it is very useful to determine how far the ES can approximate the directions. As can be seen in Figure \ref{fig:ad_off_ad_fitted}, a offset of up to $50^\circ$ from the true direction does not pose a problem to a final multi-iteration fit for a wave candidate. In other words, it is sufficient if the ES approximates the direction with a accuracy of $\pm 50^\circ$. This allows reducing the number of generations in the ES to only a few ($\approx 4$ for the tests conducted in this work).

\subsection{Application of the Evolution Strategy for the Direction}
\label{sec:es_find_freq}
As could be seen in Figure \ref{fig:ad_off}, the pit of the $\chi^2$ objective function is quite broad for offsets in the direction of the wave. Additionally, the function seems to be relatively smooth. These are two requirements for a successful operation of an evolution strategy. Hence, the idea is, to employ an evolution strategy to find good suggestions for starting points of the directions for a later nonlinear Gauss-Newton fit with all parameters. In this way the expensive multi iteration fit of the directions from multiple starting points can be avoided. The starting point for the frequency is found (or better selected) by stepping through the frequency space with a sufficiently small step size (e.g. $10^{-8}$\,Hz). For each of these frequencies, an ES which uses, initially, randomly selected values for directions as free parameters is started. The fitness function of this ES is the post fit $\chi^2$ value of the linear fit of the amplitudes. This yields the approximate starting values for all seven parameters: the approximate frequency in discrete steps, the direction from the ES and even the approximate amplitudes from the linear fit in the fitness function of the ES. These results can be sorted for the best $\chi^2$ values from all probed points, and parameters of the best candidates can be used as starting parameters for an multi iteration Gauss-Newton fit of all parameters.\\

To demonstrate the effectiveness of this method a dataset with 10162 stars has been created and a GW with the following parameters shown in Table \ref{tab:gw_parameters_experiment} has been added:
\begin{table}[H]
 \centering
 \begin{tabular}{cl}
  \toprule
  \textbf{Parameter}            & \textbf{Value}\\
  \midrule
  $h_c^+$                       & 1.0\,µas\\
  $h_s^+$                       & 0.001\,µas\\
  $h_c^\times$                  & 0.001\,µas\\
  $h_s^\times$                  & 0.001\,µas\\
  $\Omega$                      & $9.10176 \cdot 10^{-6}$\,Hz\\
  $\alpha$                      & $123.902^\circ$ \\
  $\delta$                      & $33.181^\circ$ \\
  \bottomrule
 \end{tabular}
 \caption[The values of the GW parameters which has been added to the data]{The values of the GW parameters which has been added to the data. The goal is to recover these values as exact as possible.\label{tab:gw_parameters_experiment}}
\end{table}

\begin{figure}[H]
 \centering
 \includegraphics[keepaspectratio, width=0.9\textwidth]{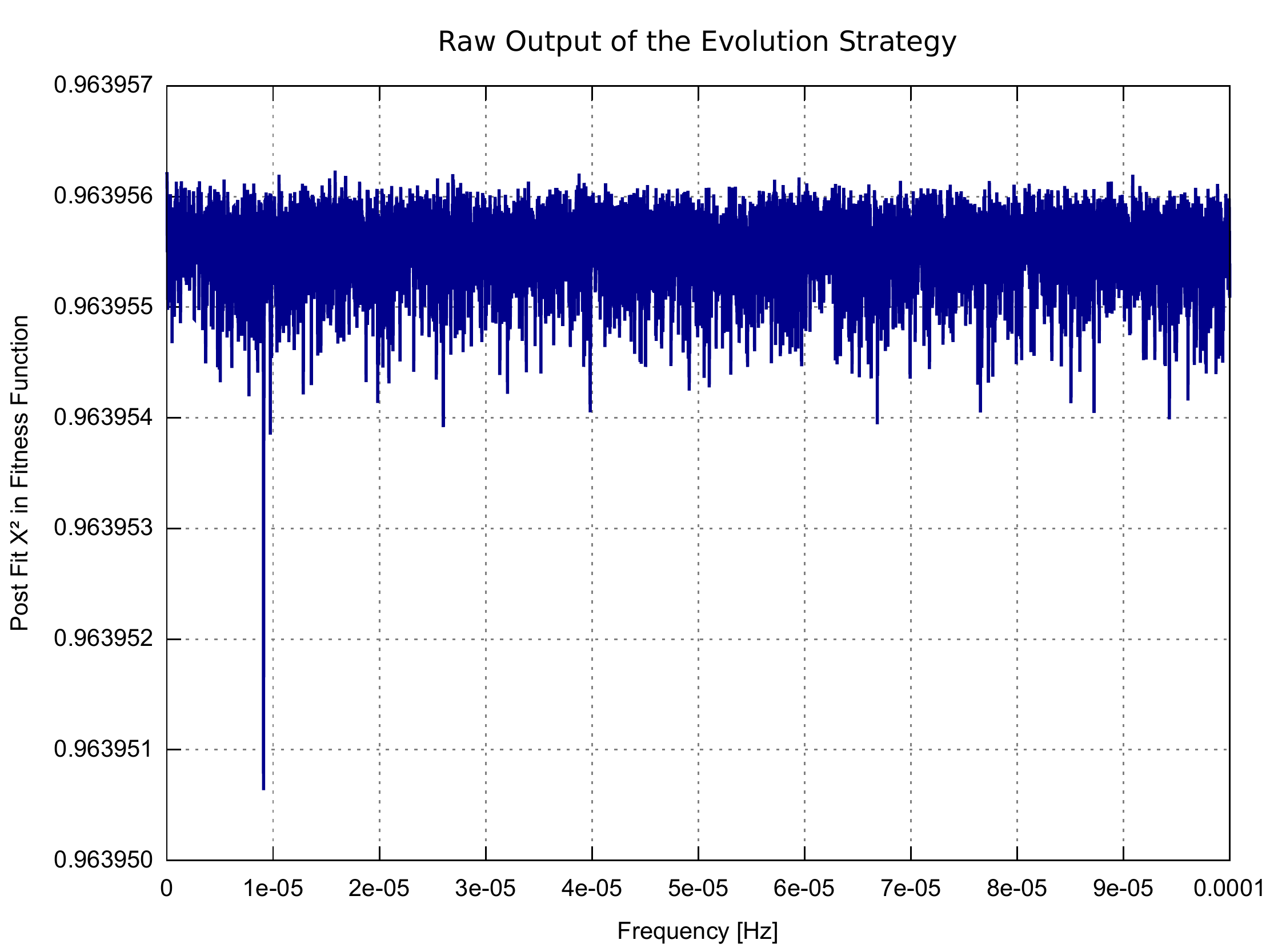}
 \caption[Value of the fitness function ($\chi^2$) of the fittest individual from the ES]{Value of the fitness function ($\chi^2$) of the fittest individual from the ES in dependence of the frequency. For the approximately correct frequency, the fitness function shows a sharp peak. The parameters of the individual from which this peak results can be used as starting points for a Gauss-Newton fit.\label{fig:ES_raw}}
\end{figure}

\begin{figure}[H]
 \centering
 \includegraphics[keepaspectratio, width=0.7\textwidth]{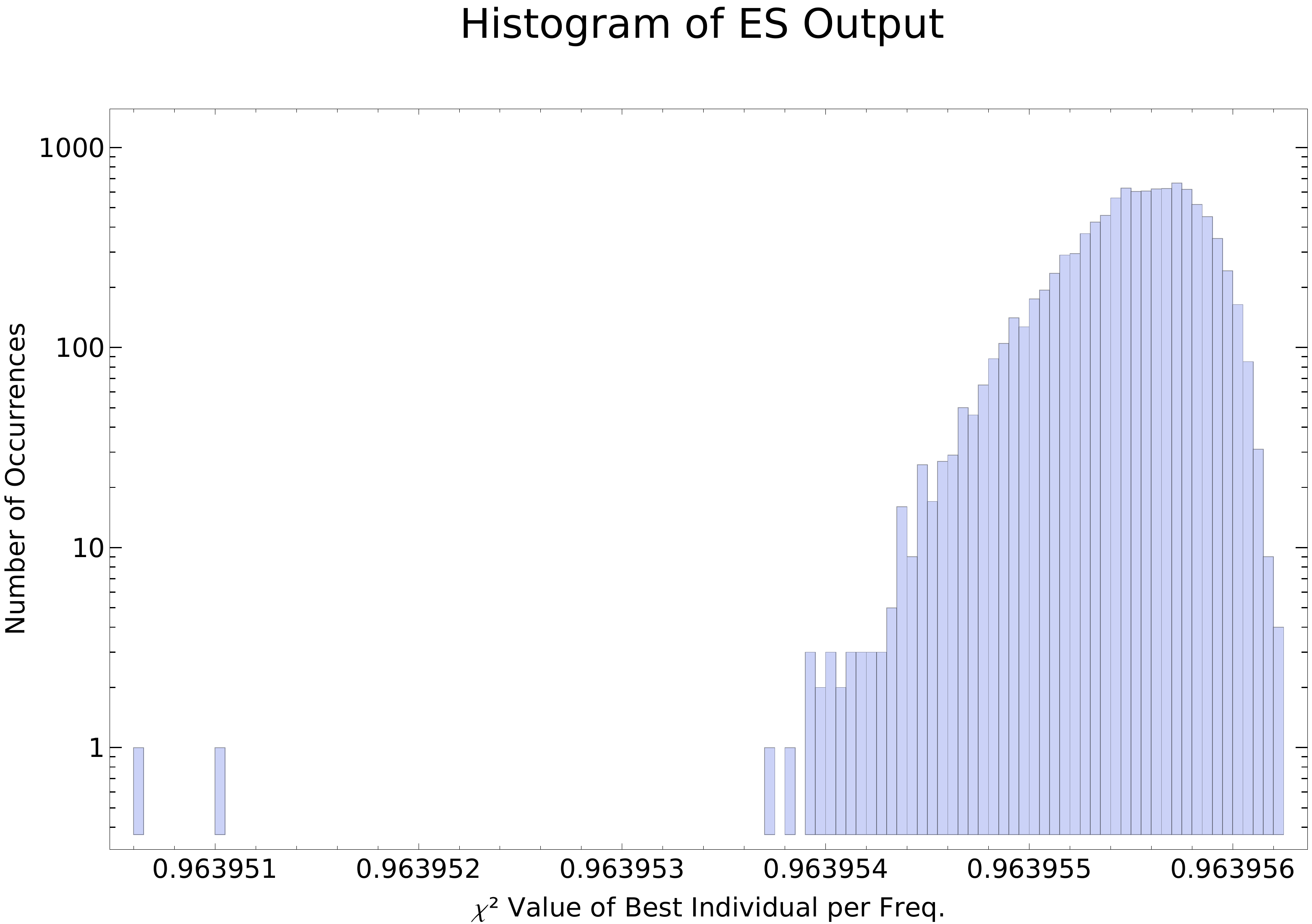}
 \caption[Distribution of the fitness values of the ES for all probed frequencies]{Distribution of the fitness values (post fit $\chi^2$ from amplitude fit) of the ES for all probed frequencies. For each frequency the post fit $\chi^2$ value yielded by the fittest individual has been used. One can clearly see the two bars in the left corner of the histogram. They correspond with the two individuals which are near the true frequency. A histogram which compares the case showed in this figure (wave added to the data) with case in which no wave was added to the data can be found in Appendix \ref{apx:histo_both}. \label{fig:ES_chi2_histo}}
\end{figure}

For each frequency, the set of parameters of the fittest individual can be saved and put in an ``high score''. After all frequency steps are evaluated, the fitness of the best individual for the approximately right frequency should be significantly higher than for the rest. This can be seen in figures \ref{fig:ES_raw} and \ref{fig:ES_chi2_histo}. Both figures show the result of the same experiment. Figure \ref{fig:ES_raw} shows the raw output of the evolution strategy, for each frequency step the best $\chi^2$ value (i.e. the quality of the fittest individual) is plotted. One can see that this function has a sharp peak near the true frequency of the wave (with $\chi^2 = 0.9639506344$). For any other frequency the fitness fluctuates around a mean of 0.9639555062.

For the use of such a peak as an indicator for a successful detection of a wave it is important to understand the significance of it. For doing so it is useful analyze the distribution of the values of the fitness function taking all frequencies into account. This has been done in Figure \ref{fig:ES_chi2_histo}, this figure shows a histogram of the best $\chi^2$ values of each frequency step. One can clearly see that the two frequency steps which are nearest to the true frequency are separated from the distribution. The nature of this distribution is not yet fully understood, with a kurtosis of 3.92642 and a skewness of -0.822654 it is clearly not normal distributed\footnote{The normal distribution has a skewness of 0.0 and a kurtosis of 3.0}. Without a wave added the statistical parameters of the data is as follows:

\begin{alignat*}{2}
 N              &= 8548800 &&= 10162~\mathrm{Stars} \nonumber \\
 r^2            &= \sum_{i=1}^{N}\left(\dfrac{O_i}{\sigma_i}\right)^2 &&= 8.24136553\cdot 10^6 \nonumber \\
 \chi^2         &= \dfrac{r^2}{N} &&= 0.96403771 \nonumber \\
 w              &= \sum_{i=1}^{N}\left(\dfrac{1}{\sigma_i^2}\right) &&= 3.95117360 \cdot 10^{25}\,\mathrm{rad}^{-2}\nonumber \\
 \mathrm{wRMS}  &= \sqrt{\dfrac{r^2}{w}} &&= 4.56705806 \cdot 10^{-10}\,\mathrm{rad}
\end{alignat*}

Assuming that the statistical significance of the peak is high enough, and that the approximate frequency has been found in this way, the ES has yielded the following approximate values for the parameters (Table \ref{tab:gw_parameters_from_ES}):

\begin{table}[H]
 \centering
 \begin{tabular}{cl}
  \toprule
  \textbf{Parameter}                 & \textbf{Value}\\
  \midrule
  $h_{c_{ES}}^+$                     & 1.110\,µas\\
  $h_{s_{ES}}^+$                     & 0.109\,µas\\
  $h_{c_{ES}}^\times$                & 0.164\,µas\\
  $h_{s_{ES}}^\times$                & 0.176\,µas\\
  $\Omega_{ES}$                      & $9.107273 \cdot 10^{-6}$\,Hz\\
  $\alpha_{ES}$                      & $113.82^\circ$\\
  $\delta_{ES}$                      & $22.87^\circ$\\
  \bottomrule
 \end{tabular}
 \caption{Approximated values for all 7 parameters from the ES at the frequency which yielded the highest fitness.\label{tab:gw_parameters_from_ES}}
\end{table}

\subsubsection{Using the ES Results for the Final Fit}
The results from the previous Section (Table \ref{tab:gw_parameters_from_ES}) are used as starting points for a nonlinear fit with all 7 parameters. The fitted parameters, errors and signal to noise ratios can be found in Table \ref{tab:gw_parameters_from_GN}. The SNR is the quotient of the fitted amplitude divided trough the error of the amplitude from the fit as defined in Equation \eqref{eq:statsig}.

\begin{table}[H]
 \centering
 \begin{tabular}{cllll}
  \toprule
  \textbf{Parameter}                 & \textbf{Value}                   &\textbf{Absolute Error}              & \textbf{SNR}          & \textbf{$\sigma$ (STD Deviation)}\\
  \midrule
  $h_{c_{Fit}}^+$                     & 1.201\,µas                      &0.200\,µas                           &$5.29$                 & 0.227\,µas\\
  $h_{s_{Fit}}^+$                     & 0.149\,µas                      &0.148\,µas                           &$0.92$                & 0.162\,µas\\
  $h_{c_{Fit}}^\times$                & 0.051\,µas                      &0.050\,µas                           &$0.25$                & 0.201\,µas\\
  $h_{s_{Fit}}^\times$                & 0.118\,µas                      &0.117\,µas                           &$0.75$                & 0.156\,µas\\
  $\Omega_{Fit}$                      & $9.10623 \cdot 10^{-6}$\,Hz     &$4.466\cdot 10^{-9}$\,Hz             &-                      & $2.886\cdot10^{-9}$\,Hz\\
                                      &                                 &                                     &                       & 5\,yr phase shift: $\pm 26.12^\circ$\\
  $\alpha_{Fit}$                      & $121.93^\circ$                  &$1.972^\circ$                        &-                      &$4.82^\circ$\\
  $\delta_{Fit}$                      & $33.71^\circ$                   &$0.529^\circ$                        &-                      &$5.10^\circ$\\
  \bottomrule
 \end{tabular}
 \caption{Results and statistics of the final multi-iteration fit for all 7 parameters.\label{tab:gw_parameters_from_GN}}
\end{table}

One can define a ``combined statistical significance'' for all 4 amplitudes together which is

\begin{equation}
 \label{eq:combined_signif}
 \varsigma = \dfrac{1}{4} \left((SNR_{h_c^+})^2 + (SNR_{h_s^+})^2 + (SNR_{h_c^\times})^2 + (SNR_{h_s^\times})^2\right)
\end{equation}

with the SNR as the ``statistical significance'' of a single amplitude, $SNR_{h_{c/s}^{+/\times}}$ defined as

\begin{equation}
\label{eq:statsig}
SNR_{h_{c/s}^{+/\times}} = \dfrac{h_{c/s_{Fit}}^{+/\times}}{\sigma_{h_{c/s_{Fit}}^{+/\times}}}~~\text{.}
\end{equation}

One can also give the probability that the fitted set of amplitudes are random by evaluating the cumulative distribution function (CDF) of the $\chi^2$ distribution with 4 degrees of freedom (denoted here as $CDF_{\chi^2}^{(4)}$).

\begin{equation}
 P_{Random} = 1 - CDF_{\chi^2}^{(4)}\left( 4\cdot \varsigma \right)
\end{equation}

For the data presented above, the combined significance is $\varsigma = 7.36$ and $P_{Random} = 6.3166\cdot 10^{-6}$.

\subsubsection{A wave with Mixed Amplitudes}
In the Sections before, a wave with only one amplitude ($h_s^+=1.0$\,µas) has been used to demonstrate the method. In this Section it is attempted to detect a wave with all amplitudes set. For this experiment the following wave parameters have been set:

\begin{table}[H]
 \centering
 \begin{tabular}{cl}
  \toprule
  \textbf{Parameter}            & \textbf{Value}\\
  \midrule
  $h_c^+$                       & 0.6\,µas\\
  $h_s^+$                       & 0.5\,µas\\
  $h_c^\times$                  & 0.5\,µas\\
  $h_s^\times$                  & 0.4\,µas\\
  $\Omega$                      & $3.29046\cdot 10^{-5}$\,Hz\\
  $\alpha$                      & $164.743^\circ$ \\
  $\delta$                      & $43.2604^\circ$ \\
  \bottomrule
 \end{tabular}
 \caption[Values of the GW parameters for mixed test]{Values of the GW parameters which have been added to the data for the test with mixed amplitudes.\label{tab:gw_parameters_mixed}}
\end{table}

Any of the amplitudes alone would not be detectable, but the total amplitude is:

\begin{align}
 h_{total} &= \sqrt{(h_{s}^+)^2 + (h_{c}^+)^2 + (h_{s}^\times)^2 + (h_{c}^\times)^2}\\
           &= \sqrt{(0.6\,\text{µas})^2 + (0.5\,\text{µas})^2 + (0.5\,\text{µas})^2 + (0.4\,\text{µas})^2} \nonumber\\
           &= 1.00995\,\text{µas}
\end{align}

The approximate values given by the ES are shown in Table \ref{tab:es_result_diff_amps}.

\begin{table}[H]
 \centering
 \begin{tabular}{clcl}
  \toprule
  \textbf{Parameter}            & \textbf{Value}\\
  \midrule
  $h_{c_{ES}}^+$                       & 0.374\,µas\\
  $h_{s_{ES}}^+$                       & 0.285\,µas\\
  $h_{c_{ES}}^\times$                  & 0.386\,µas\\
  $h_{s_{ES}}^\times$                  & 0.757\,µas\\
  $\Omega_{ES}$                      & $3.28928\cdot 10^{-5}$\,Hz\\
  $\alpha_{ES}$                      & $150.55^\circ$ \\
  $\delta_{ES}$                      & $52.53^\circ$ \\
  \bottomrule
 \end{tabular}
 \caption[Approximated values from the ES for the mixed test]{Approximated values from the ES for the test with mixed amplitudes.\label{tab:es_result_diff_amps}}
\end{table}

Using these approximated values as starting points for a multi iteration Gauss-Newton fit, the following final results can be achieved (table \ref{tab:final_results_diffamps}):

\begin{table}[H]
 \centering
 \begin{tabular}{cllll}
  \toprule
  \textbf{Parameter}                 & \textbf{Value}              &\textbf{Absolute Error}   & \textbf{SNR}          & \textbf{$\sigma$ (STD Deviation)}\\
  \midrule
  $h_{c_{Fit}}^+$                     & 0.544\,µas                 &0.056\,µas                     &$3.28$                  & 0.166\,µas\\
  $h_{s_{Fit}}^+$                     & 0.437\,µas                 &0.063\,µas                     &$2.23$                  & 0.185\,µas\\
  $h_{c_{Fit}}^\times$                & 0.401\,µas                 &0.099\,µas                     &$1.71$                  & 0.235\,µas\\
  $h_{s_{Fit}}^\times$                & 0.658\,µas                 &0.258\,µas                     &$2.64$                  & 0.249\,µas\\
  $\Omega_{Fit}$                      & $3.29013 \cdot 10^{-5}$\,Hz&$3.3\cdot 10^{-9}$\,Hz         &-                      & $3.319\cdot10^{-9}$\,Hz\\
                                      &                            &                          &                       & 5\,yr phase shift: $\pm 30.0^\circ$\\
  $\alpha_{Fit}$                      & $157.23^\circ$             &$0.046^\circ$                   &-                      &$5.52^\circ$\\
  $\delta_{Fit}$                      & $49.20^\circ$              &-$0.014^\circ$                  &-                      &$6.51^\circ$\\
  \bottomrule
 \end{tabular}
 \caption{Results and statistics from the mixed amplitude test of the final multi iteration fit for all 7 parameters.\label{tab:final_results_diffamps}}
\end{table}

As one can see in Table \ref{tab:gw_parameters_from_GN}, if only one amplitude is set to 1.0\,µas, this amplitude alone can be fitted with a statistical significance over $5\sigma$. Here, this is not the case, all $h_{c/s}^{+/\times}$ together produce a wave with almost the same amplitude, but their components for themselves cannot be recovered with such a high significance. The combined significance $\varsigma$, however, is somewhat lower with $\varsigma = 6.58$ compared to the test in which only one amplitude have been set to 1\,µas. The probability that these amplitudes are fitted randomly is $P_{Random} = 2.7529 \cdot 10^{-5}$.

\subsubsection{Extrapolation for the Real Dataset}
In the previous Sections it has been shown that a 1\,µas GW ($GW_{10162} = 1.0$\,µas) can be detected with over $5\sigma$ significance in an artificial dataset with 10162 stars. To extrapolate this number to the full and real dataset, the statistical weights of the observations have to be known. They are: for the artificial observations with 10162 stars $W_{10162} = 9.6278\cdot 10^2\,\text{µas}^{-2}$ and the expected values for the real data is $W_{\mathrm{Full}} = 3.87\cdot 10^6\,\text{µas}^{-2}$. Since the statistical weight is defined as,

\begin{equation}
 W = \sum_{i=1}^{N} \dfrac{1}{\sigma_i^2}
\end{equation}

one can scale the results of the 10162 star dataset to the full ($GW_{\mathrm{Full}}$) dataset by:

\begin{equation}
 GW_{\mathrm{Full}} = GW_{10162}\cdot \sqrt{\dfrac{W_{10162}}{W_{\mathrm{Full}}}} = 1.0\,\text{µas} \cdot 1.5773\times 10^{-2}~~\text{.}
\end{equation}

Hence, one can state that GWs with an astrometric effect of 0.0158\,µas might be detectable with Gaia. To put this in relation, it is useful to know the maximal theoretical limit to determine a constant parameter with all Gaia data is 0.00051\,µas. It is worth noting, that $W_\mathrm{Full}$ can easily change by a factor of two or even more in both directions, depending on the actual performance of Gaia.

\chapter{Feasibility for Real Mission Data, Summary and Outlook}
\label{chap:Feasibility-real-mission}
\section{Considerations Regarding Complexity}
The complexity of the search for a GW in the way it is performed in this work depends on 3 parameters: the number of observations ($N$), the number of probed frequencies ($F$) and the number of Gauss-Newton fits (i.e. ES individuals) per frequency ($G$) needed to determine the direction of the wave. One can write the complexity of the search as:

\begin{equation}
 \label{eq:complexity_search}
 \mathcal{O}(N\cdot F \cdot G)~~\text{.}
\end{equation}

The runtime can be approximated similarly to the complexity with $T = N \cdot F \cdot \alpha G$, $\alpha$ is a scaling factor for $G$, since it is possible to fit the directions or the frequency for each ES individual too. In the current implementation $\alpha = 1$. If the frequency is fitted too $\alpha$ would be $\alpha = 1.18 \cdot I$ where $I$ is the number of iterations necessary to converge, and, if all 7 Parameters are fitted $\alpha = 2.1\cdot I$. The numbers (1.18 and 2.1) are the factors by which one iteration is slower than in the case if only the amplitudes are fitted. These factors can be obtained by comparing the results from the profiling, as done in Section \ref{sec:profiling}, Table \ref{tab:vt_trace_300k}.\\

It is clear that the number of observations is fixed. With the approach used in this work, to probe the frequency range in steps of fixed width, also the number of frequency steps is fixed. The only way to lower the runtime is to lower the parameter $\alpha$ by using the ES (which internally fits only the amplitudes) to get approximations for the directions. One consideration was to fit also the frequency in these ES steps. Neglecting the slightly higher runtime per Gauss-Newton iteration, the problem with this approach is, that 3-5 iterations are necessary to fit the frequency with the other parameters fixed. As figure \ref{fig:freq_determination_omfit} shows, this could be used to increase the step size to $4\cdot 10^{-8}$\,Hz, which is hardly a gain considering that the runtime per step is raised by a factor of 3-5 because of the multiple iterations necessary.\\

The same considerations can be made for the directions, the directions can be fitted with the Gauss-Newton method using a couple of uniformly distributed starting points for $\alpha$ and $\delta$, one of the fits with the starting points near the optimum (compare to Figure \ref{fig:ad_off_ad_fitted}) might probably converge. To be sure that one of the points converge, at least 5 starting points should be used (compare to Figure \ref{fig:ad_off_ad_fitted}). Depending on offset in direction, the fit would require 3-15 iterations of which each is 2.1 times more expensive than just fitting the amplitudes. This also is slower than to use the linear fit for the amplitudes in the ES.\\

One possible approach which might lower the complexity is the use of vectorial spherical harmonics (VSH) \cite{Klioner-VSH} to determine the wave direction and to decide if a wave is detected at a given frequency. Different frequencies have to be probed with this method too, but for any frequency a decision in the fashion ``no-wave detected''/``wave detected at this frequency'' can be made directly. In addition the exact direction of the wave is yielded by this method. The VSH method requires a linear least-squares fit to determine the parameters of the vector field. Depending on the order of the VSH, the number of parameters to fit is substantially higher than in the method used in this work. A performance gain might come from the fact that only one such fit is necessary per frequency, and that the design matrices can be build with less computational expenses.\\

The generation of VSH design matrices has been implemented in GaiaGW. The use of VSH for direction determination has not been tested during this work.

%

\section{Approximation of Hardware Requirements}
\label{sec:approx_hardware}

The runtime of a full search depends on the number of frequency steps, which in turn depend on how broad the frequency range is. This frequency range can only be constrained downwards. Since the planed mission duration is 5 years, waves with longer periods than 10 years can be considered as a lower limit. An upper limit of the frequency range cannot be given before it is known how the instrument behaves. Such a upper limit can only be determined from extensive analysis of real data. However, from practical considerations one could formulate a rule of thumb (although it is highly speculative): signals with periods significantly longer than the spin period of Gaia might be easier to detect. The spin period of Gaia is 6 hours, depending on the interpretation of ``significantly longer'' that period could---for instance---be 24 hours.\\

For the 300k star dataset, the average runtime for one frequency step was $13.315\,\mathrm{s}$ on 1024 CPU cores. Assuming that one can weak-scale to millions of CPUs, one can make the following extrapolations based on the current code and the current architecture. The total CPU-time for the 300k stars dataset is 13634.56\,s per frequency step, the real dataset with 1 billion stars would hence require $\approx 4.54485\cdot 10^7$\,CPU-seconds for one frequency step.\\

\begin{figure}[htb]
 \centering
 \includegraphics[keepaspectratio,width=0.9\textwidth]{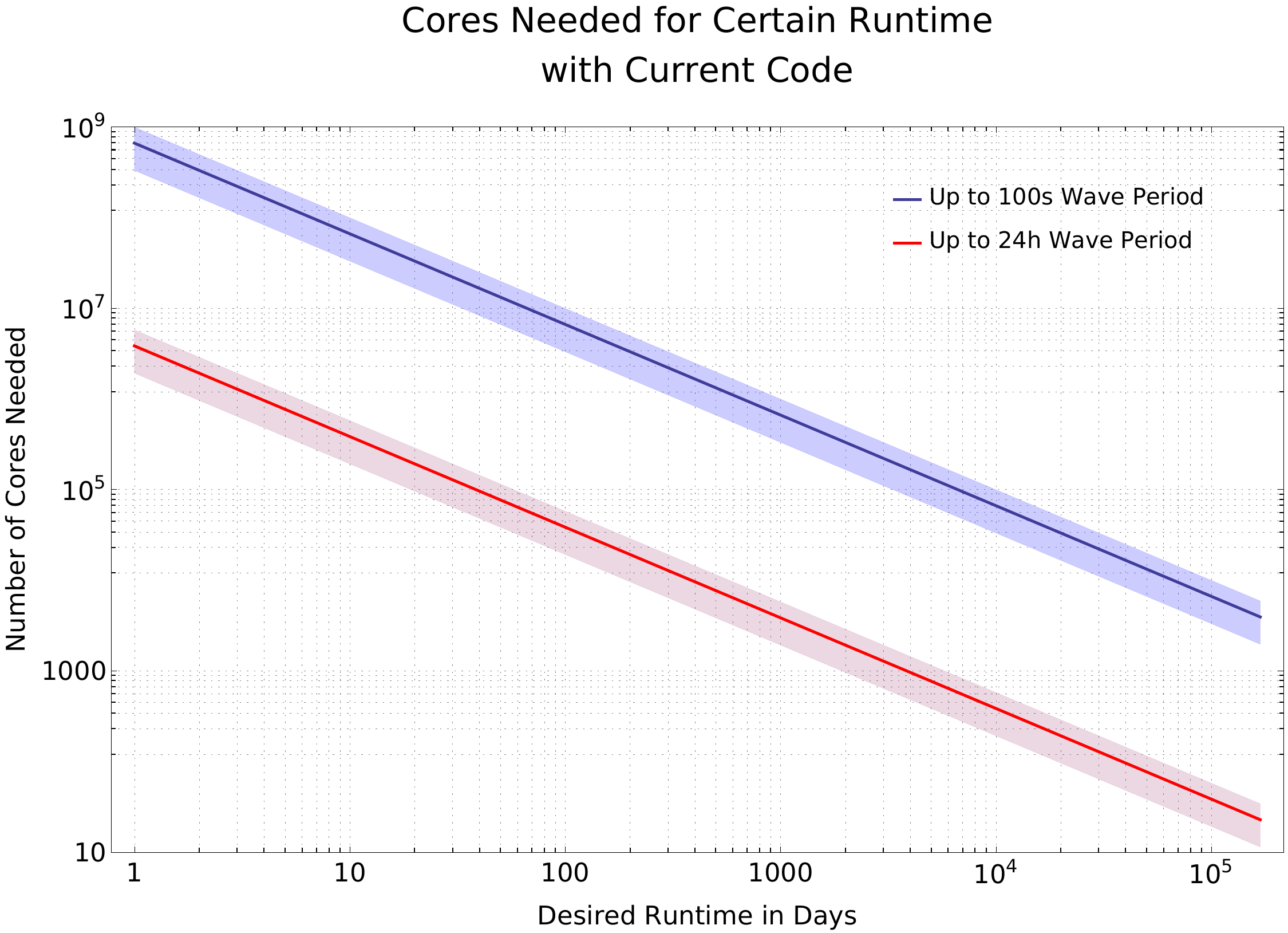}
 \caption[Approximation of the number of CPU cores necessary to perform a full search]{Approximation of the number of CPU cores necessary to perform a full search with the current code, on current hardware, in a specific time frame. Two scenarios are shown, in red for a search up to a wave period of 24\,h, in blue for a search up to a period of 100\,s. The $\pm 50\%$ bands are given as an orientation.\label{fig:cores_for_runtime}}
\end{figure}

Considering the frequency range of $10\,\mathrm{yr} \leq P \leq 24\,\mathrm{h}$, $\approx 7.3\cdot 10^{3}$ frequency steps of $1\cdot 10^{-8}$\,Hz would be necessary to cover the whole range. The overall CPU time is hence $\approx 3.3\cdot 10^{11}$\,s. Figure \ref{fig:cores_for_runtime} shows what an extrapolation of how many cores for a specific runtime would be needed. Two scenarios are shown in this figure, one in is for $10\,\mathrm{yr} \leq P \leq 24\,\mathrm{h}$, and one for $10\,\mathrm{yr} \leq P \leq 100\,\mathrm{s}$. 

These extrapolations can be considered as extremely pessimistic. The code is prototypical and has a high potential for performance optimizations. The performance evaluations using PAPI counters (Section \ref{sec:vampirtrace}) show a utilization of the CPU of approximately 50\% of the theoretical peak performance and the memory organization can be made more cache friendly. Usage of GPUs is worth being considered, which should give a factor of 2 to 8 too. Furthermore, it is also conceivable that mathematical ``optimizations'' regarding the model and the search strategy are possible. The figures can hence be considered as worst-case scenarios.

\section{Related Work}
\label{sec:rel_work}
The Gauss-Newton method for least-squares fitting is based on basic linear algebra subroutines (BLAS). A lot of highly parallel packages exist for performing linear algebra operations. A particular notable is the package PETSc \cite{petsc-efficient}. PETSc is a C++ framework, which is designed to be used for highly parallel PDE solvers. It is not limited to BLAS and provides in addition an extensive set of solvers, preconditioners and data containers.\\

A parallel implementation of the ES-CMA method has been presented by Müller et. al. in \cite{MuellerpCMALib}. The software library is called \texttt{pCMAlib}, it is written in Fortran 90 and uses MPI to synchronize multiple instances of the ES. The quality function can be parallelized by the user, independent from the library. The quality function must be independently evaluable by each process, it is not envisaged that the quality function can only be evaluated by all processes together.\\

Approaches using Map-Reduce (hadoop) for least-squares fitting seem to exist, but no reliable scientific papers could be found.

\section{Summary and Outlook}
\label{sec:outlook}
The motivation for this work was to study the feasibility of the search for signals of gravitational waves in high-precision astrometric data. For such a search, a large amount of data must be processed and fitted to a complex and substantially nonlinear model. The two main questions regarding the feasibility are: can such a search be done on existing or future HPC systems? And, how far under the noise can a signal be hidden and still being detected?\\

To answer these questions, different mathematical methods for model fitting have been presented in this work and evaluated regarding the usability for the GW search and regarding the usability in massive parallel HPC systems. As the result of this evaluation a hybrid approach (linear search + evolution strategy + gradient method) has been proposed to determine the existence of a wave as well as their parameters. A linear search was used to search the frequency space for evidence of a wave at a given frequency. During this search the hybrid least-squares fitter approximates values for the wave directions as well as the amplitudes of a possible wave at the frequency. The final determination of wave parameters was done by a multi-iteration, ``classical'', nonlinear Gauss-Newton least-squares fitter.\\

A prototype which uses these methods has been implemented and tested. This prototype allowed---for the first time---a successful detection of a gravitational wave signal in simulated data and the recovering of its parameters with sufficiently good accuracy. It has been shown that the prototype scales up to the maximal number of CPUs which have been available for tests (4096 at the ZIH machine Taurus), and will scale most probably far beyond this number. It can be concluded that it is feasible to conduct a search on the real dataset with the help of much larger HPC systems than those used in the tests.

\subsection*{Outlook}
A lot of development and research is needed to implement a search with real mission data. But, the prototype presented in this work is a first step and can act as a demonstrator and the basis for further work. A future search-software must be able to deal with additional effects, such as a much more complicated noise model, and presumably other wave models too. The prototype can also be used away from this Gaia-specific purpose, since it provides a generic interface which is not limited to this problem. It is conceivable that the combination of algorithms used in this work can be used to find periodic signals in other data as well.

\appendix

\chapter{Appendices}
\section{Pictures of the Gaia Satellite}
\label{sec:apx:gaiaSatPictures}
\begin{figure}[H]
 \centering
 \includegraphics[keepaspectratio, width=0.9\textwidth]{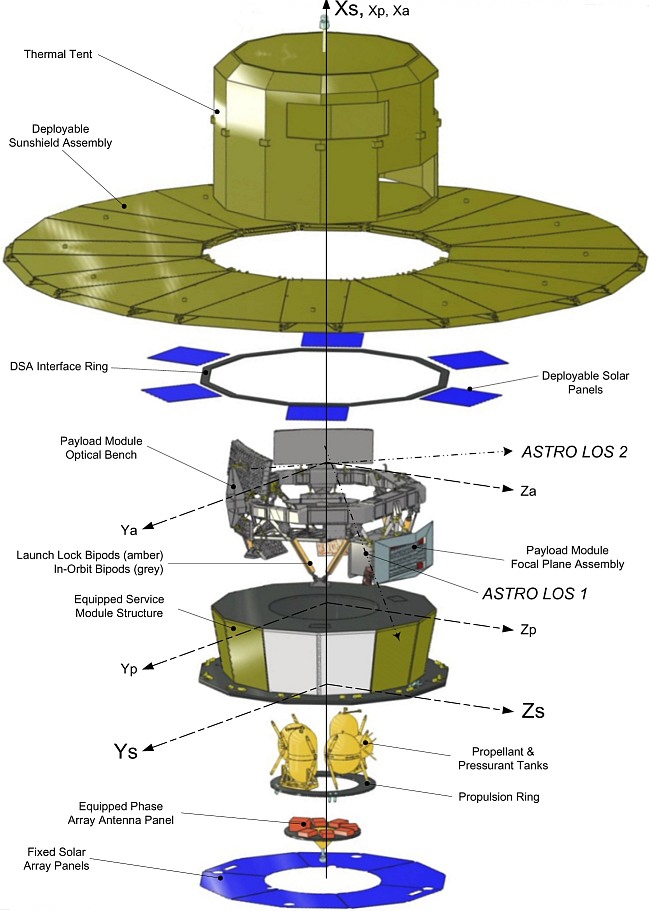}
 \caption[Explosion schematic of Gaia]{Explosion schematic of Gaia. Picture taken from: \url{http://sci.esa.int/jump.cfm?oid=45337}, credit and copyright: EADS Astrium \label{fig:satPictureExploded}}
\end{figure}

\begin{figure}[H]
 \centering
 \includegraphics[keepaspectratio, width=0.95\textwidth]{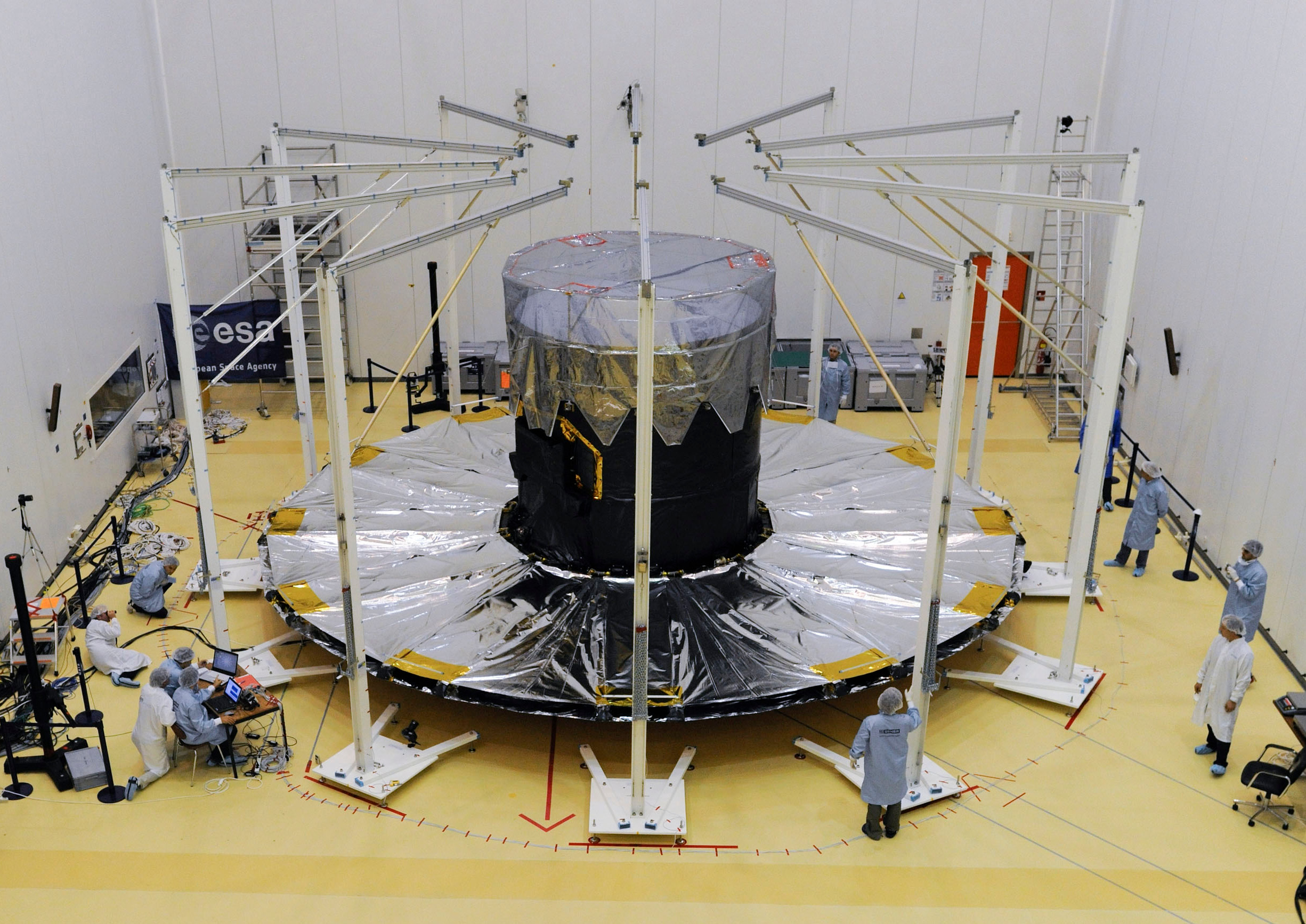}
 \caption[Picture of the Gaia satellite in the integration building.]{Gaia with the deployed sun shield during testing at the spaceport in Kourou. Picture taken from \url{http://sci.esa.int/jump.cfm?oid=53112}, credit and copyright: 10 October 2013, ESA / M. Pedoussaut \label{fig:satPicture}}
\end{figure}

\newpage

\section{ES Game Card Symbols}
\label{sec:apx-es-game-cards}
\begin{longtable}{m{0.18\textwidth}m{0.7\textwidth}}
 \toprule\\
 Game Symbols																				& Description \\
 \midrule\\
 \includegraphics[keepaspectratio,width=0.15\textwidth]{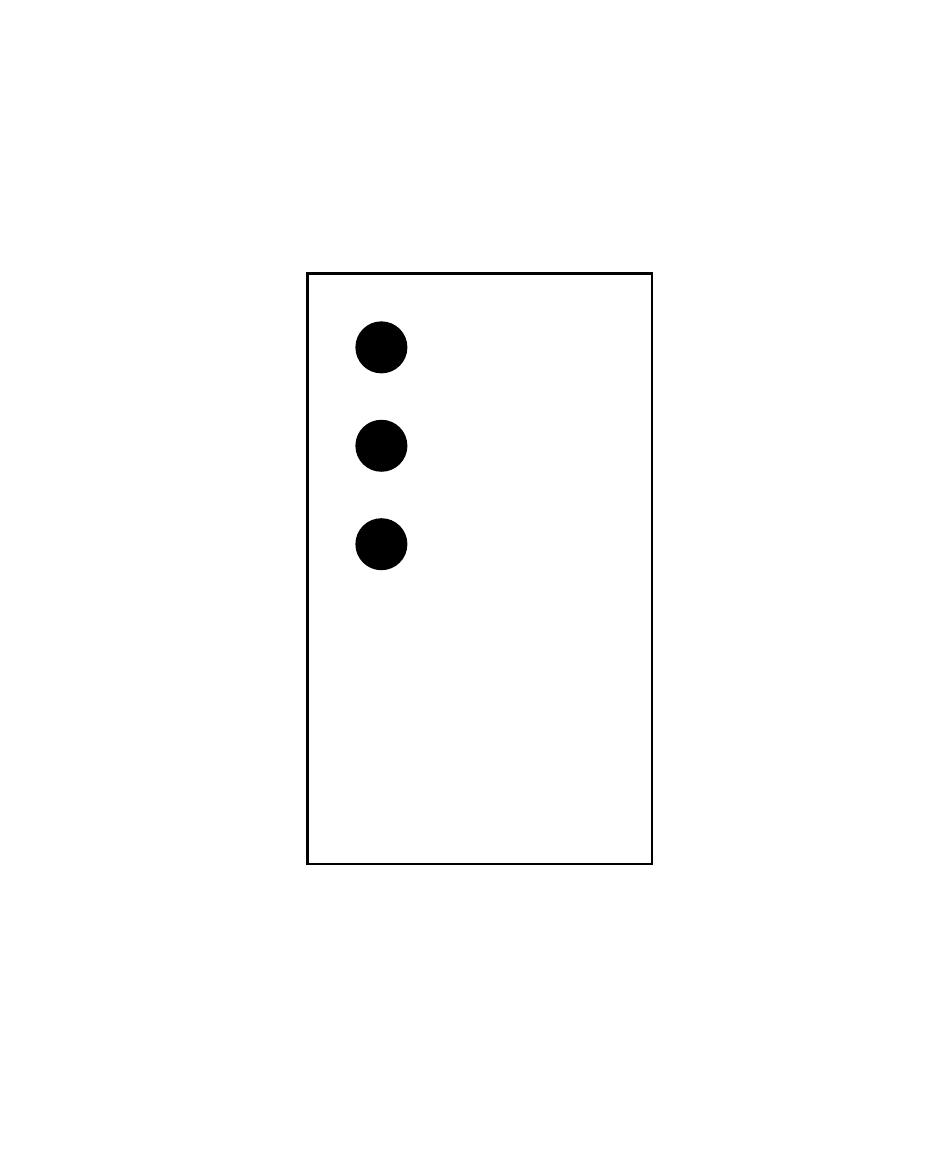}				& One individual, with its parameters (in nature the DNA).\\
 \includegraphics[keepaspectratio,width=0.15\textwidth]{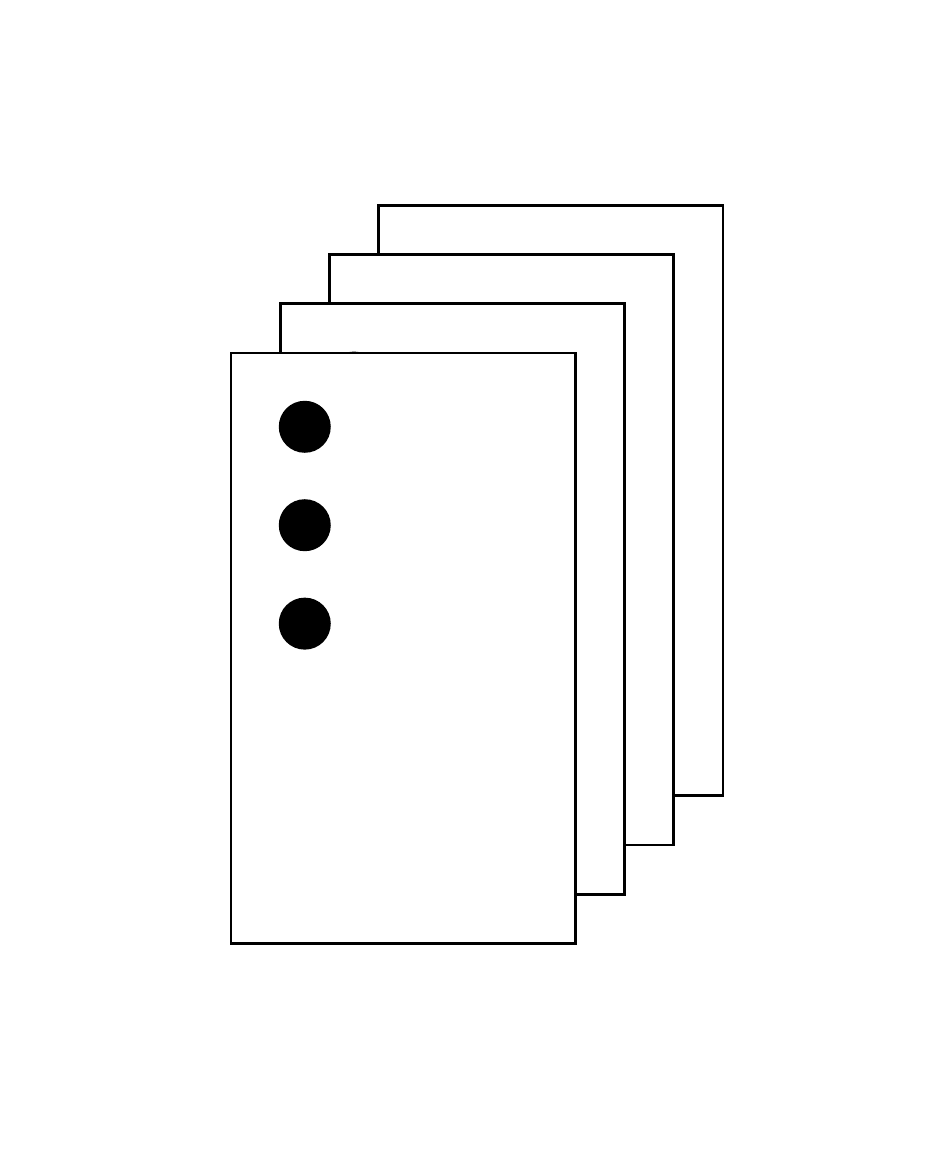}				& A population of individuals.\\
 \includegraphics[keepaspectratio,width=0.15\textwidth]{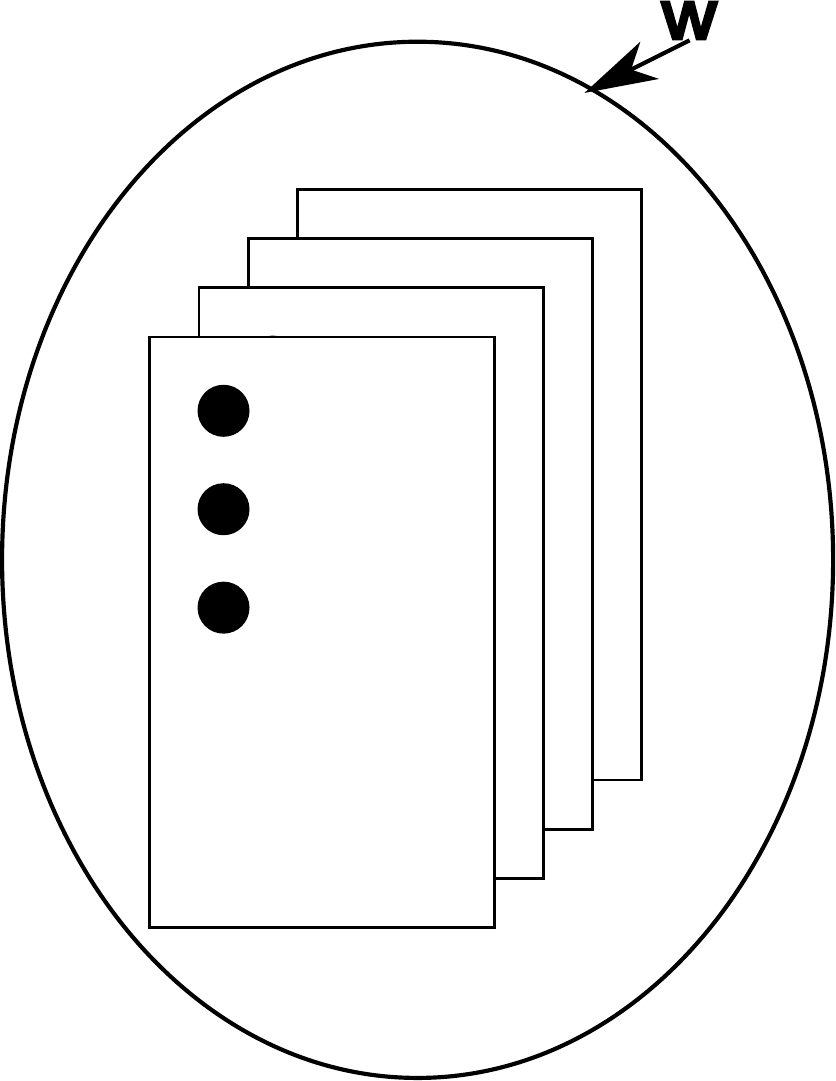}				& Random selection from the population pool. W individuals get selected if one population is in the circle. If more than one population is enclosed, W populations get selected. The random variable is normally distributed by default.\\
 \includegraphics[keepaspectratio,width=0.15\textwidth]{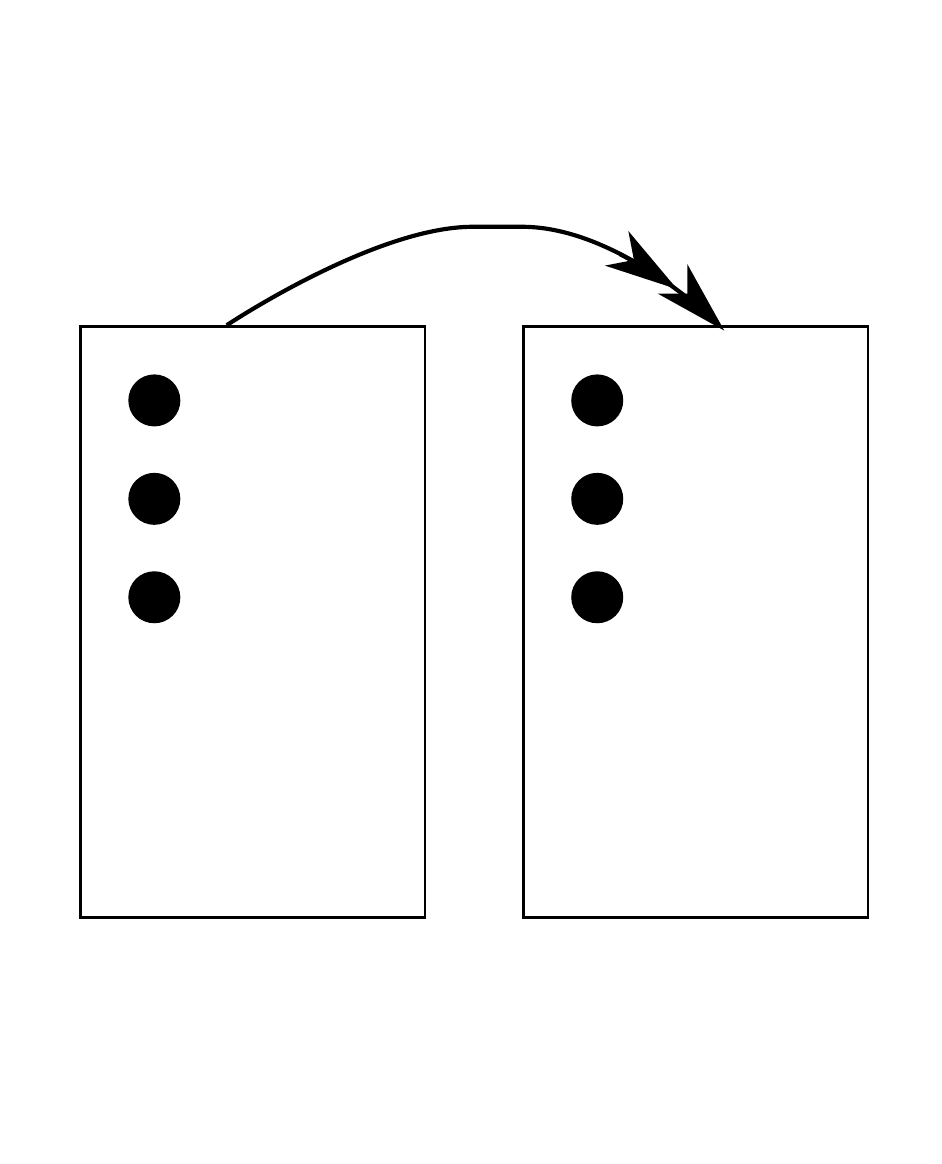}				& Duplication of a individual. This symbol is often used together with the mutation, since offspring is most of the time a copy of the parent but slightly mutated.\\
 \includegraphics[keepaspectratio,width=0.15\textwidth]{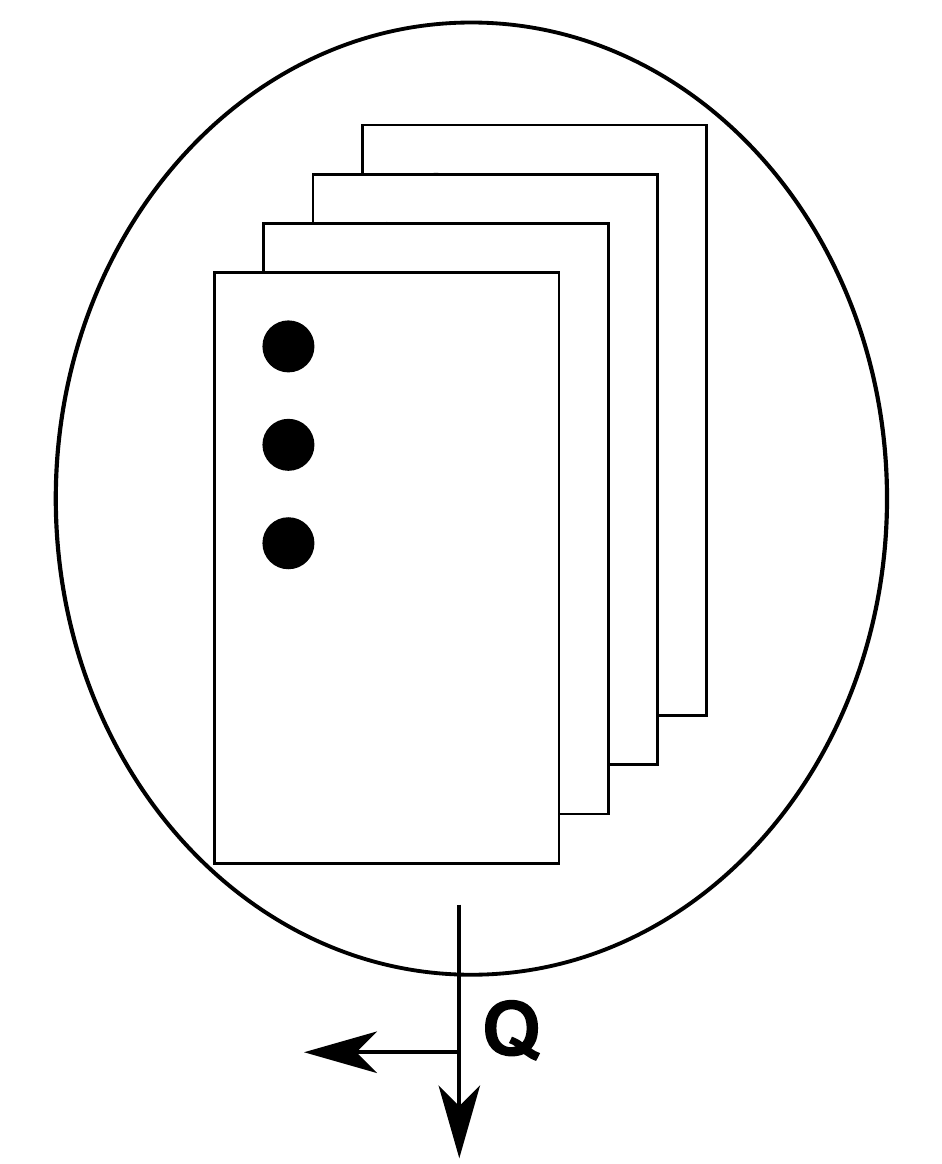}				& The best individual according to the quality function $Q$ gets selected from the population. If more than one population is enclosed, the best population gets selected with all its individuals.\\
 \includegraphics[keepaspectratio,width=0.15\textwidth]{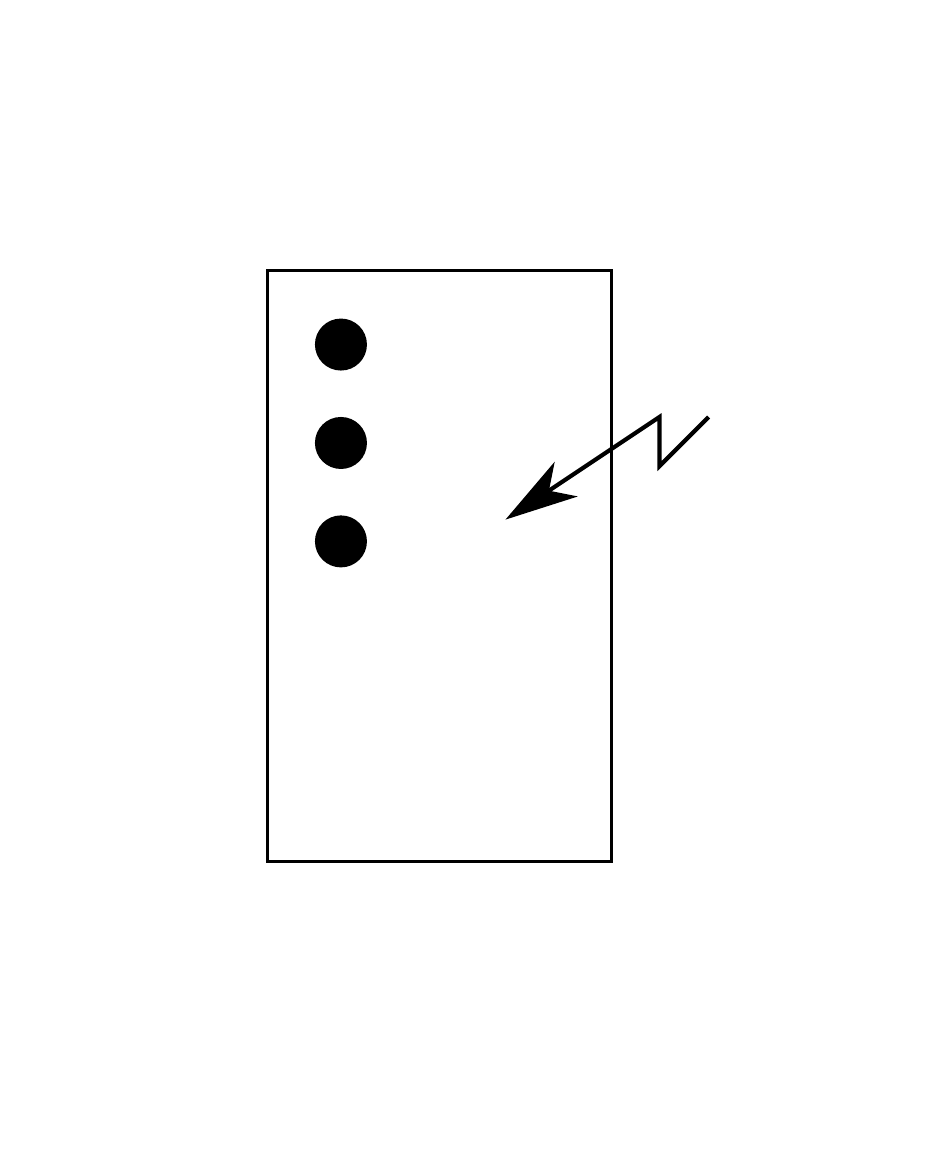}					& A individual gets mutated, an arbitrary number of parameters get changed in a random way. The randomness can be influenced, for instance with damped in every step.\\
 \includegraphics[keepaspectratio,width=0.15\textwidth]{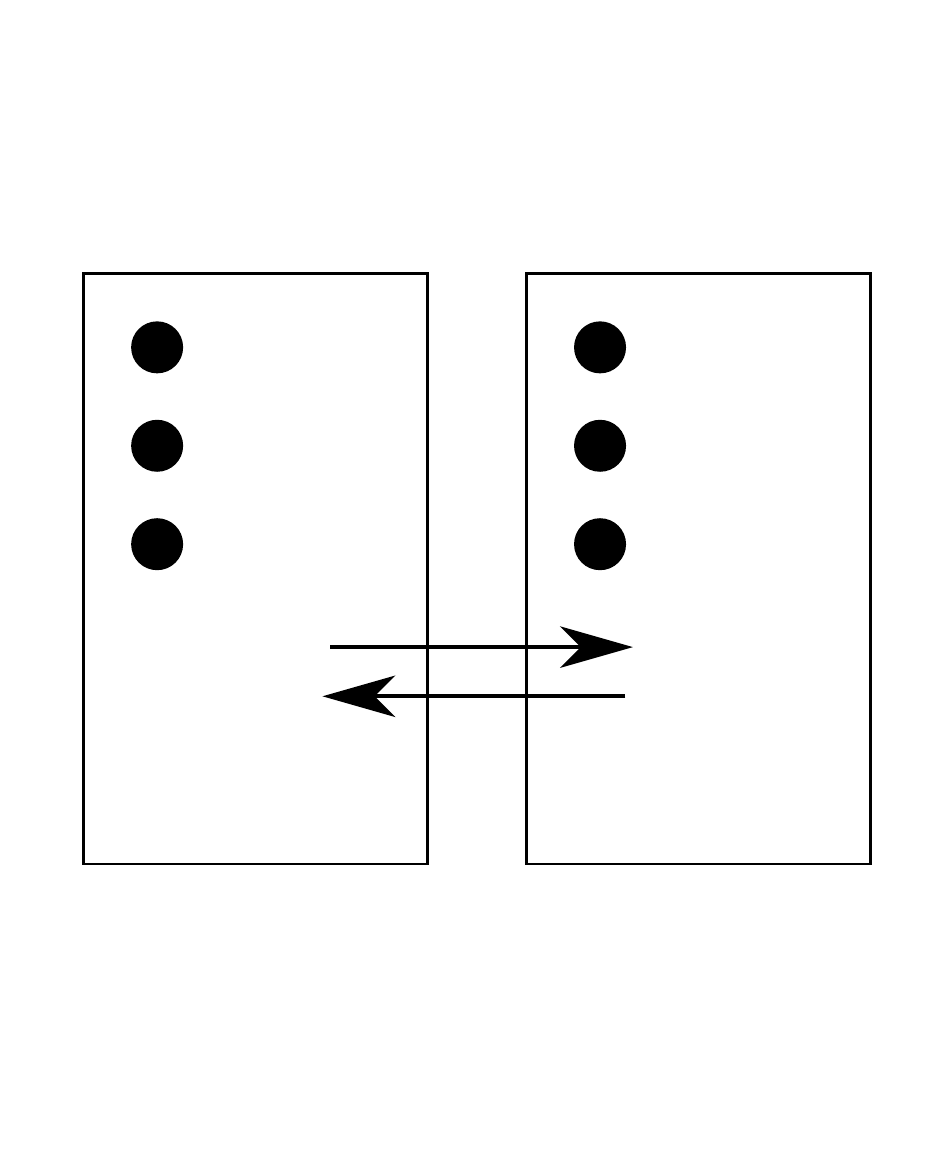}			& Recombination: Two individuals mixing there parameters. This can be compared to sexual reproduction. A recombination function can, and should, be given.\\
 \includegraphics[keepaspectratio,width=0.15\textwidth]{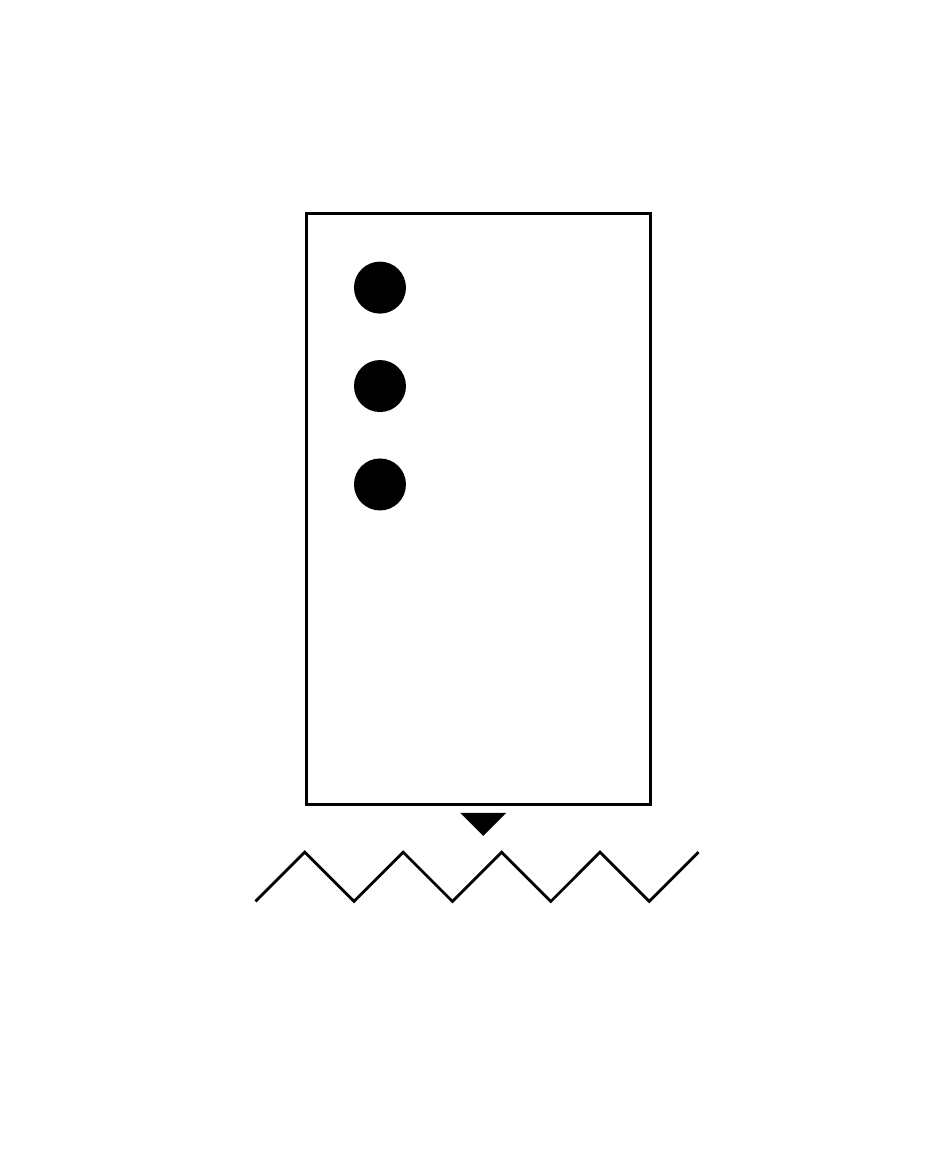}				& The parameters of the individuals (noted on the playing cards) get realized. This marks the transition from genotype to phenotype. This is closely related to the quality function $Q$, in almost all cases we will consider the quality of the phenotype. The quality of the genotype can not be evaluated easily. So, in order to measure the quality, the phenotype of one individual's genotype (parameters) has to be established.\\
 \includegraphics[keepaspectratio,width=0.15\textwidth]{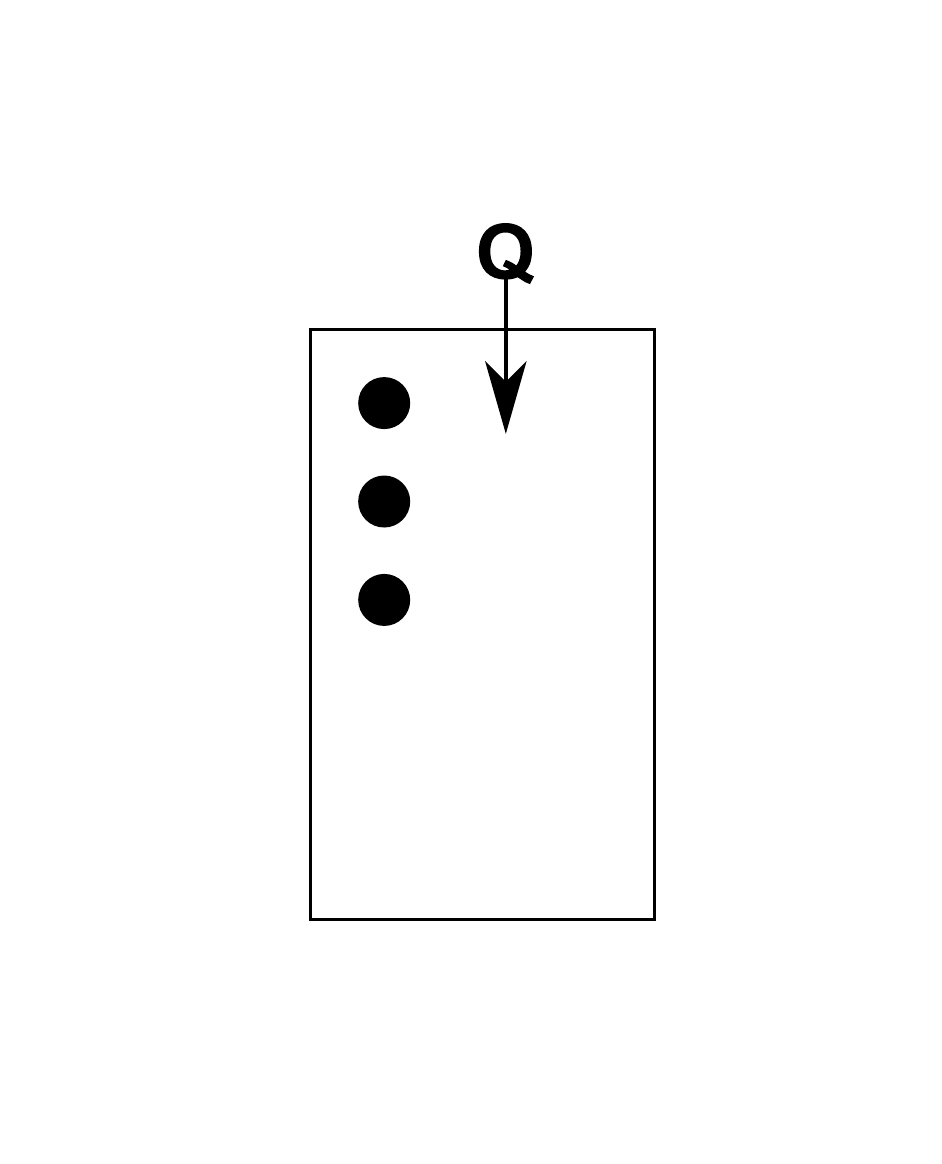}					& The quality of the individual gets noted on the card. This is just for saving this value, a selection is not done, but it can be later done using this value.\\
 \includegraphics[keepaspectratio,width=0.15\textwidth]{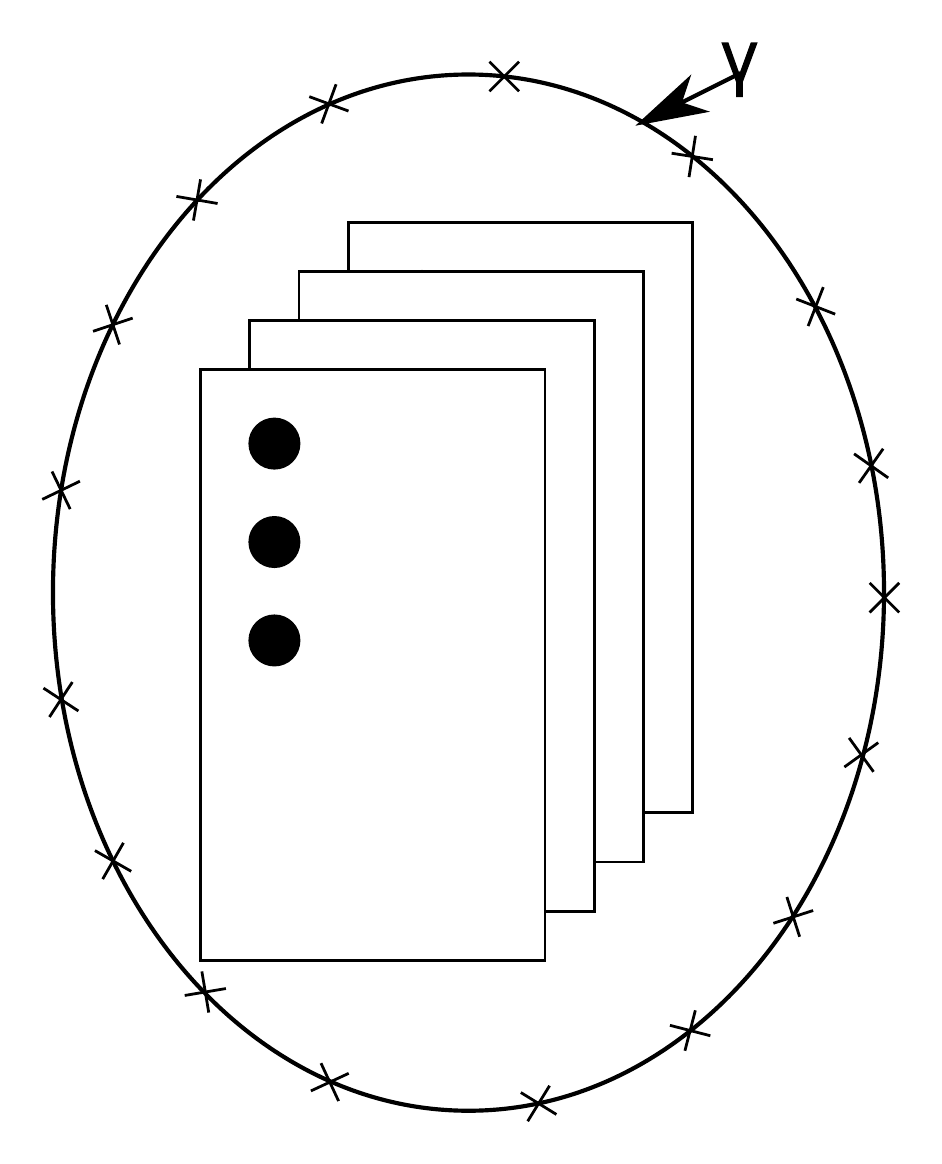}				& A population of individuals is isolated for $\gamma$ steps, or time. Processes in the encircled area  are completely independent for $\gamma$ steps.\\
 \bottomrule\\
\caption{Game symbols for the basic evolution strategy, shortened taken from: \cite{rechenberg1581evolutionsstrategie}.} \label{tab:ES-game-symbols} \\
\end{longtable}

\newpage

\section{Nonlinear Model Structure}
\label{sec:apx:model}
\begin{lstlisting}[style=CA,caption={The \texttt{nonlinearModel} type is used to pass informations about a (nonlinear) model to the fitting routines.},label={lst:apx:model}]
typedef struct nonlinearModel {
    int numFreeParameters;        // # parameters to fit
    int *fixedParams;             // bit mask to prevent updating during fit
    double *paramFactors;         // factors for design matrices, for
                                  // numerical stability
    double *freeParametersValues; // values of the parameters
} nonlinearModel;
\end{lstlisting}

\section{Data Index Structure}
\label{sec:apx:index}
\begin{lstlisting}[style=CA,caption={Data index structure, here all ``meta-data'' for the data values are stored},label={lst:apx:index}]
typedef struct dataIndex {
    uint64_t size;         // number of datapoints per process
#ifdef GAIAGW
    Observation *idxObs;   // Gaia specifc data field
#endif
    double *idxDbl;        // generic vector of observations
} dataIndex; 
\end{lstlisting}

\section{Fit Result Structure}
\label{sec:apx:fitresult}
\begin{lstlisting}[style=CA,caption={The \texttt{FitResult} type holds all statistical informations from a Gauss-Newton fit},label={lst:apx:fitresult}]
typedef struct FitResult {
    int n;                        // size of arrays
    double *resultParameters;     // model parameters fitted (size = n)
    double *startParameters;      // pre fit, starting points
    double *lastRefinement;       // refinement during last iteration
    double *formalStdDeviations;  // SQRT(f_ii) (size = n)
    double *postFitStdDeviations; // formal std-deviations * Post Fit Sigma
                                  // == SQRT(f_ii) * Chi^2_postfit (size = n)
    double *correlationMatrix;    // formal correlation matrix (size = n*n)
                                  // = n_ji / (sqrt(n_ii) * sqrt(n_jj))
    double *SVDvector;            // result of SVD
    double condNumber;            // Condition number max(svd)/min(svd)
    double ChiSqPF;               // post fit Chi Square
    double wPreFitRMS;            // weighted pre fit RMS
    double wPostFitRMS;           // weighted post fit RMS

    double preFitResidue;         // r^2 = SUM (O-C/sigma)^2
    double postFitResidue;        // n^2 = r^2 - B.s
    double sigmaWeight;           // what we call "w"
    long long unsigned int numDataPoints;
                                  // number of ALL datapoints over all CPUs
} FitResult;
\end{lstlisting}


\newpage

\section{Full Profile}
\begin{table}[H]
 \centering
 \begin{tabular}{lrrr}
 \toprule
                                & \textbf{Incl. Time}            & \textbf{Incl. Time}          & \textbf{Incl. Time}\\
  \textbf{Function}                      & \textbf{4 Parameters}              & \textbf{5 Parameters}            & \textbf{7 Parameters} \\
  \midrule
    vecVecDot3d         &582.119        &682.192        &967.058\\
    vecScale3d          &606.227        &718.682        &1046.72\\
    vecVecAdd3d         &489.311        &587.89         &979.075\\
    matMatMul3d         &626.956        &628.307        &2006.11\\
    matScale3d          &328.504        &495.833        &883.274\\
    matVecMul3d         &261.713        &316.461        &474.361\\
    matTranspose3d      &218.124        &216.576        &472.742\\
    matMatAdd3d         &56.4742        &111.327        &494.792\\
    \midrule
    capitalP            &126.8          &130.084        &257.367\\
    smallP              &147.741        &153.409        &160.663\\
    getRz               &-              &-              &271.018\\
    getRx               &-              &-              &189.777\\
    localTriad          &-              &-              &136.159\\
    PGW\_getPhase        &100.655        &99.7269        &100.681\\
    computeCorrection   &3711.1         &4459.17        &4453.7\\
    getEHVFromAD2       &333.378        &338.531        &347.196\\
    getCelestCoordFromVec&186.885       &187.132        &188.754\\
    getDeltaUALACFromTheta&443.744      &531.784        &717.109\\
    \midrule
    getPGW\_d\_hTimesSin  &935.53         &934.447        &932.744\\
    getPGW\_d\_hTimesCos  &936.502        &943.212        &938.625\\
    getPGW\_d\_hPlusSin   &961.402        &961.184        &955.095\\
    getPGW\_d\_hPlusCos   &989.173        &996.333        &1004.8\\
    getPGW\_d\_Omega      &-              &1264.44        &1281.78\\
    getPGW\_d\_aGWdGW     &-              &-              &7932.27\\
    getPGWDistortion    &2091.68        &2086.52        &2080.54\\
    getPGWALDistortion  &2660.99        &2656.3         &2649.33\\
    getPGWJacobian      &6326.12        &7787.81        &16182.4\\
    \midrule
    getSumSqSigmaGaia   &58.3727        &58.3684        &58.3666\\
    nonlinearGaussNewtonFit&9977.37     &11753.2        &20645.4\\
    MPI\_Allreduce       &856.066        &1166.4         &1651.42\\
  \bottomrule
 \end{tabular}
 \caption{Detailed profiling results\label{apx:tab:vt_trace_300k}}
\end{table}

\newpage

\section{ES Output Histogram}
\begin{figure}[H]
\centering
\includegraphics[keepaspectratio,width=0.95\textwidth]{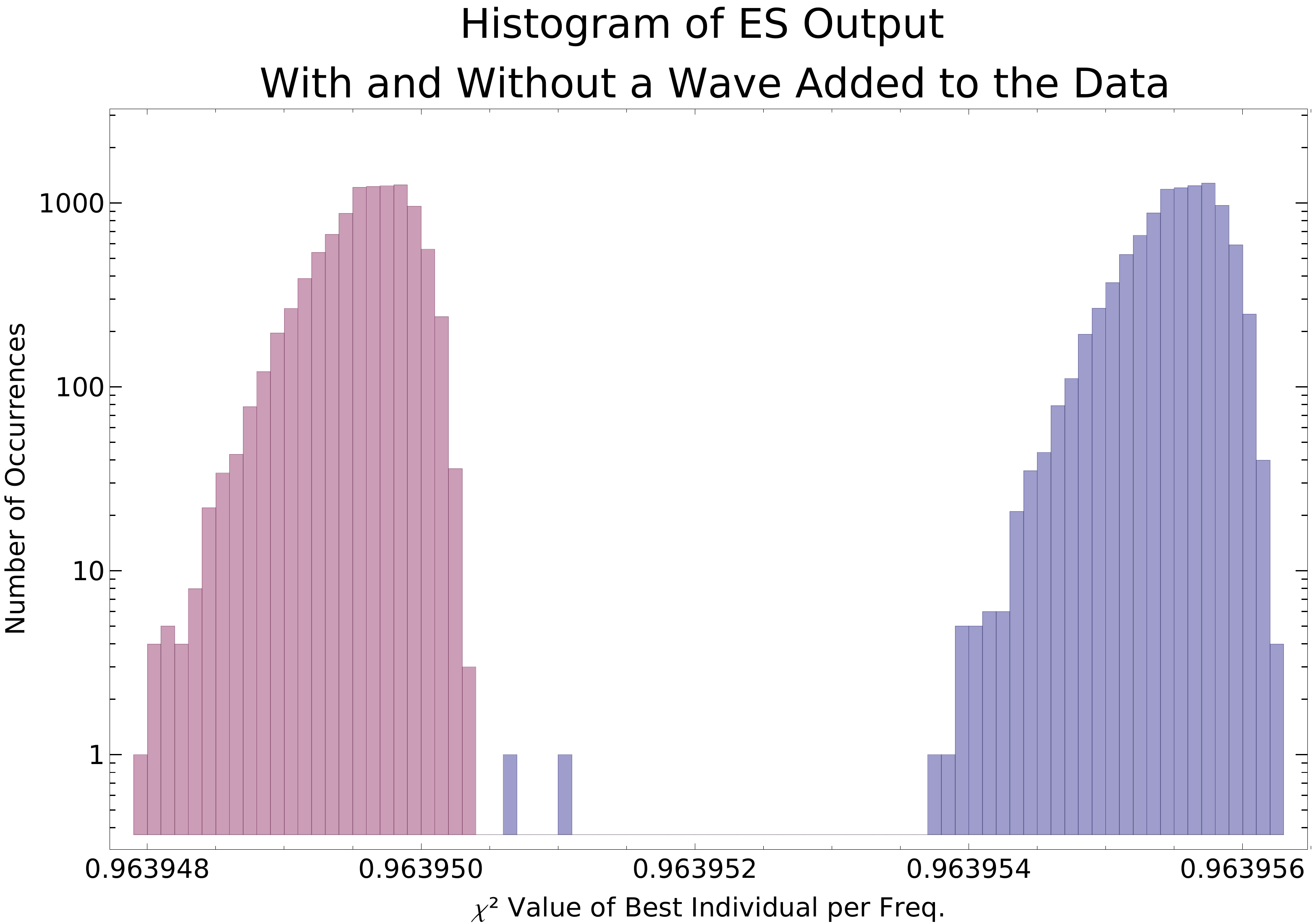}
\caption[Histogram of ES output with and without wave added to data]{Histogram of the ES output ($\chi^2$-value) for the best individuals per frequency. Right, blue: the histogram for the ES run on data with a wave added (same as Figure \ref{fig:ES_chi2_histo}). Left, red: the histogram for the ES run on data without a wave. \label{apx:histo_both}}
\end{figure}

\newpage

\chapter{Bibliography}

\chapter{Acknowledgments}
I would particularly like to thank Prof. Sergei A. Klioner, Prof. Wolfgang E. Nagel and Dipl.-Inf. Thomas William for making this thesis possible. I am very grateful to Prof. Nagel for making it once again possible to work interdisciplinary and for providing the HPC systems to conduct this work. I am also very grateful to Prof. Klioner for the great support and guidance during this work and for the many hours of discussions and explanations. Many thanks also for the opportunity to join the Gaia group at the Gaia launch event at ESOC. My special thanks also go to Thomas William for mentoring me, thank you for the support, the fruitful discussions and all the help.

I would also like to thank the team of the Lohrmann-Observatory, especially the Gaia team. Many thanks to Hagen Steidelmüller for adapting AGISLab to my needs. Many thanks also to Alex Bombrun and Enrico Gerlach for the help with mathematical problems and the many insightful discussions.

I would like to thank everyone from the ``1038'' student laboratory for the many years of friendship and mutual help. I learned so much from you all. I also owe my thanks to Michael Kluge for the explanations regarding Lustre and IO in general, and Daniel Molka for the help with PAPI.

Finally, I owe my most profound thanks to my family and my fiancée Mandy for their relentless support, even in the most difficult times.

\end{document}